\documentclass[
  journal=largetwo,
  manuscript=research-paper, 
  year=202x,
  volume=YY,
]{cup-journal}

\usepackage{amsmath}
\usepackage[nopatch]{microtype}
\usepackage{booktabs}
\usepackage{subcaption}
\usepackage{subcaption}
\DeclareUnicodeCharacter{0084}{}
\usepackage{hyperref} 
\usepackage{amsmath,amssymb}
\hypersetup{colorlinks,citecolor=blue,linkcolor=blue,urlcolor=blue}
\usepackage{orcidlink}

\title{Evolutionary Map of the Universe: A pilot survey to detect high Galactic latitude pulsars in variance images with ASKAP}

\author{A. Ahmad\,\orcidlink{0000-0002-0457-3661}}
\affiliation{School of Science, Western Sydney University, Locked Bag 1797, Penrith, NSW 2751, Australia}
\email[A. Ahmad]{22031320@student.westernsydney.edu.au}

\author{S. Dai\,\orcidlink{0000-0002-9618-2499}}
\affiliation{Australia Telescope National Facility, CSIRO, Space and Astronomy, P.O. Box 76, Epping, NSW 1710, Australia}
\alsoaffiliation{School of Science, Western Sydney University, Locked Bag 1797, Penrith, NSW 2751, Australia}

\author{E. Lenc\,\orcidlink{0000-0002-9994-1593}}
\affiliation{Australia Telescope National Facility, CSIRO, Space and Astronomy, P.O. Box 76, Epping, NSW 1710, Australia}

\author{M. D. Filipovi\'{c}\,\orcidlink{0000-0002-4990-9288}}
\affiliation{School of Science, Western Sydney University, Locked Bag 1797, Penrith, NSW 2751, Australia}

\author{B. S. Koribalski\,\orcidlink{0000-0003-4351-993X}}
\affiliation{Australia Telescope National Facility, CSIRO, Space and Astronomy, P.O. Box 76, Epping, NSW 1710, Australia}
\alsoaffiliation{School of Science, Western Sydney University, Locked Bag 1797, Penrith, NSW 2751, Australia}

\author{S. Johnston\,\orcidlink{0000-0002-7122-4963}}
\affiliation{Australia Telescope National Facility, CSIRO, Space and Astronomy, P.O. Box 76, Epping, NSW 1710, Australia}

\author{G. Hobbs\,\orcidlink{0000-0003-1502-100X}}
\affiliation{Australia Telescope National Facility, CSIRO, Space and Astronomy, P.O. Box 76, Epping, NSW 1710, Australia}

\author{S. W. Duchesne\,\orcidlink{0000-0002-3846-031}}
\affiliation{ Australia Telescope National Facility, CSIRO, Space and Astronomy, PO Box 1130, Bentley WA 6102, Australia}

\author{S.~Lazarevi\'c\,\orcidlink{0000-0001-6109-8548}}
\affiliation{School of Science, Western Sydney University, Locked Bag 1797, Penrith, NSW 2751, Australia}
\alsoaffiliation{Australia Telescope National Facility, CSIRO, Space and Astronomy, P.O. Box 76, Epping, NSW 1710, Australia}
\alsoaffiliation{Astronomical Observatory, Volgina 7, 11060 Belgrade, Serbia}

\author{J. T. Bai,\orcidlink{0000-0002-1052-1120}}
\affiliation{Institute for Gravitational Wave Astronomy, Henan Academy of Sciences, Zhengzhou 450046, Henan, People’s Republic of China}

\author{L.~Toomey\,\orcidlink{0000-0003-3186-3266}}
\affiliation{Australia Telescope National Facility, CSIRO, Space and Astronomy, P.O. Box 76, Epping, NSW 1710, Australia}

\author{N. D. R. Bhat\,\orcidlink{/0000-0002-8383-5059}}
\affiliation{International Centre for Radio Astronomy Research, Curtin University, Bentley, WA 6102, Australia}

\author{D. A. Leahy\,\orcidlink{0000-0002-4814-958X}}
\affiliation{Department of Physics and Astronomy, University of Calgary, 2500 University Drive NW, Calgary, AB T2N 1N4, Canada}

\author{A. M. Hopkins\,\orcidlink{0000-0002-6097-2747}}
\affiliation{School of Mathematical and Physical Sciences, 12 Wally's Walk, Macquarie University, NSW 2109, Australia}
\alsoaffiliation{Macquarie University Astrophysics and Space Technologies Research Centre, Sydney, NSW, Australia}

\author{T. Zafar\,\orcidlink{0000-0003-3935-7018}}
\affiliation{School of Mathematical and Physical Sciences, 12 Wally's Walk, Macquarie University, NSW 2109, Australia}
\alsoaffiliation{Macquarie University Astrophysics and Space Technologies Research Centre, Sydney, NSW, Australia}

\author{S. F. Rahman\,\orcidlink{0000-0001-9414-175X}}
\affiliation{Syed Babar Ali School of Science and Engineering, Lahore University of Management Sciences, Lahore, Pakistan}


\begin{document}

\begin{abstract}

It has been proposed that radio pulsars can be distinguished from other point-like radio sources in continuum images by their unique interstellar scintillation signatures. Using data from the Australian Square Kilometre Array Pathfinder (ASKAP) Evolutionary Map of the Universe (EMU) survey, we conducted a pilot survey of radio pulsars at high Galactic latitude regions via the variance imaging method. Out of approximately 59,800 compact radio sources detected in a $\sim$480 square degree survey area, we identified 20 highly variable sources. Among them, 9 are known pulsars, 1 is a known radio star, 1 is an ultra-long period source, 3 are radio star candidates, and the remaining 6 are pulsar candidates. Notably, we discovered two strongly scintillating pulsars: one with a period of 85.707\,ms and a dispersion measure (DM) of 19.4\,cm$^{-3}$\,pc, and another with a period of 5.492\,ms and a DM of 29.5\,cm$^{-3}$\,pc. In addition, a third pulsar was discovered in the variance images, with a period of 14.828\,ms and a DM of 39.0\,cm$^{-3}$\,pc. This source shows a steep radio spectrum and a high degree of circular polarisation.
These results underscore the strong potential of variance imaging for pulsar detection in full EMU and future radio continuum surveys planned with Square Kilometre Array (SKA).
\end{abstract}

\section{Introduction}
%
%

Radio signals from pulsars undergo multipath propagation effects as they travel through the ionised interstellar medium (ISM) due to turbulence and electron density fluctuations. This leads to several observable effects, including scattering and scintillation -- the modulation of the pulsar's intensity across time and frequency. In the strong scattering regime -- typical for pulsars observed at $\lesssim$GHz frequencies -- two scintillation time-scales arise: the diffractive ($\tau_{\rm dif}$) and refractive ($\tau_{\rm ref}$) interstellar scintillation (DISS and RISS, respectively). 
While the time-scales of RISS are typically long ($\sim$months), at $\sim$1\,GHz, the time-scales of DISS are on the order of minutes, with characteristic bandwidths of a few MHz. As a result, DISS is particularly useful for identifying pulsars in radio continuum surveys conducted at $\sim$1\,GHz with integration times ranging from one to several hours.
Comprehensive reviews of interstellar scattering theory and related observations are provided by \citet{ric90} and \citet{nar92}.

Previously, it has been proposed that DISS of radio pulsars can be used to distinguish them from other compact radio sources in radio continuum surveys. For example, \citet{djb16} proposed to use the variance imaging technique to identify potential pulsar candidates in radio continuum observations by leveraging their DISS-induced flux density variability in time- and frequency-resolved radio images.
The sensitivity of variance imaging depends critically on matching the image resolution to the scintillation timescales and bandwidths of pulsars. If the time and frequency resolution are too low, DISS is smeared out; if too fine, background noise dominates the variance due to increased noise level in the subintegrations and channels. By optimising the number of subintegrations and channels, variance imaging provides a promising complement to traditional pulsar searches, enhancing the discovery potential in large-scale radio continuum surveys~\citep{dai+17}. 
More recently, \citet{salal+24} demonstrated a technique by stacking phased visibilities to form dynamic spectra and measuring their scintillation parameters using the upgraded Giant Metrewave Radio Telescope (uGMRT). 

One of such deep and all-sky surveys is Evolutionary Map of the Universe \citep[EMU;][]{andrew+25} which is being conducted by the Australian Square Kilometre Array Pathfinder (ASKAP) telescope. The capability of finding new pulsars with ASKAP surveys has been readily demonstrated by the discovery of pulsars originally identified as highly polarised radio sources \citep{kaplan+19,wang+24,wus+25}, pulsars associated with supernova remnants (SNRs) and pulsar wind nebulae (PWNe) \citep[e.g.,][]{ahmad+25,sanja+24}, and highly scattered pulsars \citep[e.g.,][]{wang+24,ldj+24}. Variance imaging with EMU will be more sensitive than current pulsar surveys at high Galactic latitudes and is expected to discover $\sim\!40$ new millisecond pulsars (MSPs) and $\sim\!30$ new normal pulsars \citep{dai+17}. 
Image-based pulsar searches have also used spectral signature targeting steep-spectrum sources, polarisation, and variability in time using the Murchison Widefield Array (MWA) and MeerKAT continuum surveys~\citep[e.g.,][]{frm+18, ian+23, frail+24, sett+25, manto+25}.

The ASKAP telescope is also conducting the Variables and Slow Transients survey specifically to search for radio transients and variable sources on minute timescales \citep[VAST,][]{2013PASA...30....6M,2021PASA...38...54M,tao+23}. Variability studies on the first EMU Pilot survey show that the full EMU survey has great potential to identify such transients and variable sources \citep{wng+23}. 
More recently, several ultra-long period (ULP) sources~\citep{hzb+22,hrm+23}, with periods of several minutes and repeating bursts of coherent radio emission, have also been reported 
with ASKAP \citep[e.g.,][]{caleb+24,dobie+24,lcm+25,akr+25}. It is expected to discover more in the full EMU survey. 

In this paper, we present a first pilot survey of pulsars using variance imaging with the EMU data sets. We focus on high Galactic latitude regions where radio pulsars typically have lower dispersion measures (DM), resulting in broader scintillation bandwidths and longer scintillation timescales. As demonstrated by \citet{djb16}, the sensitivity of variance imaging is maximised when the time and frequency resolution match the pulsar's scintillation properties. This allows us to use lower time and frequency resolutions for variance image generation in these regions, significantly reducing the computational complexity and enabling efficient pilot studies over a wide area of the sky.
The organisation of this paper is as follows: in Section~\ref{sec1}, we describe ASKAP and EMU surveys and Murriyang, the Parkes radio telescope's observations and data analysis, followed by the description of variance imaging and source detection pipeline in Section~\ref{obs}. In Section~\ref{results}, we present our results, including the discovery of two new pulsars, summarise the properties of known test pulsars and new candidates. In Section~\ref{discuss}, we conclude and discuss the false alarms and potential problems, future improvements in the variance imaging pipeline, and application of this approach on future and ongoing radio continuum surveys, including the Galactic plane.
\section{Observations}\label{sec1}

\subsection{ASKAP observations}
ASKAP is a radio interferometer consisting of 36 dishes, each with 12\,m diameter, located at Inyarrimanha Ilgari Bundara, CSIRO's Murchison Radio-astronomy Observatory, Western Australia~\citep{hotan21}. Each dish is equipped with a phased array feed, which forms 36 dual polarisation beams on the sky, providing ASKAP with a wide field of view of $\sim$30 square degrees.
The EMU is an all-sky survey which aims to make deep radio continuum maps covering the whole southern sky $-90^{\circ}\leq DEC \leq +10^{\circ}$ \citep{andrew+25}. The survey is being conducted at a central frequency of 943.5\,MHz with a bandwidth of 288\,MHz. The total integration time for a typical observation is 10\,h, reaching a sensitivity of 20--30\,$\mu$Jy\,beam$^{-1}$. 
Visibilities are recorded to capture full-Stokes information in 10\,s integrations across 1\,MHz-wide channels.

The data processing was done with the ASKAPsoft package \citep{askap19}, including procedures such as calibration, flagging, and image generation. For each observation with a unique schedule block ID (SBID), ASKAPsoft produces calibrated visibilities for individual 36 primary beams in the form of complex values as a function of time, frequency, baseline, and correlations and stored in a measurement set (MS) format. A corresponding full-Stokes radio continuum image for the entire field of view is also produced. Total intensity catalogues and noise maps are generated with \textit{Selavy} \citep{whit+12}, and data are automatically uploaded to the CSIRO ASKAP Science Data Archive (CASDA\footnote{\url{https://research.csiro.au/casda/}}) and made publicly available after completing scientific validation.

As explained earlier, this paper focuses on high Galactic latitude regions. 16 EMU tiles at high Galactic latitudes were used for analysis (Table~\ref{tab:askap}), containing 31 known pulsars within the total field of view. We successfully detected 27 pulsars from 16 tiles within a DM range of $\sim$ 3--152\,pc\,cm$^{-3}$.
{Of the four pulsars not detected by EMU, three have measured flux densities at 1.4\,GHz. All three pulsars (J0555--7056, J0456--7031 and J1749--4931) have flux densities $\lesssim0.1$\,mJy. Their flux densities are below the $\sim5\sigma$ sensitivity threshold of their corresponding EMU continuum images, explaining their non-detection. PSR~J1216--50 is a rotating radio transient, and only single pulses have been detected~\citep{bbj+11}.}

\begin{table*}
\caption{16 high Galactic latitude EMU tiles are listed in this table. Columns include schedule block ID (SBID), EMU tile name, central coordinates (RAJ $\&$ DECJ),  Galactic longitude and latitude (Gl $\&$ Gb), number of compact sources in the tiles (N$_{\rm c}$), known pulsars along with respective DMs taken from ATNF Pulsar Catalogue \citep{psrcat+05}, and pulsar flux densities at 943.5\,MHz from EMU. Each tile spans a duration of 10 hours, except the SB 61946, with a total observing duration of five hours. Tiles without a listed pulsar contain no known pulsars in their field of view.}     
\label{tab:askap} 
\centering     
\setlength{\tabcolsep}{7pt}
\renewcommand{\footnoterule}{}  
\begin{tabular}{l c c c c c c c c c}
\hline\hline                   
SBID & Tile name & RAJ & DECJ & Gl & Gb & N$_{\rm c}$ & PSR & DM & S$_{943.5}$ \\
   & & (hms) & (dms)  & (degree)  & (degree)   &   &  & (pc\,cm$^{-3}$)  & (mJy) \\
\hline
49990 & EMU\_0242-55 & 02:41:59.408 & --55:43:42.413 & 275.71 & -55.16 & 4519 & J0255--5304 & 17.10 & 12.96 $\pm$ 0.10 \\

46978 & EMU\_0526-73 & 05:26:23.646 & --73:39:39.342 & 284.92 & --31.88 & 4159 & J0536--7543 & 18.6 & 31.70 $\pm$ 0.26 \\

 & &  & & & &  & J0555--7056 & 72.9 & \ldots \\

 & & & & & &  & J0540--7125 & 29.4 & 0.73 $\pm$ 0.07 \\

 & & & & & &  & J0456--7031 & 100.3 & \ldots \\

72176 & EMU\_0941-75 & 09:41:15.275 & --75:45:58.717 & 291.98 & --17.12 & 4190 & J0904--7459 & 49.20 & 8.60 $\pm$ 0.08 \\ 

72194 & EMU\_1215-46 & 12:14:59.709 & --46:30:21.660 & 296.42 & 15.90 & 4077 & J1231--4609 & 66.55 & 1.04 $\pm$ 0.08 \\

& &  & & & &  & J1232--4742 & 26 & 1.33 $\pm$ 0.09 \\

64419 & EMU\_1216-51 & 12:16:21.393 & --51:07:18.395 & 297.33 & 11.36 & 4162 & J1216--50 & 110 & \ldots \\

& &  & & & &  & J1207--5050 & 50.64 & 1.02 $\pm$ 0.06 \\

73186 & EMU\_1241-41 & 12:41:32.129 & --41:52:52.083 & 300.95 & 20.05 & 4033 & J1240--4124 & 43.22 & 0.61 $\pm$ 0.05 \\

70283 & EMU\_1249-51 & 12:49:05.029 & --51:07:20.395 & 302.55 & 11.74 & 4015 & J1244--5053 & 109.95 & 0.47 $\pm$ 0.05 \\

&  & & & &  & & J1236--5033 & 105.01 & 0.76 $\pm$ 0.06 \\

62581 & EMU\_1415-46 & 14:14:59.709 & --46:30:19.660 & 317.58 & 13.98 &  & J1405--4656	& 13.88 & 1.15 $\pm$ 0.07 \\

50539 & EMU\_1513-69 & 15:13:48.295 & --69:29:26.829 & 314.90 & --9.10 & 4792 & J1502--6752 & 152.2 & 1.55 $\pm$ 0.05 \\

&  & & & & & & J1507--6640 & 130.2 & 1.54 $\pm$ 0.08 \\

&  & & & &  & & J1456--6843 & 8.6 & 152.36 $\pm$ 0.83 \\

74230 & EMU\_1708-60 & 17:08:33.478 & --60:19:44.170 & 330.03 & -11.82 & 4286 & J1705--6135 & 95 & 0.80 $\pm$ 0.06 \\

&  & & & &  &  & J1706--6118 & 76.13 & 1.06 $\pm$ 0.05 \\

&  & & & &  & & J1704--6016 & 54 & 14.84 $\pm$ 0.06 \\

&  & & & &  & & J1648--6044 & 106.19 & 0.72 $\pm$ 0.06 \\

&  & & & &  & & J1717--5800 & 125.21 & 0.76 $\pm$ 0.07 \\

54105 & EMU\_1742-55 & 17:41:59.408 & --55:43:48.417 & 336.4 & -13.14 & 4114 & J1749--5605 & 58 & 1.43 $\pm$ 0.06 \\

&  & & & & &  & J1733-5515 & 83.9 & 0.77 $\pm$ 0.07 \\

63789 & EMU\_1743-51 & 17:43:37.757 & --51:07:20.395 & 340.65 & -11.07 & 4092 & J1749--4931 & 55.3 & \ldots \\

&  & & & & &  & J1732--5049 & 56.82 & 2.80 $\pm$ 0.10 \\

&  & & & & &  & J1757--5322 & 30.80 & 1.40 $\pm$ 0.09 \\

&  & & & & & & J1744--5337 & 109.59 & 1.35 $\pm$ 0.07 \\

&  & & & & & & J1740--5340A & 71.8 & 0.72 $\pm$ 0.06 \\

74235 & EMU\_1749-60 & 17:49:42.049 & --60:19:44.170 & 332.72 & --16.19 & 3735 & \ldots & \ldots & \ldots \\

74275 & EMU\_1830-60 & 18:30:50.621 & --60:19:44.170 & 334.78 & --20.90 & 4702 & J1833--6023 & 35 & 2.41 $\pm$ 0.05 \\

74410 & EMU\_1844-46 & 18:44:59.709 & --46:30:21.660 & 349.36 & --18.27 & 4569 & \ldots & \ldots & \ldots \\

61946 & EMU\_2212-04B & 22:12:32.239 & --04:40:35.835 & 56.43 & --46.01 & 3063 & J2222--0137 & 3.28 & 0.63 $\pm$ 0.16 \\
\hline
\end{tabular}
\label{tab:tab1}
\end{table*}
\begin{figure*}
\begin{center}
\includegraphics[width=\textwidth]{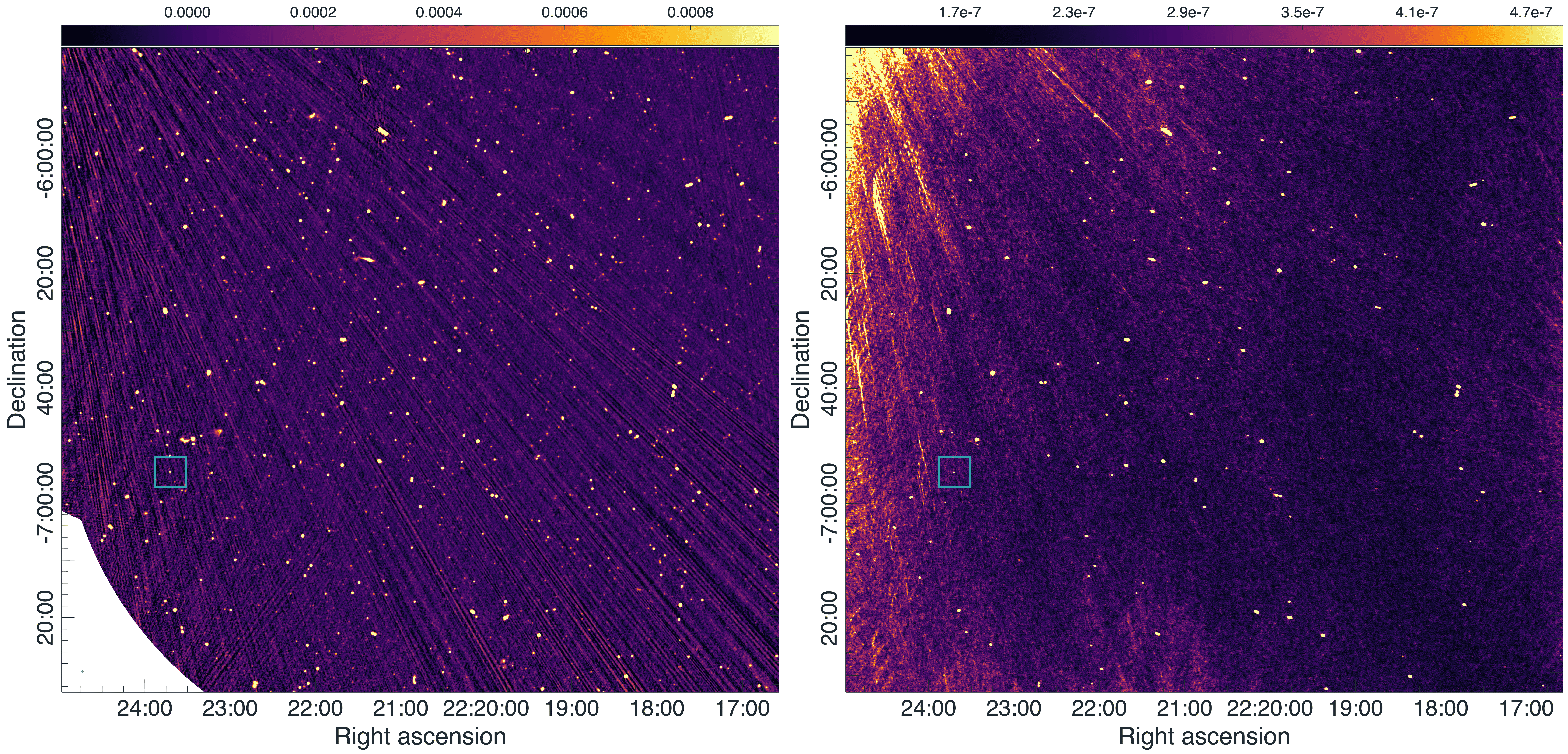}
\end{center}
\caption{Left: A portion of the radio-continuum image of the tile SB 61946 at 943.5\,MHz. The field of view is 2$\times$2 deg$^{2}$. The white region at the bottom-left corner lies outside the FWHM for the primary beam, which is approximately $1.5^\circ$. The beam size of the radio image is 15\,arcsec $\times$ 15\,arcsec and shown at the bottom left corner. The color bar is in units of Jy\,beam$^{-1}$.  Right: The corresponding variance image (beam 05) containing the new pulsar J2223--0654 in a suqare box. The region of enhanced variance at the top left corner is due to the sidelobes of an out-of-beam bright source.}
\label{fig_psr}
\end{figure*}

\subsection{Murriyang observations}
\label{sec:pks}

We conducted follow-up observations of pulsar candidates identified in variance images using Murriyang's Ultra-Wideband Low (UWL) receiver in conjunction with the Medusa backend~\citep{hmd+20}. The total integration time ranged from 0.5\,hr to 4\,h and only the total intensity was recorded. Data were recorded in pulsar search mode with 2-bit sampling every 64\,$\mu$s within 0.125\,MHz wide frequency channels, totalling 26,624 channels covering 704 to 4032\,MHz. 
Since pulsars detected in variance images show strong scintillation, we searched for their periodicity over a wide bandwidth from 1344\,MHz to 3264\,MHz. This frequency range was chosen to avoid strong radio frequency interference (RFI) at lower frequencies and maximise our chance of detecting scintillating pulsars. The periodicity search was performed using a pulsar searching pipeline based on the \textsc{PRESTO} software package \citep{ransom01}. Periodic signals were searched within a DM range of 0--500\,pc\,cm$^{-3}$ in the Fourier domain. Candidates with a signal-to-noise ratio threshold of $>8$ were folded and inspected visually. 

Follow-up observations of confirmed pulsar discoveries were carried out with the coherently dedispersed search mode using the UWL system. The total intensity was recorded with 2-bit sampling every 64\,$\mu$s and 1\,MHz frequency resolution, totalling 3328 channels covering 704 to 4032\,MHz. Search mode data were folded using the \textsc{DSPSR} software package~\citep{dspsr} with a sub-integration length of 30\,s. Folded pulse profiles were then processed and calibrated with the \textsc{PSRCHIVE} software package~\citep{psrchive}. Each observation was visually examined using the \texttt{pazi} program to remove RFI and then averaged in time to form an averaged pulse profile. The pulse time of arrivals (ToAs) was measured for each observation using the \textsc{pat} routine of \textsc{PSRCHIVE}. Timing analysis was carried out using the \textsc{TEMPO2} software package~\citep{tempo2}.

\section{Variance imaging and source detection}\label{obs}
\subsection{Imaging pipeline}
Our data processing pipeline involves several steps to produce final products (dynamic spectra) from calibrated visibilities. All data processing has been done on OzSTAR\footnote{\url{https://supercomputing.swin.edu.au}} computing facility with Common Astronomy Software Applications \citep[CASA;][]{casa+22} package in combination with \textsc{Python} scripts.  
{The details of each tile are given in Table~\ref{tab:tab1}. We downloaded the visibility data comprising 384 MS (24 beams per tile) from CASDA to the Ozstar for further processing. We processed each beam independently, and the steps of our dedicated processing are described further.}

\subsubsection{Correcting measurement set}

The processing of calibrated visibility data using the CASA package requires the MS to be corrected for beam phase centre and flux density scale. ASKAP visibilities, processed with the ASKAPsoft pipeline, are stored on a per-beam basis. The MS are phased to the centre of their respective beams; however, non-ASKAPsoft packages such as CASA do not account for the FEED table, which specifies the offset of each beam from the centre of the 36-beam mosaicked tile.
ASKAPsoft also defines the Stokes parameters based on the total flux density of orthogonal correlations (e.g., I = XX + YY), whereas CASA uses Stokes definitions based on the average flux density (e.g., I = (XX + YY)/2). Therefore, a beam offset and flux density scale correction was applied to each MS prior to CASA imaging using the command line script \texttt{dstools-askap-preprocess} of \textsc{DSTOOLS} \citep{josh+25}.

\subsubsection{Image creation}
To clean and make snapshot images from corrected visibility data, we employed the \textit{tclean} functionality of CASA using a default version of \textit{CLEAN} deconvolver \citep{hogbom74}. 
For each beam, we choose a frequency resolution of 24\,MHz and a time resolution of 20\,minutes, producing 360 images for a typical 10\,h observation.
We did not subtract a sky model from the calibrated visibilities, nor did we include higher-order spectral terms in the imaging. We performed cleaning with 10,000 iterations using Briggs weighting with robustness of 0.5 to obtain a good balance between sensitivity and resolution \citep{brigs+95}. We chose an image cell size of 2\,arcsec, an image size of 4000$\times$4000 pixels to include as many of the neighbouring sources to reduce the side-lobes effect, and a widefield gridding option. A 4000$\times$4000 pixel image corresponds to a square field of view (FoV) of approximately 2.2$^{\circ}$, which is about 1.5 times the Full Width Half Maximum (FWHM) of primary beam at 943.5\,MHz, requiring the processing of only 24 beams to fully cover the whole tile\footnote{Beams 01, 03, 07, 09, 13, 15, 19, 21, 25, 27, 31, 33 are not processed.}. The baselines shorter than 100\,m were also flagged to suppress diffuse radio emission. In total, we generated $\sim$129,800 images for all 384 MS. The typical residual rms of resulting images is about 400--600\,$\mu$Jy\,beam$^{-1}$.

\subsubsection{Variance image creation}
Variance images are generated by calculating the variance of radio flux densities for each image pixel across time and frequency. For each EMU tile, we generated 24 variance images. An example of a variance image is shown in Fig.~\ref{fig_psr}.

\begin{figure*}
\centering
\includegraphics[width=0.498\linewidth]{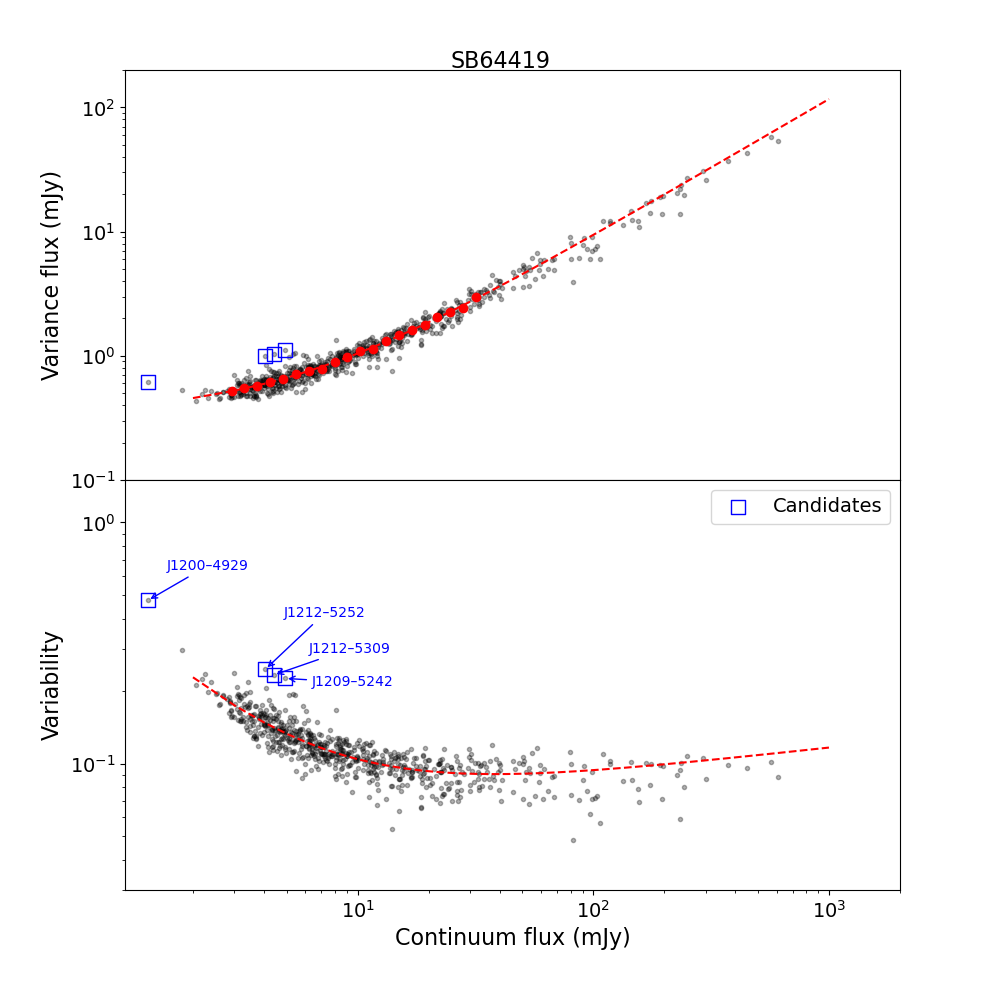}
\includegraphics[width=0.498\linewidth]{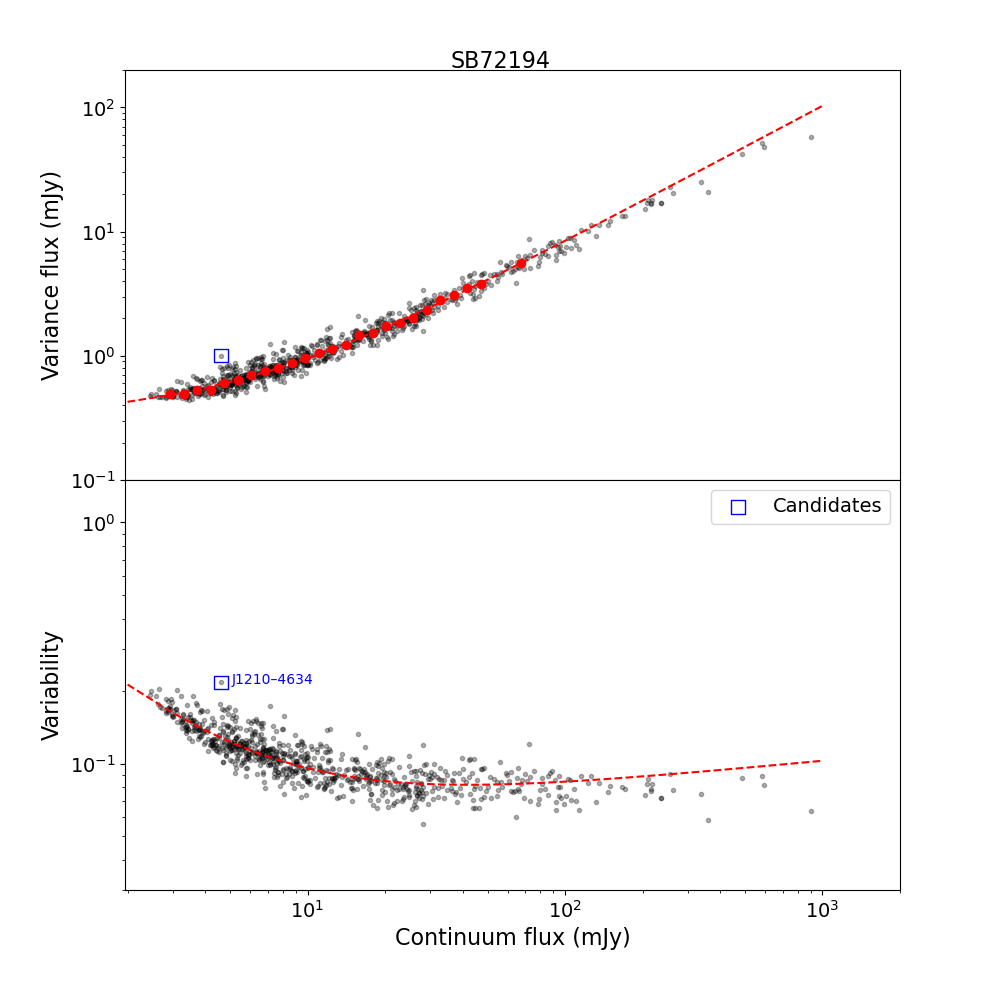}
\includegraphics[width=0.498\linewidth]{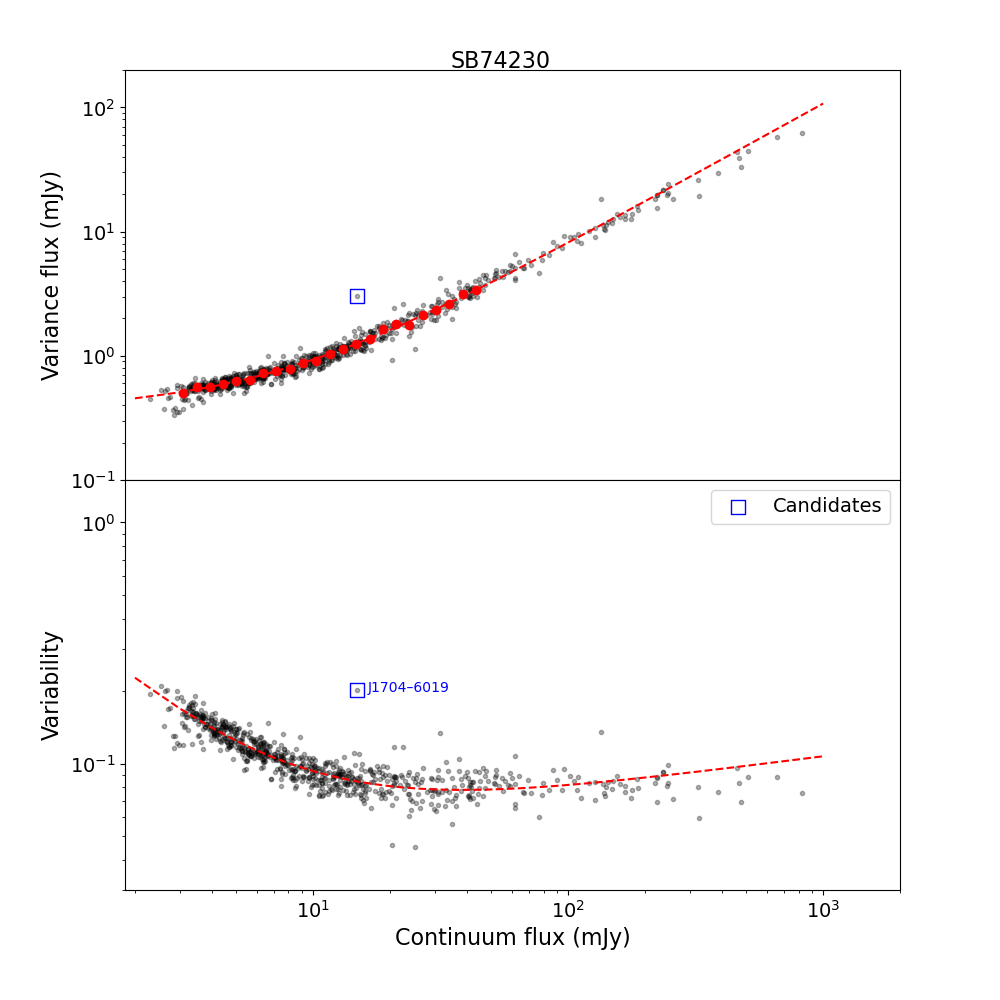}
\includegraphics[width=0.498\linewidth]{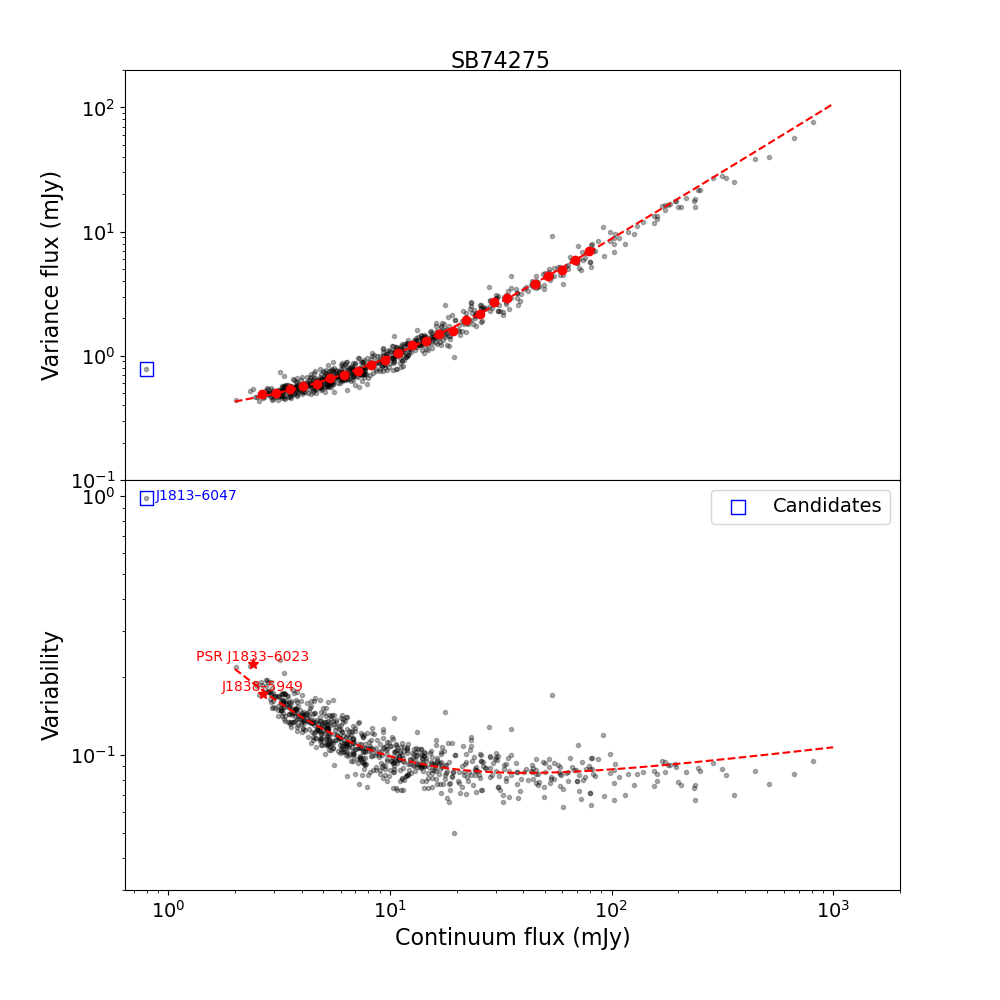}
\caption{Distribution of sources in variance images for tiles where pulsar and radio star candidates are detected. Known pulsars (with tag PSR) and candidates are shown as blue squares. The sources labelled as red are detected in variance images but not picked up by our 5$\sigma$ threshold explained in Section~\ref{sec:select}.}
\label{fig_var}
\end{figure*}
\begin{figure*}
\ContinuedFloat
\centering
\includegraphics[width=0.498\linewidth]{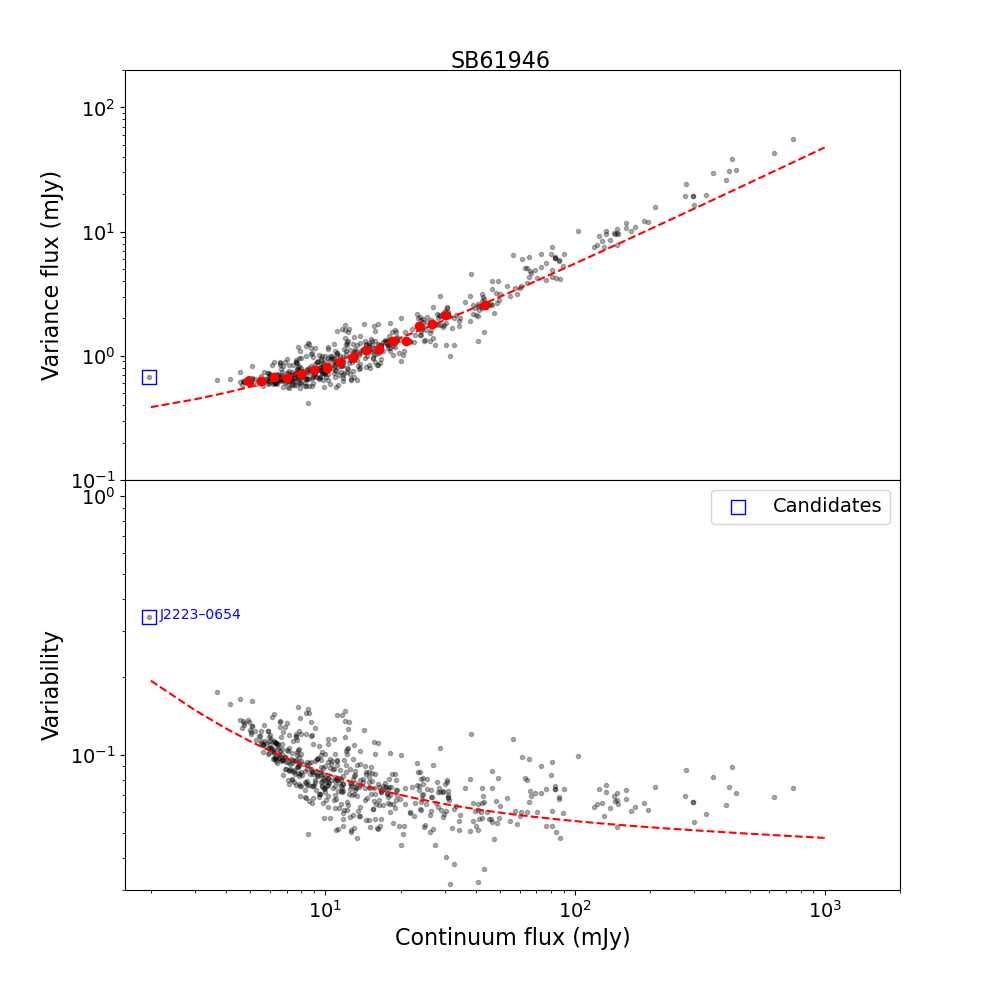}
\includegraphics[width=0.498\linewidth]{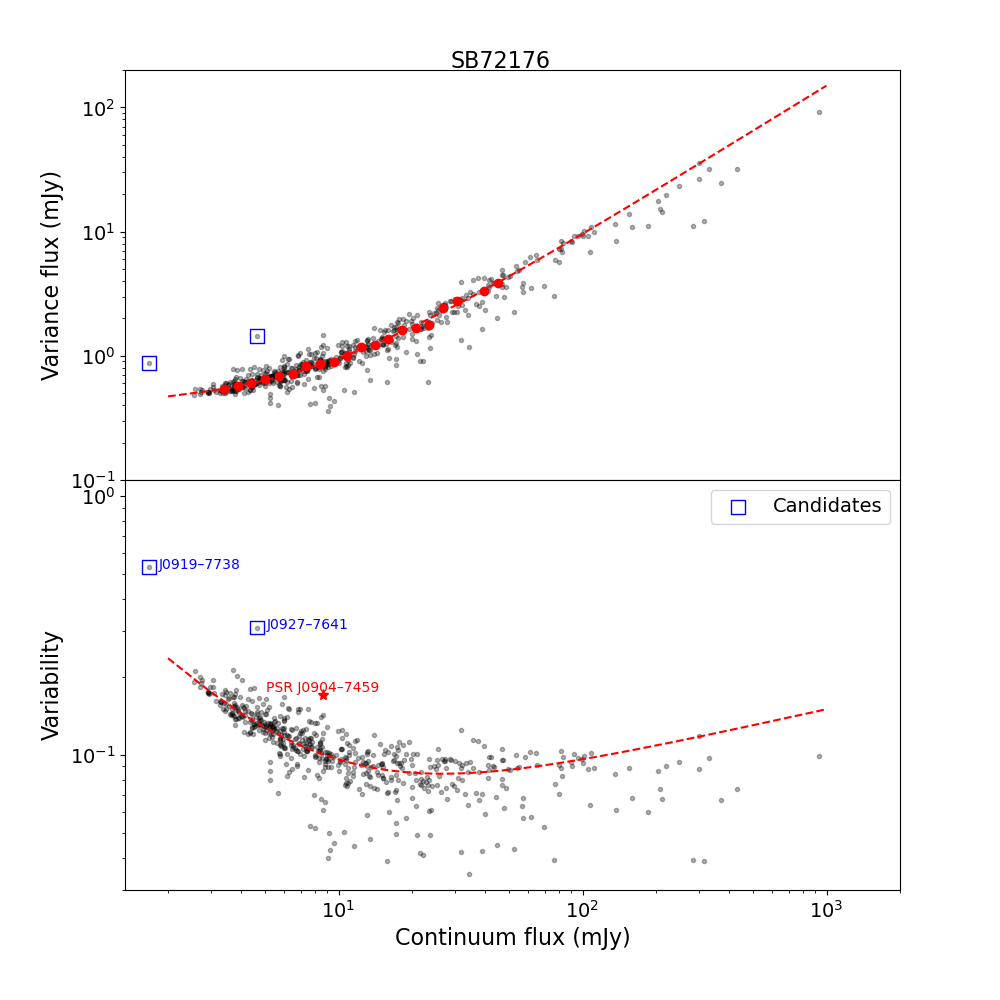}
\end{figure*}
\subsection{Candidate selection}
\label{sec:select}

To identify pulsar candidates in variance images, we first select compact radio sources in the \emph{Selavy} catalogues, which is produced by the EMU pipeline. We employ a \emph{compactness} criterion defined as the ratio of the integrated flux density (S$_{\rm int}$) to the peak flux density (S$_{\rm p}$) of the sources in the EMU continuum image. Specifically, sources with \emph{compactness} $>$ 1.5 and/or with multiple \emph{Selavy} components are regarded as extended and excluded from our further analysis. In addition, weak sources with $S_{\rm int}/\sigma_{\rm rms} \leq 5$ and the sources with flux density uncertainties exceeding 10\% are also excluded. For each EMU tile, the number of compact sources selected by these criteria is given in Table~\ref{tab:askap}. 

The \emph{Selavy} catalogues generated by the EMU pipeline exclude sources close to the edge of tiles. While the edges of tiles often show reduced sensitivity, we found that strong compact radio sources can be well detected in variance images. Therefore, to avoid missing potentially interesting variable sources in these regions, we manually inspected the variance images of the outermost beams using Cube Analysis and Rendering Tool for Astronomy \citep[CARTA;][]{carta+21}. For compact sources that can be clearly detected in variance images, we measured their continuum flux density using CARTA and included them in our following analysis. 

For each selected compact radio source, we fitted a two-dimensional Gaussian distribution
to its variance flux density in the variance image. This enables us to account for variable sizes of sources in the variance image and reject artefacts. We define the size of a source in the variance image as $r=(\sigma_{\rm x}+\sigma_{\rm y})/2$, where $\sigma_{\rm x}$ and $\sigma_{\rm y}$ are the standard deviation in $x$ and $y$ direction, respectively.
If the best fit centre of the Gaussian distribution is offset from the position of the continuum source by more than the size of the source ($r$) or the size of the source is larger than 10 pixels ($r>10$), we consider this source in the variance image as an artefact. 
The variance flux density ($S_{\rm v}$) of a source is measured by integrating pixel values within a contour at 10\% of the peak of the Gaussian distribution. The RMS noise ($\sigma_{\rm v}$) of the background is measured within an annulus region $3<r<6$ centred at the source position. Finally, we select sources with $S_{\rm v}/\sigma_{\rm v}\geq5$ as reliable detections in the variance image.

Sources showing up in the variance image consist of potential pulsars, variable sources, steep-spectrum sources, and extremely bright sources. However, we expect scintillating pulsars to show the highest level of variability in variance images. In Fig.~\ref{fig_var}, the top panel shows the variance flux density as a function of continuum flux density for all sources detected in the variance image for an EMU tile. To identify pulsar candidates, we divided the continuum flux density range for each EMU tile into 50 logarithmic bins. For bins containing more than 10 sources, we computed the median variance flux density (red solid circles in Fig.~\ref{fig_var}) and fitted a power-law to these medians as a function of continuum flux density. Sources deviating by more than $5\sigma$ from the best-fit curve were selected as candidates.
%
In Fig.~\ref{fig_var}, the bottom panel shows the variability~(R) defined as the ratio of variance flux (S$_{\rm v}$) to continuum flux densities (S$_{\rm 943.5}$) for all sources detected in the variance images, with pulsar and pulsar candidates highlighted by blue squares. We carefully analysed the corresponding cutout images of all pulsar candidates and excluded any suspicious sources (e.g., extended, image artifacts near bright sources, etc). 
In addition to applying the variability threshold, we generated circular polarisation cutout images for all sources detected in the variance maps. Sources exhibiting significant circular polarisation ($|V|/I > 5\%$) were identified as pulsar candidates for follow-up observations.
\begin{figure*}
\centering
\includegraphics[width=0.19\linewidth]{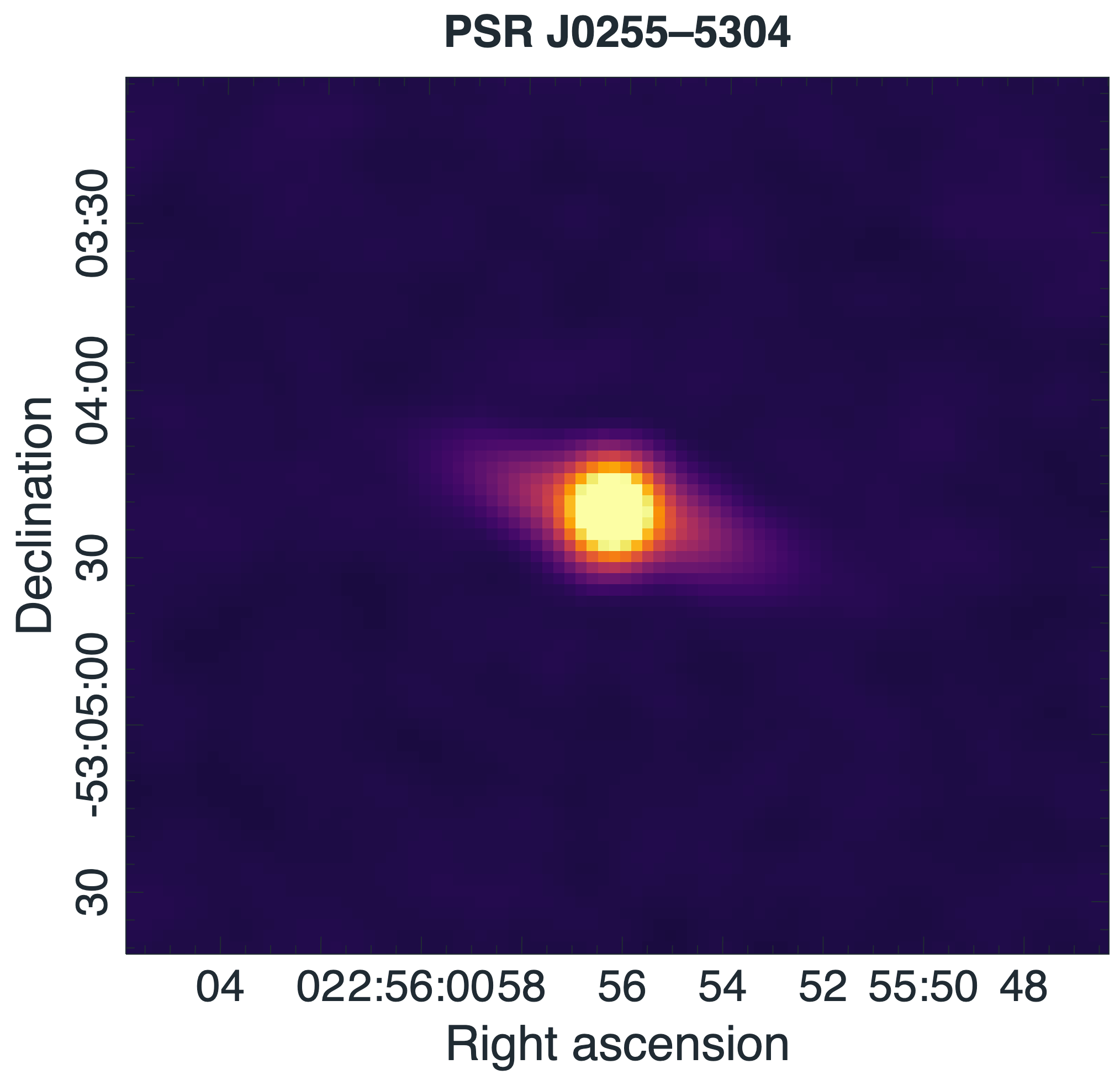}
\includegraphics[width=0.26\linewidth]{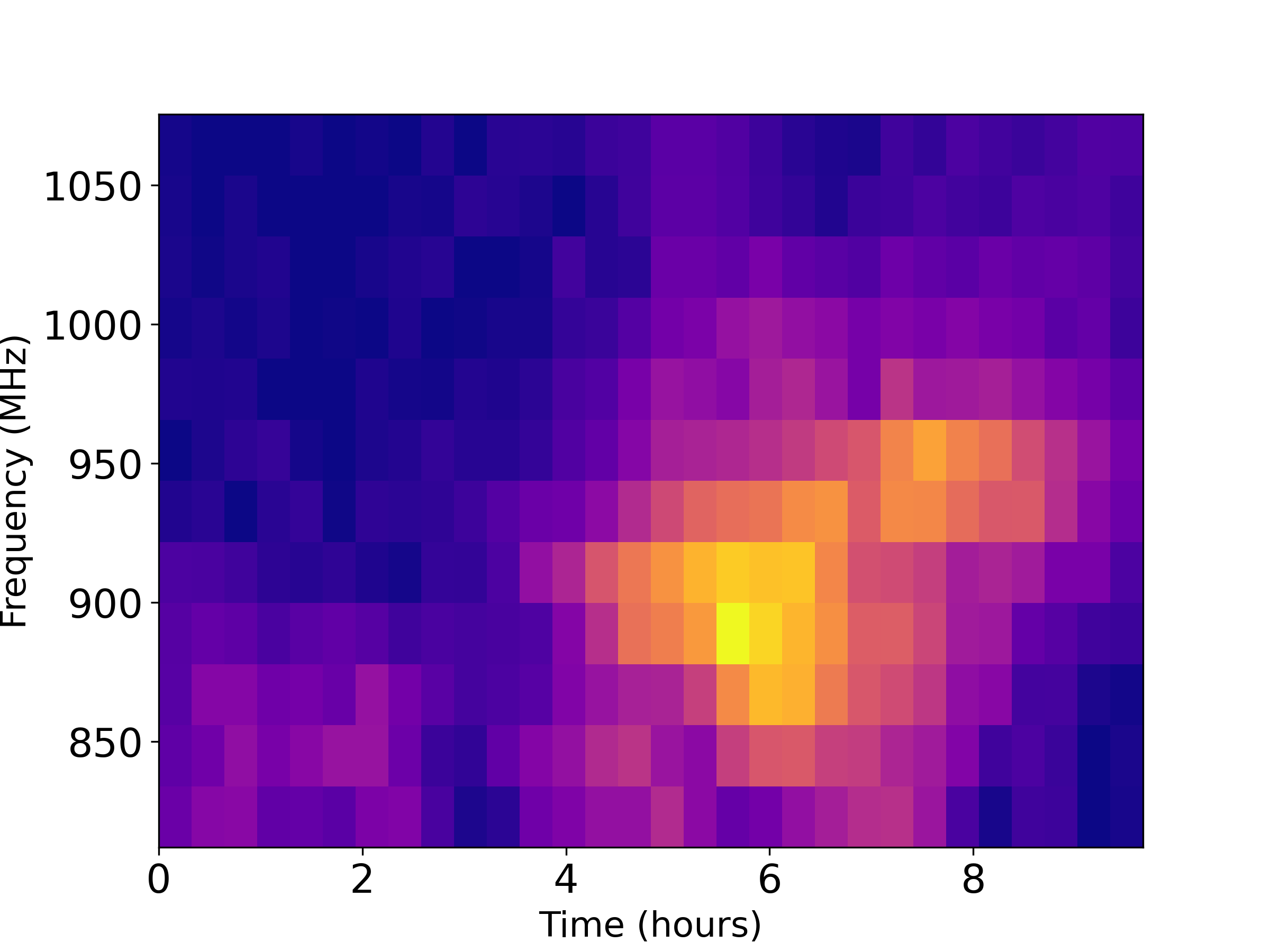}
\includegraphics[width=0.19\linewidth]{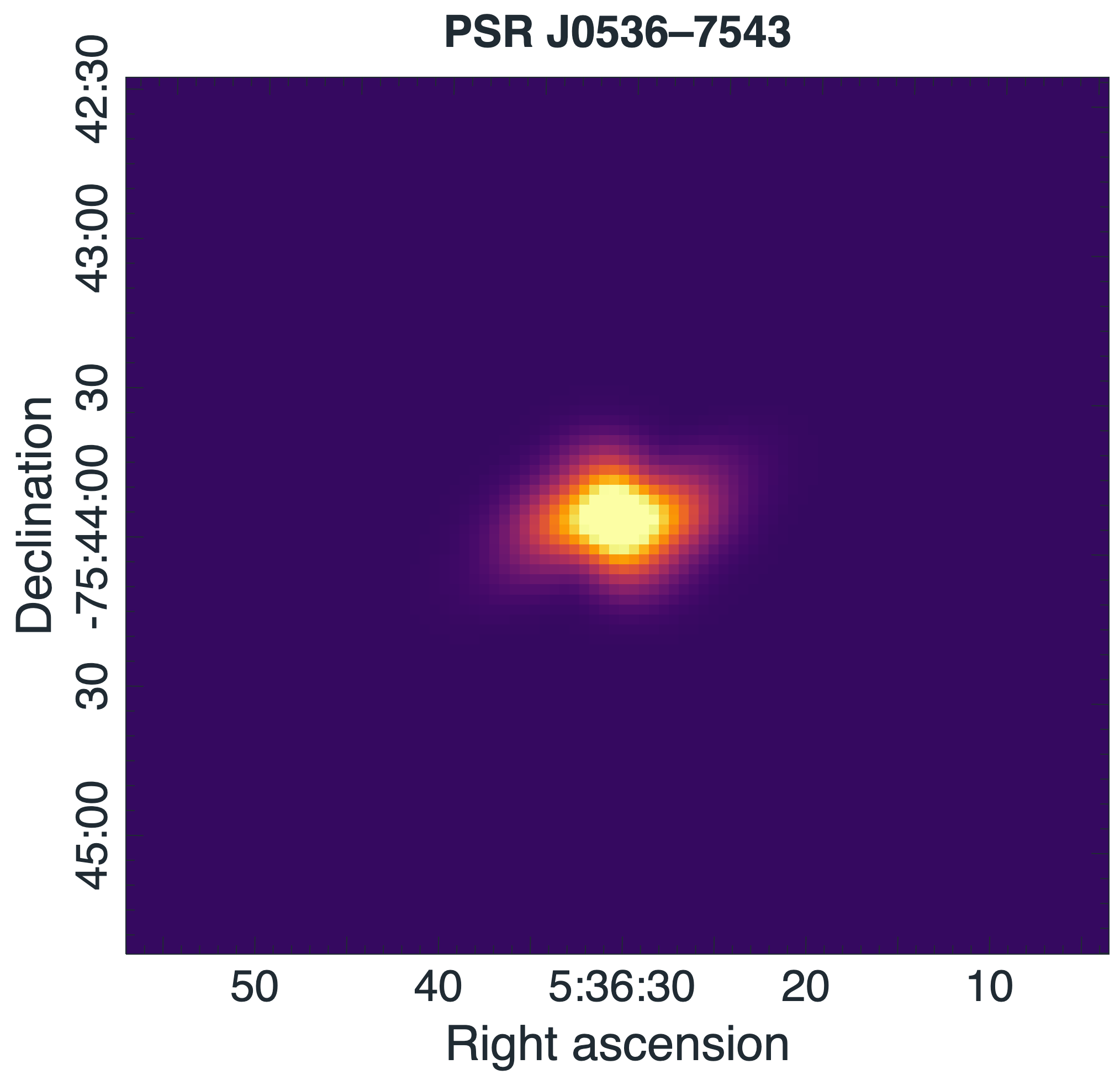}
\includegraphics[width=0.26\linewidth]{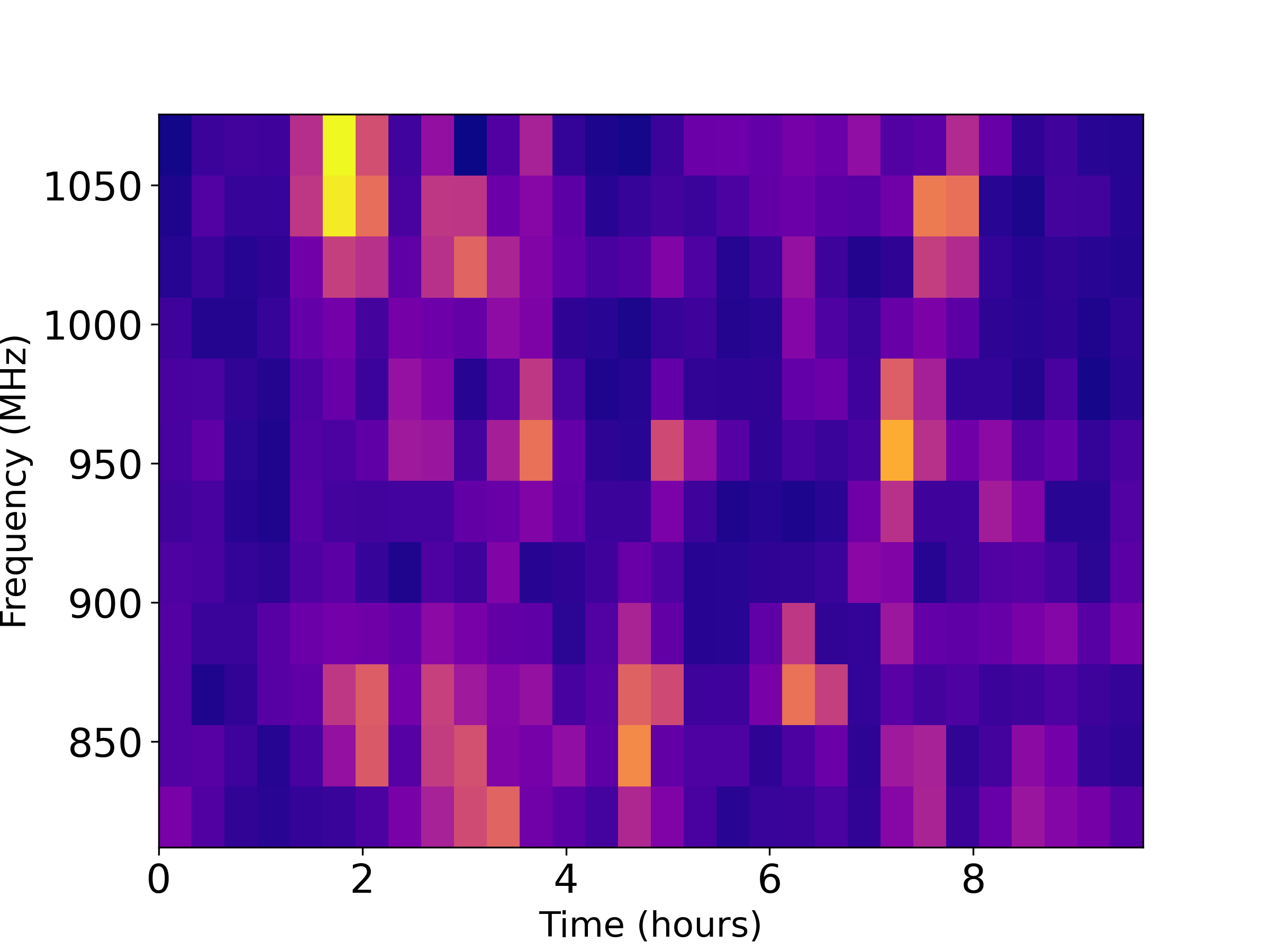}
\includegraphics[width=0.19\linewidth]{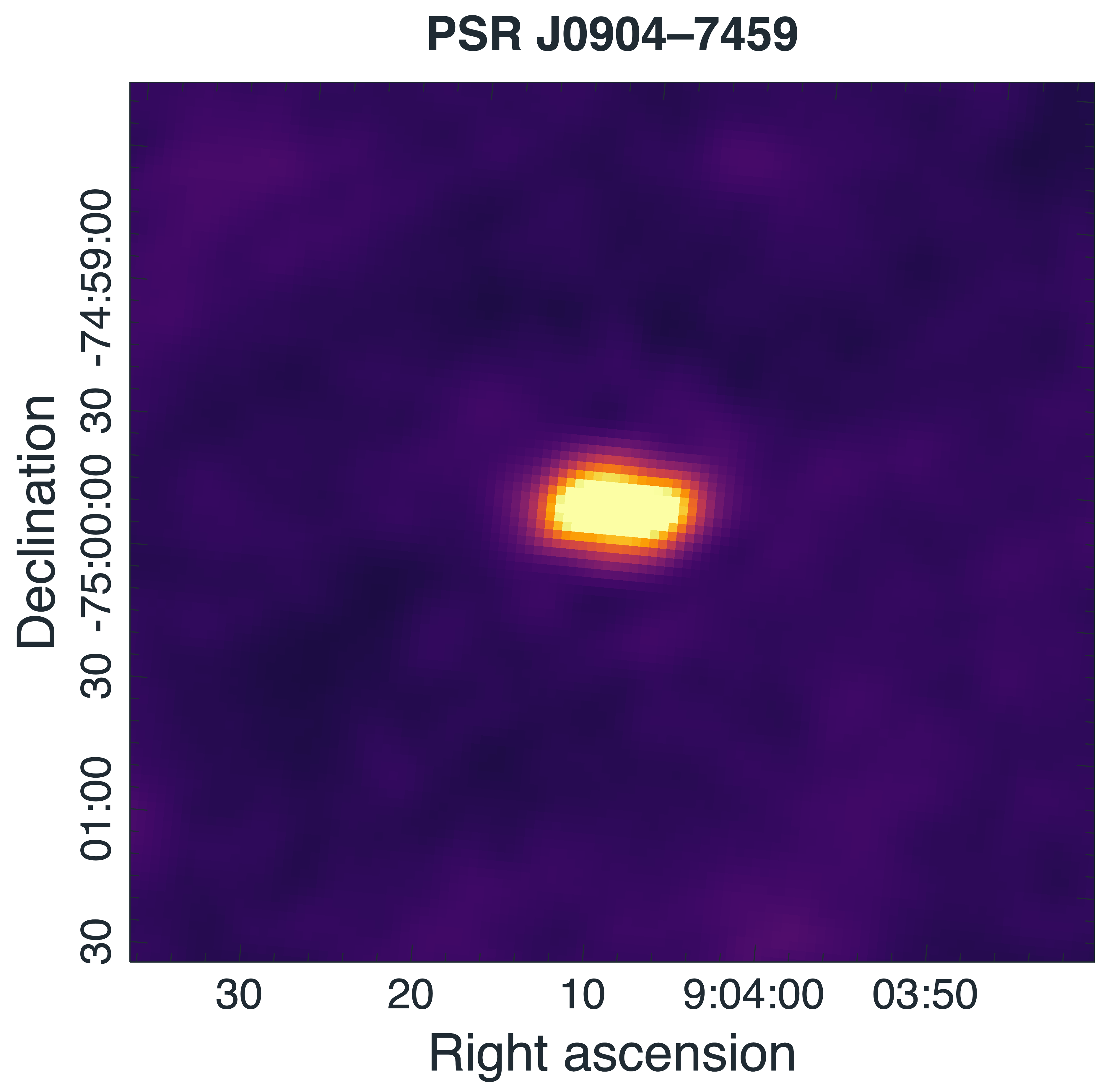}
\includegraphics[width=0.26\linewidth]{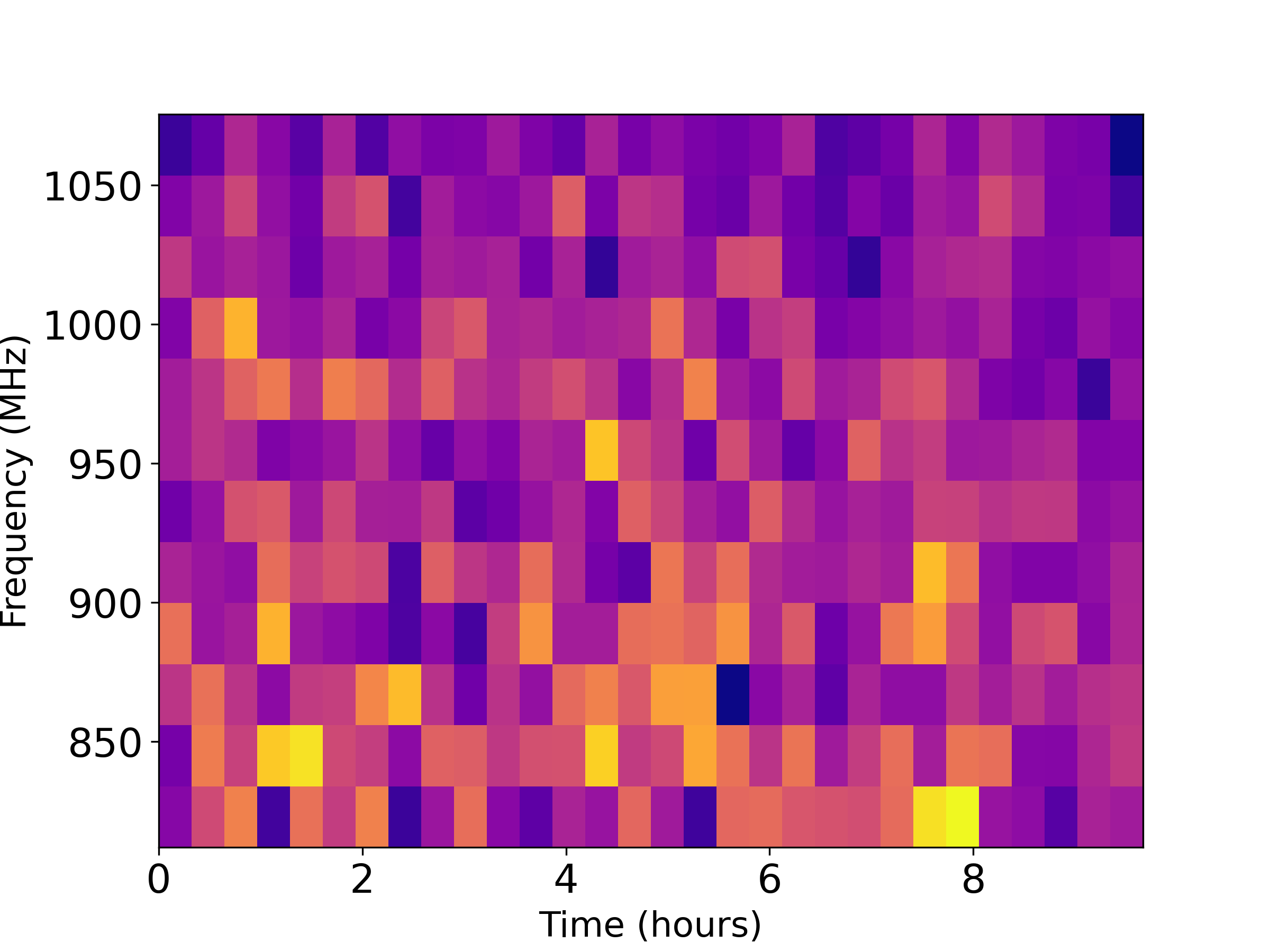}
\includegraphics[width=0.19\linewidth]{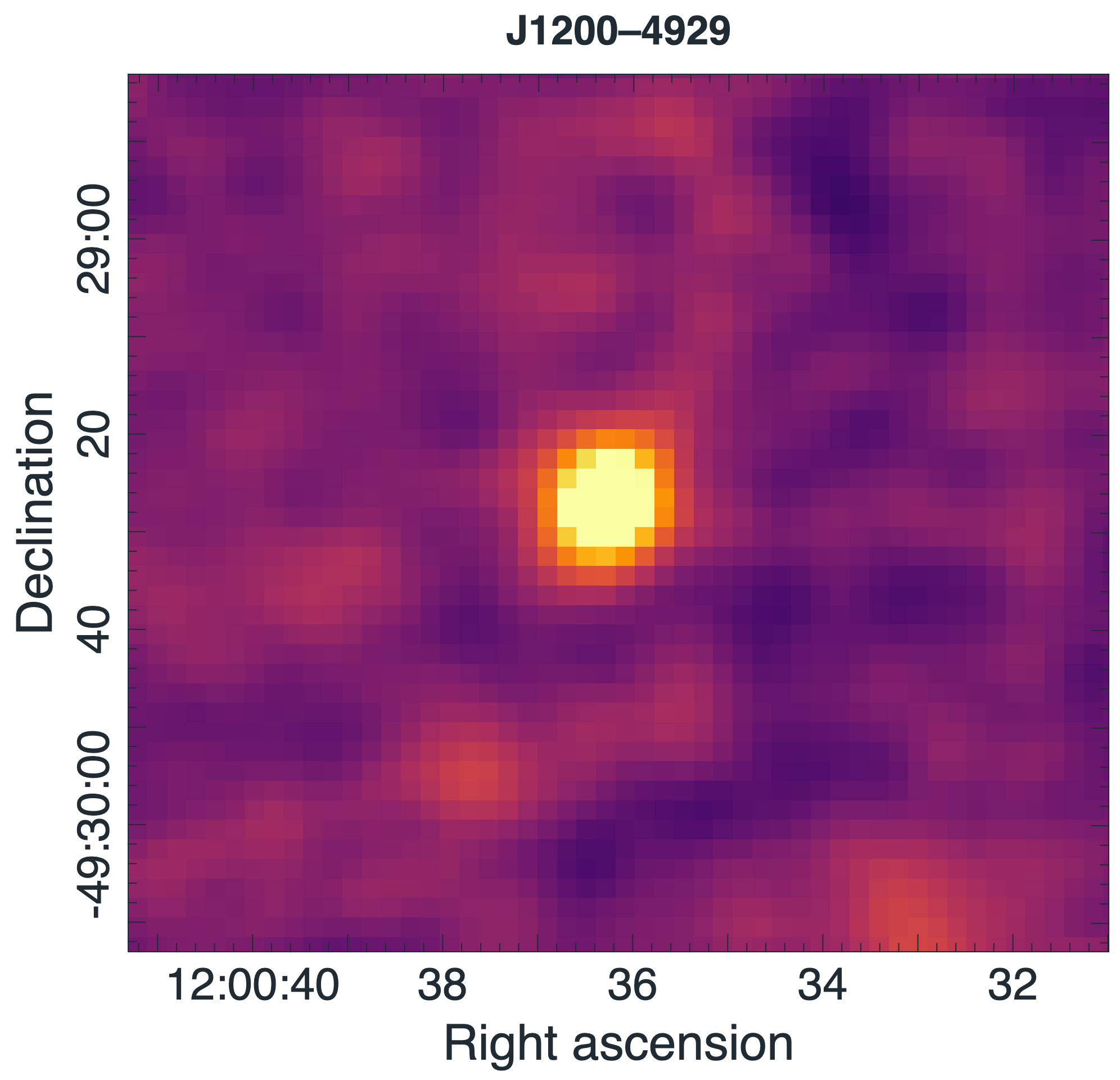}
\includegraphics[width=0.26\linewidth]{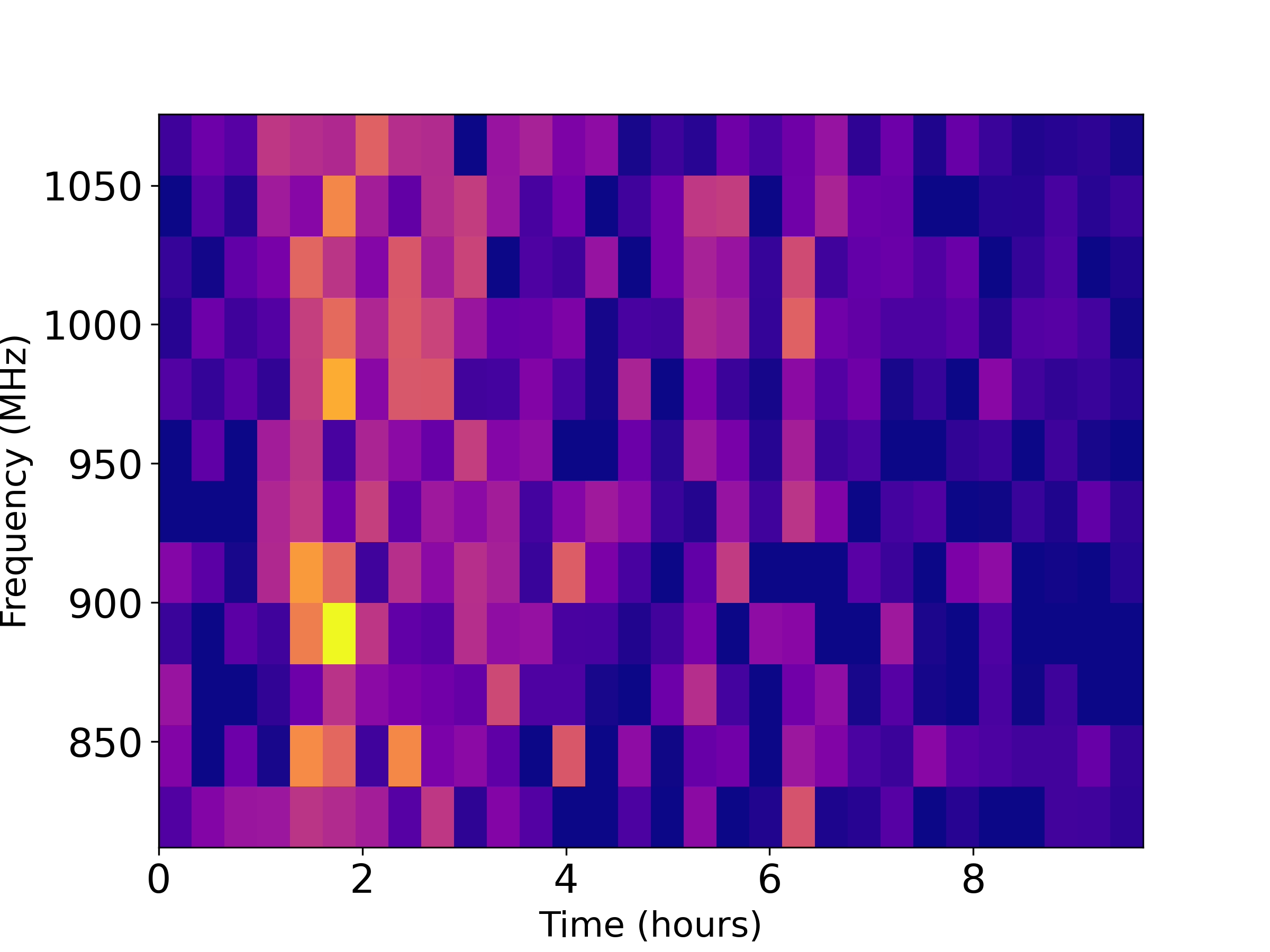}
\includegraphics[width=0.19\linewidth]{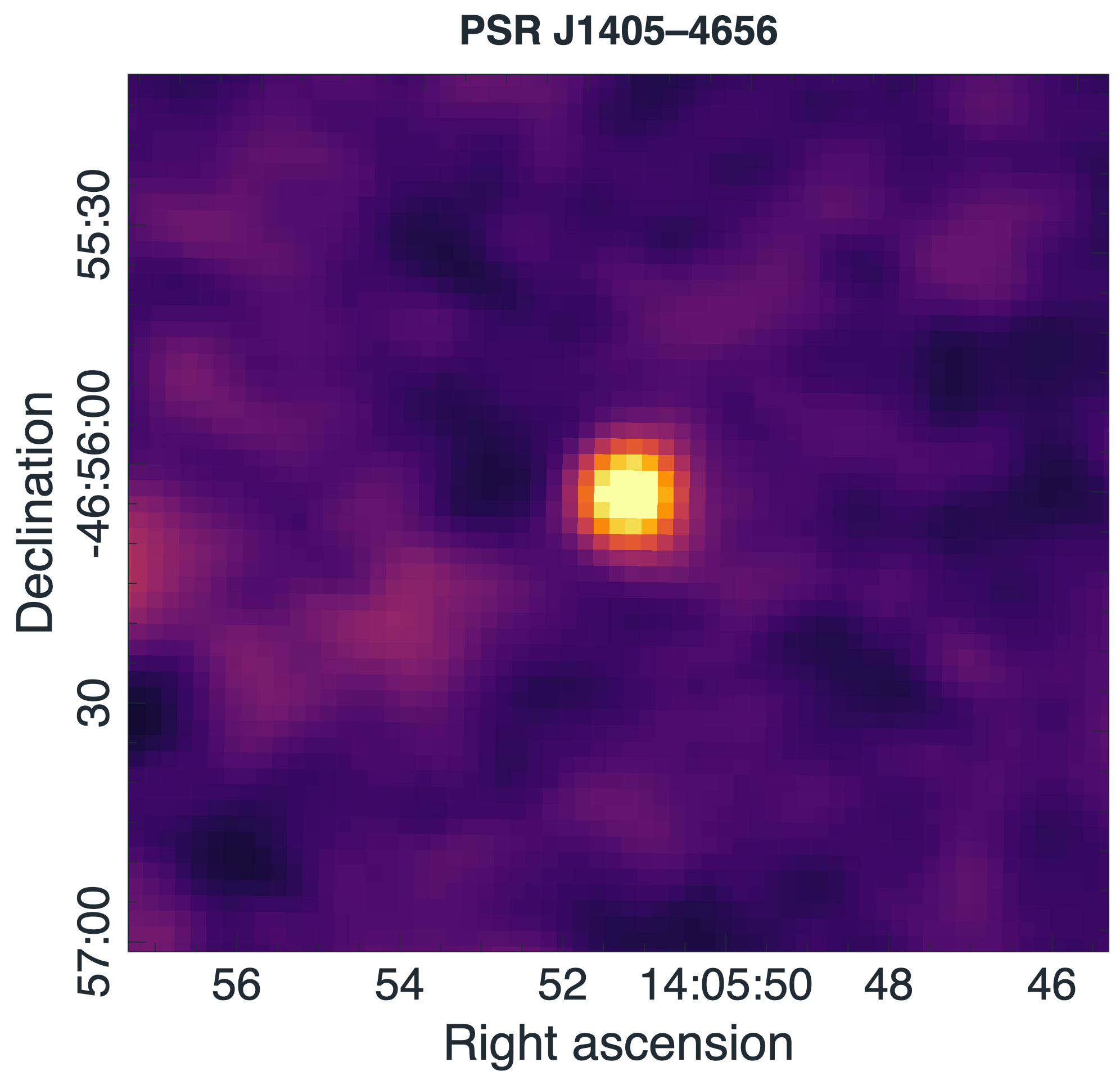}
\includegraphics[width=0.26\linewidth]{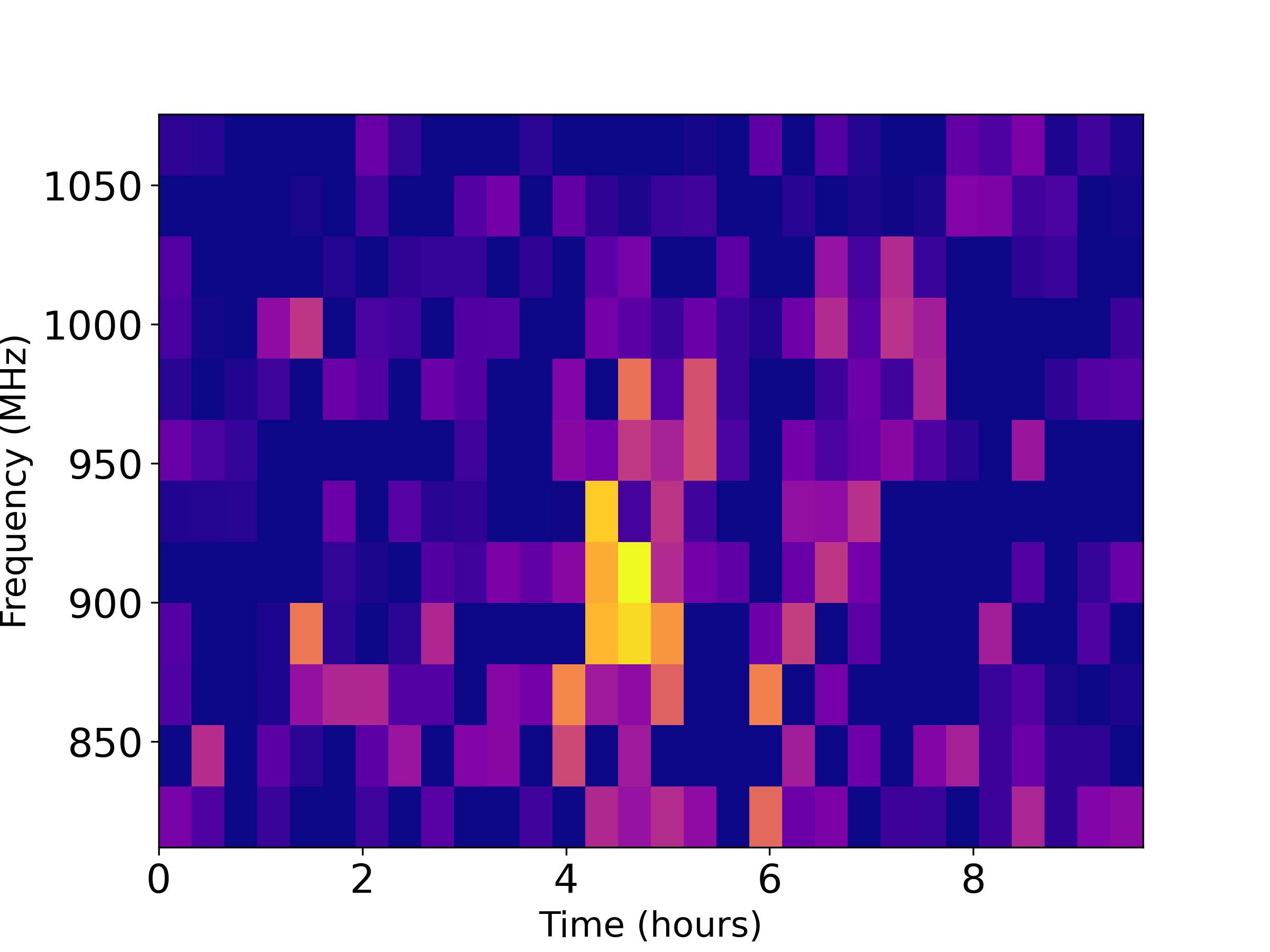}
\includegraphics[width=0.19\linewidth]{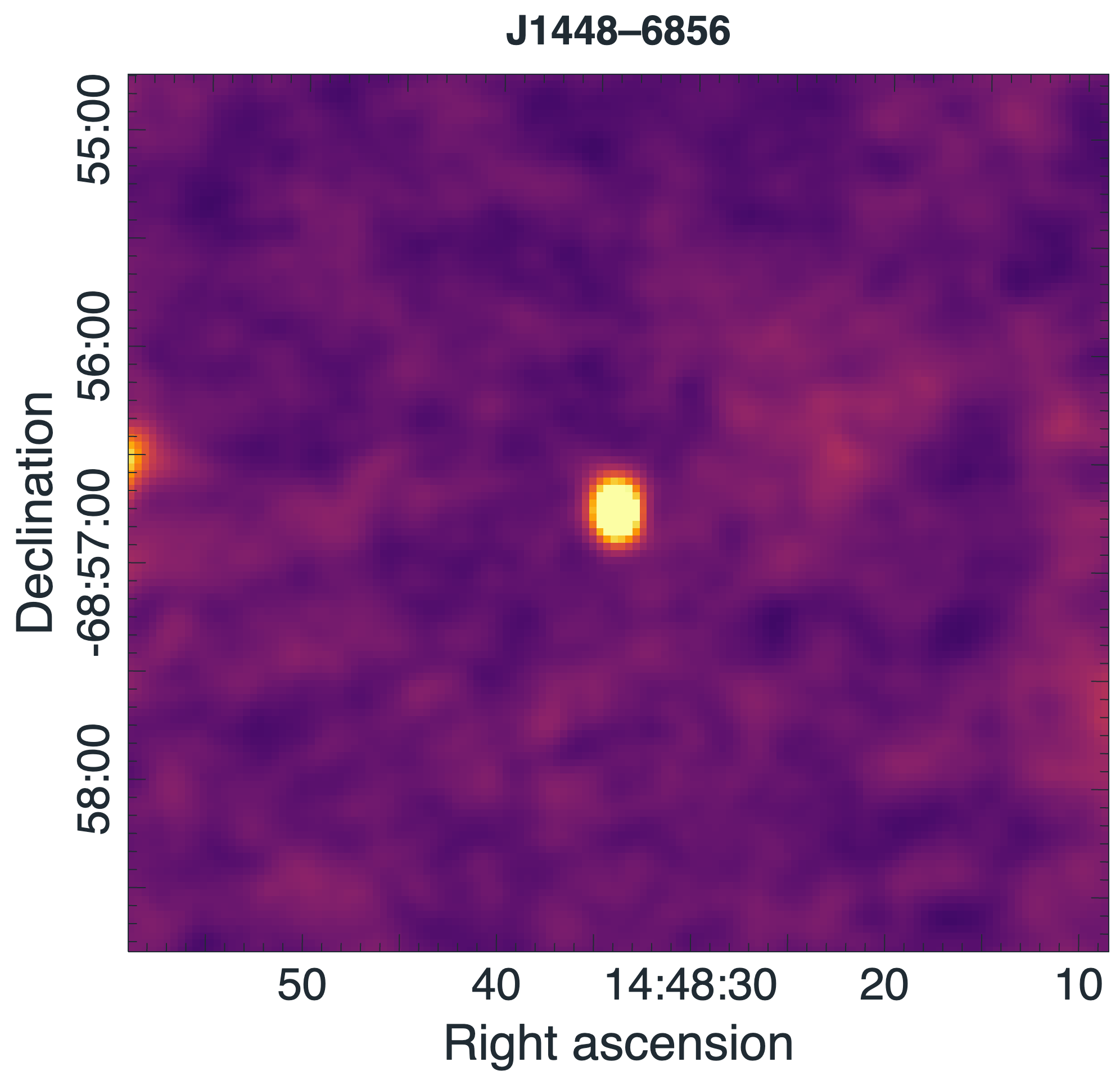}
\includegraphics[width=0.26\linewidth]{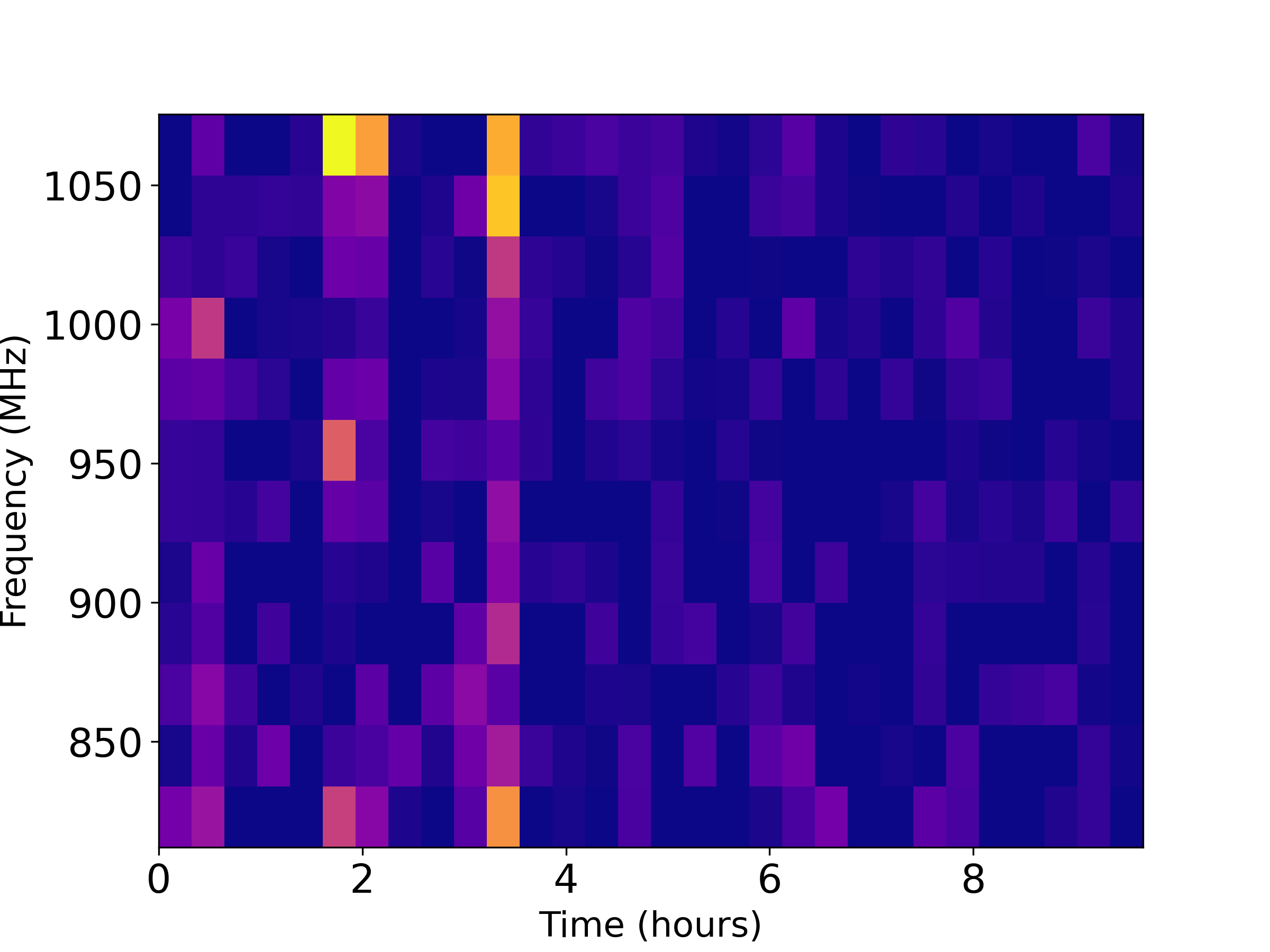}
\includegraphics[width=0.19\linewidth]{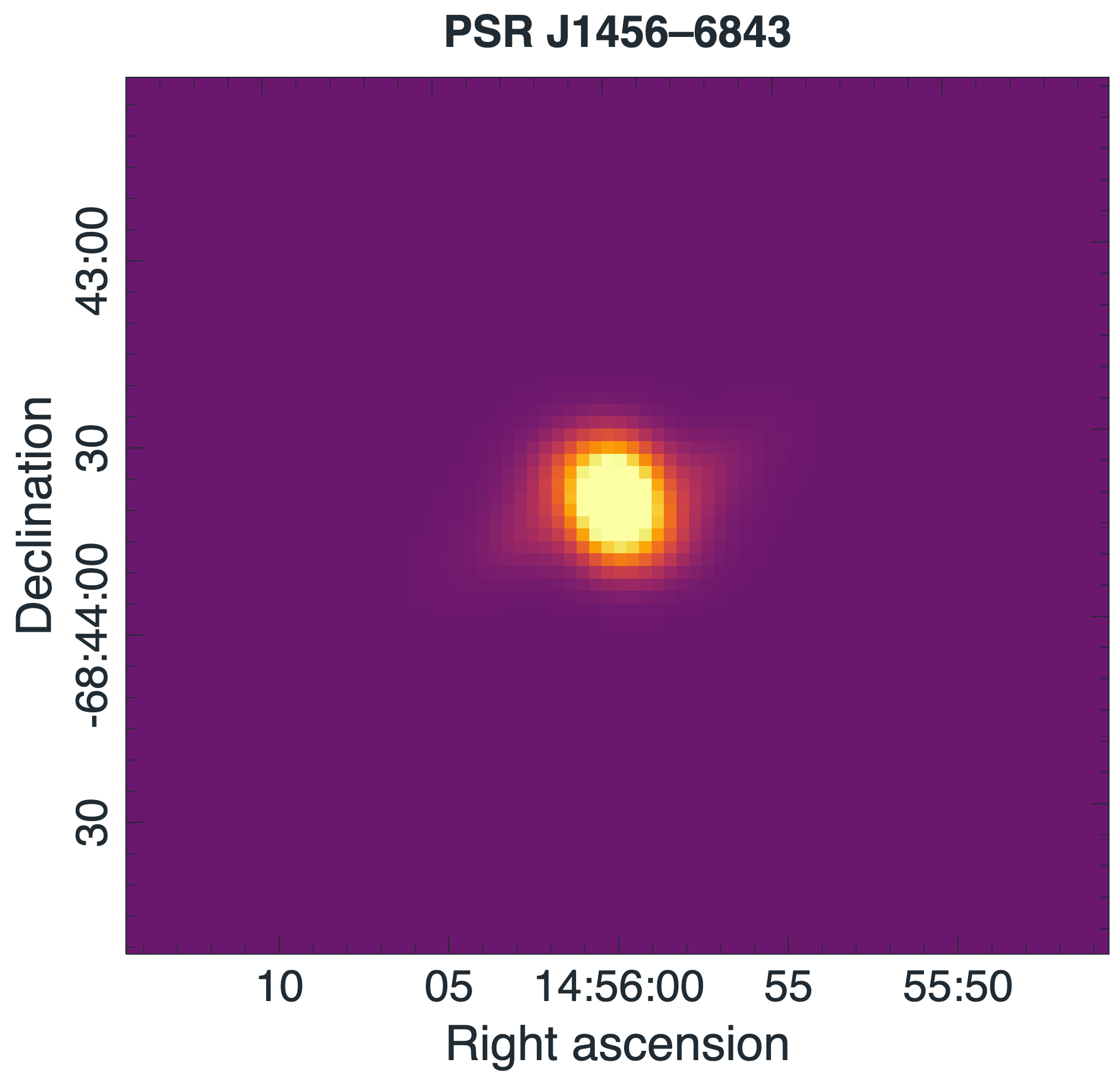}
\includegraphics[width=0.26\linewidth]{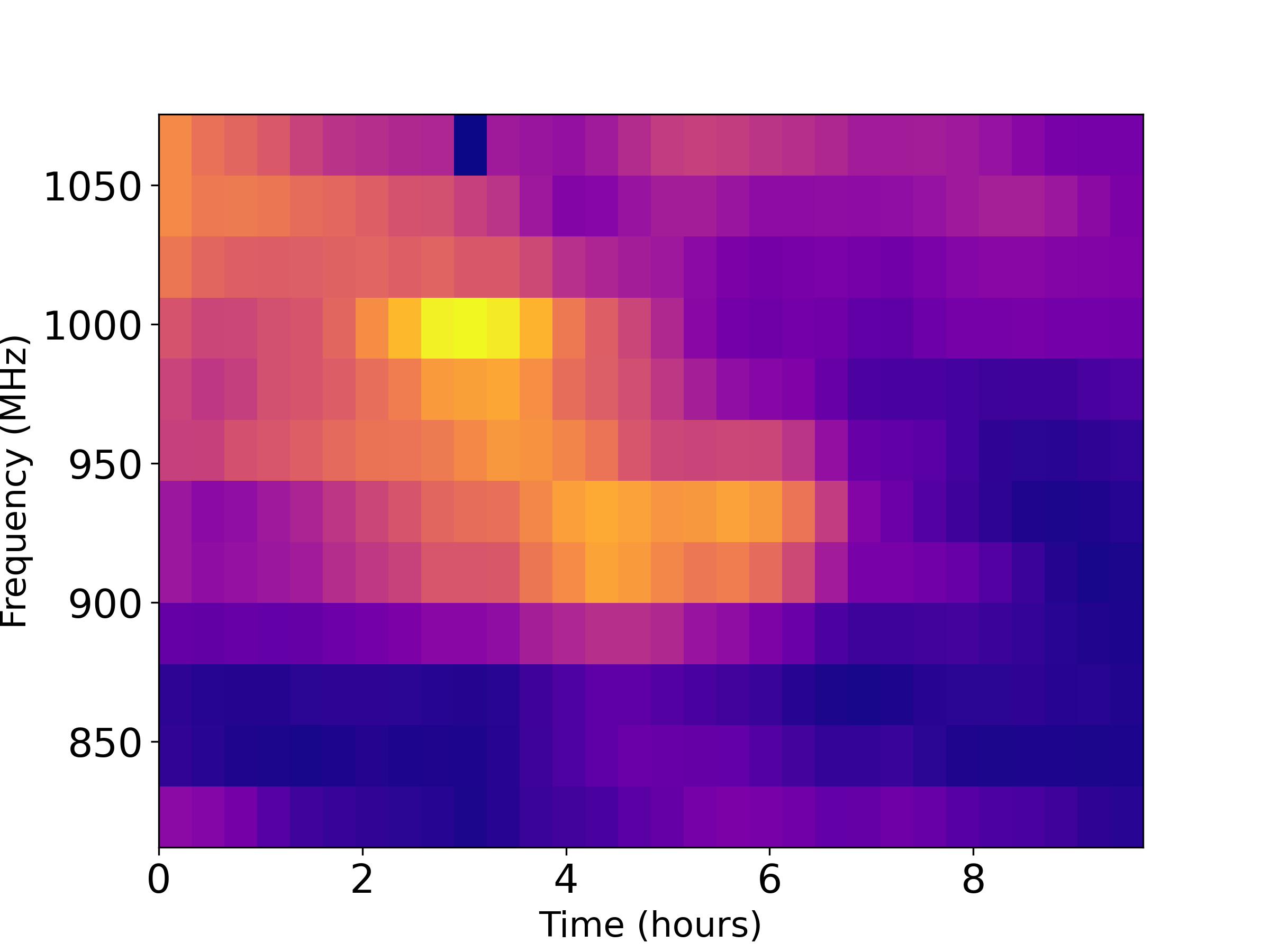}
\includegraphics[width=0.19\linewidth]{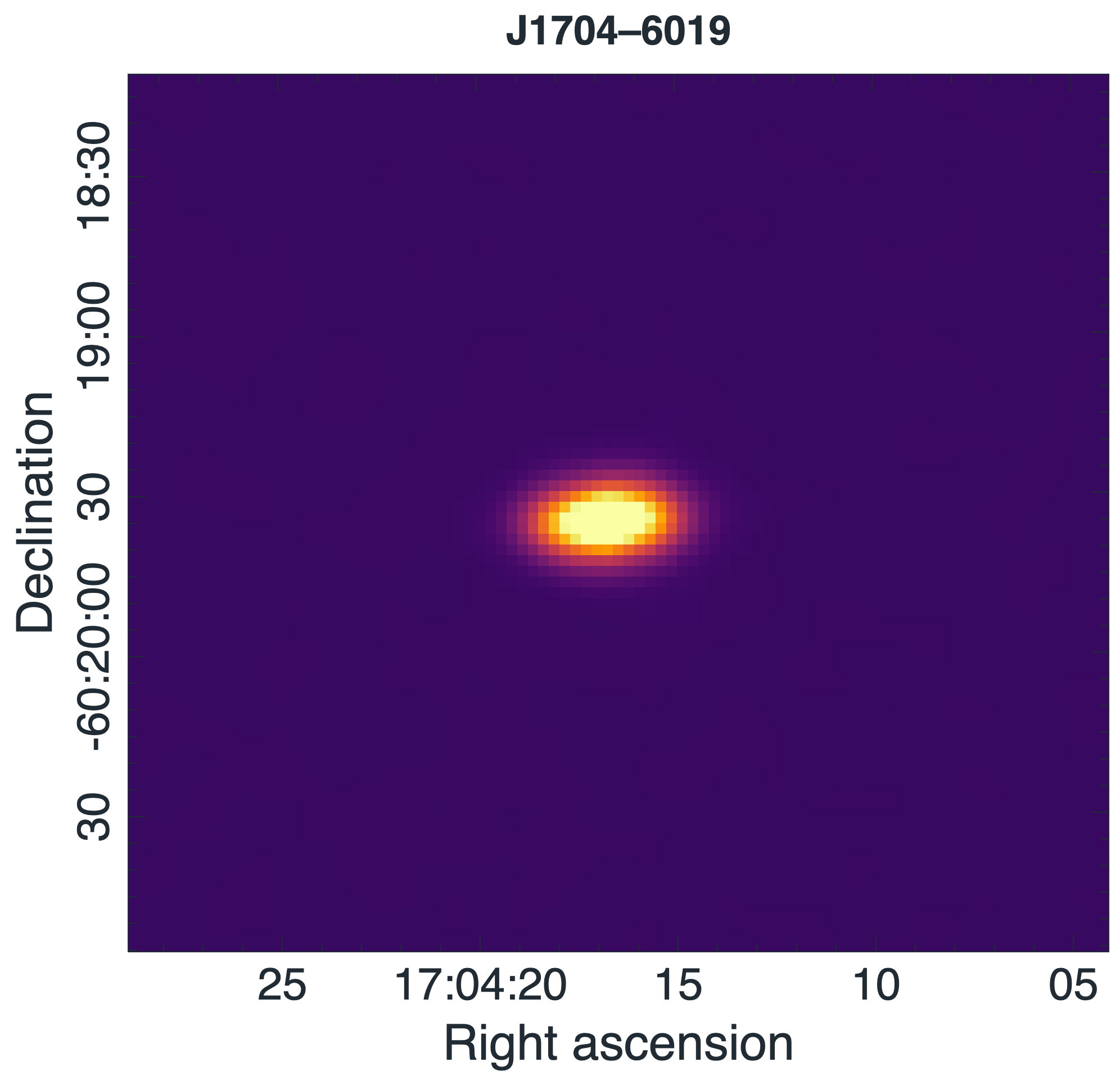}
\includegraphics[width=0.26\linewidth]{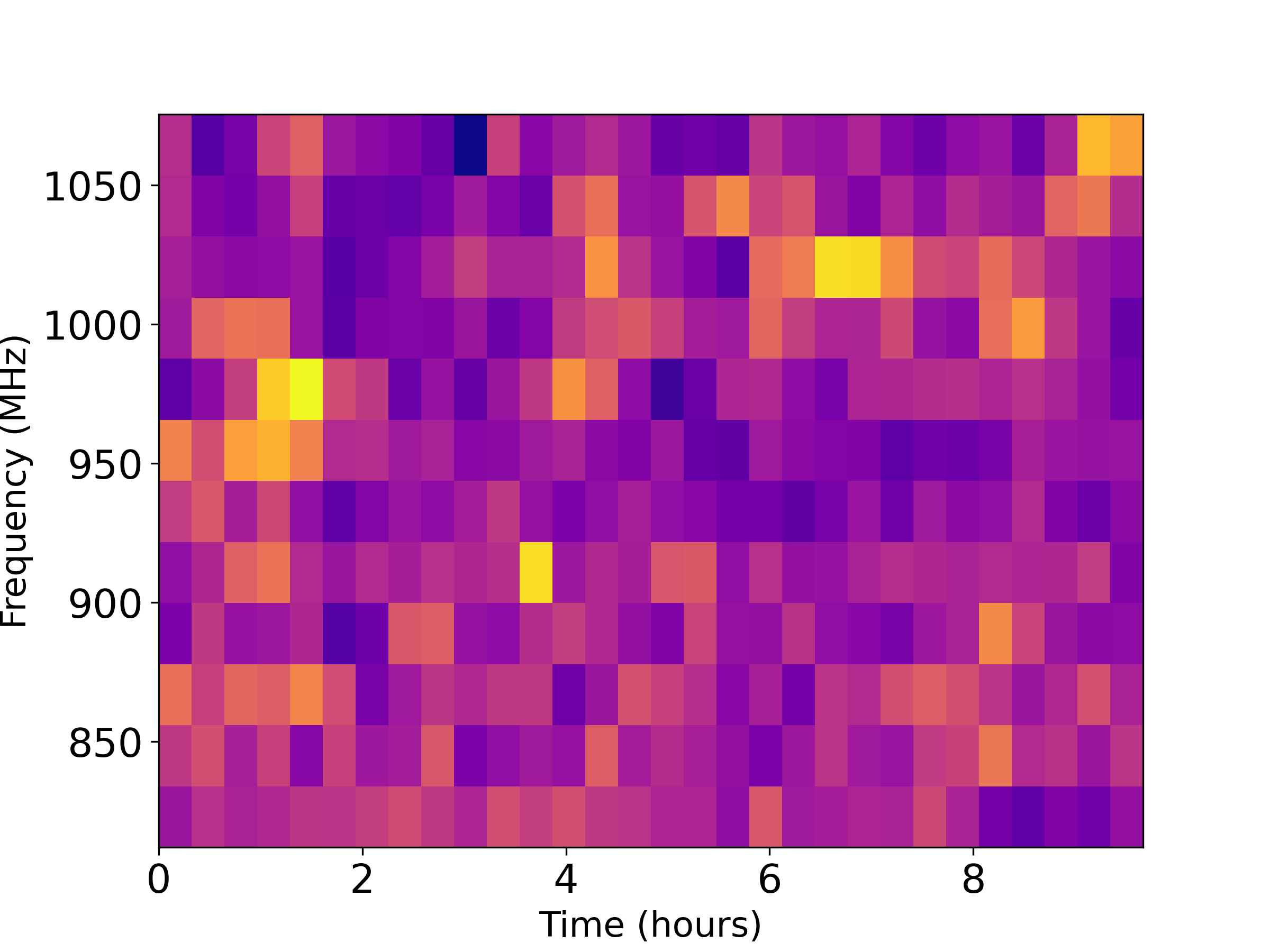}
\includegraphics[width=0.19\linewidth]{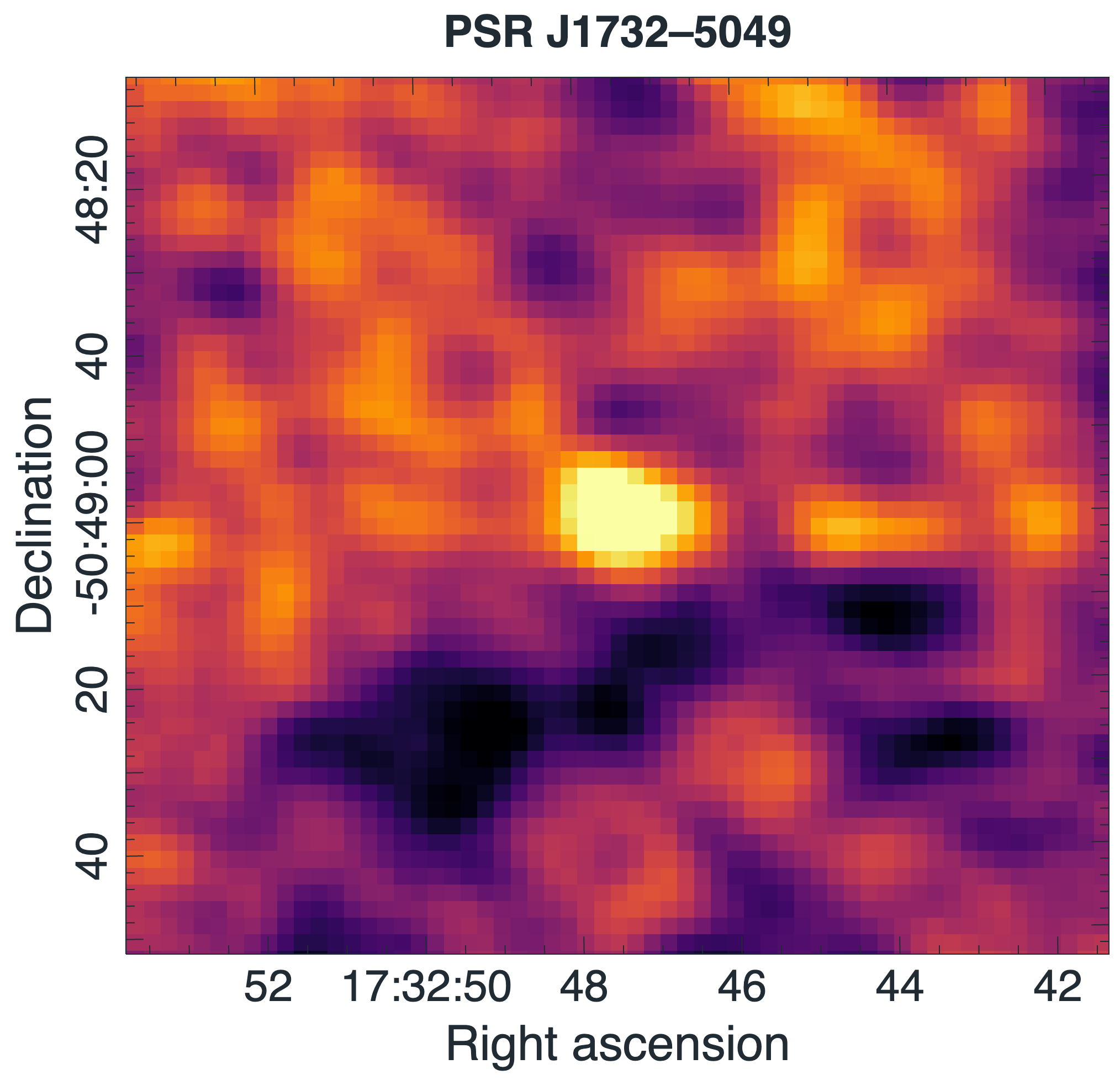}
\includegraphics[width=0.26\linewidth]{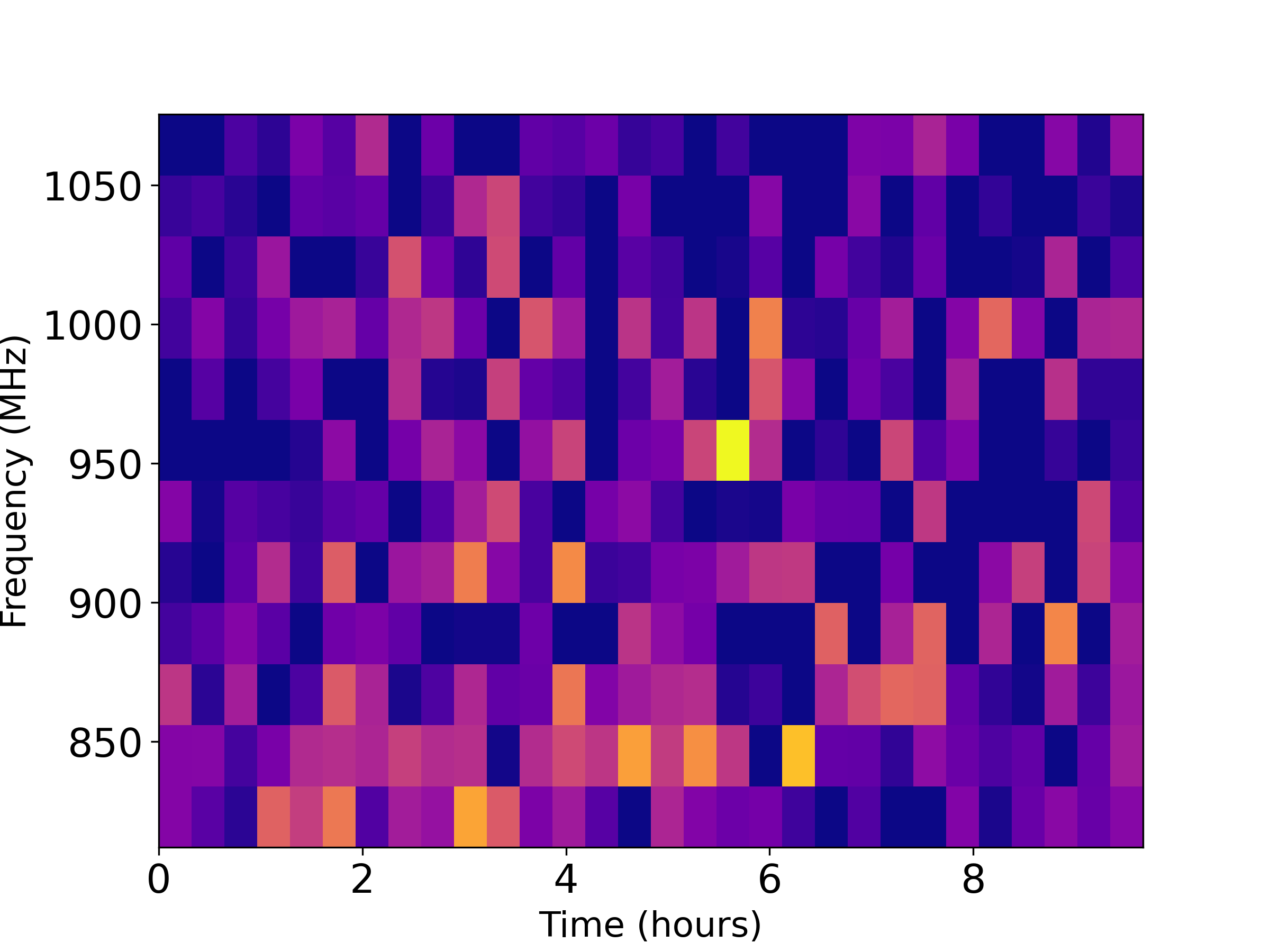}
\includegraphics[width=0.19\linewidth]{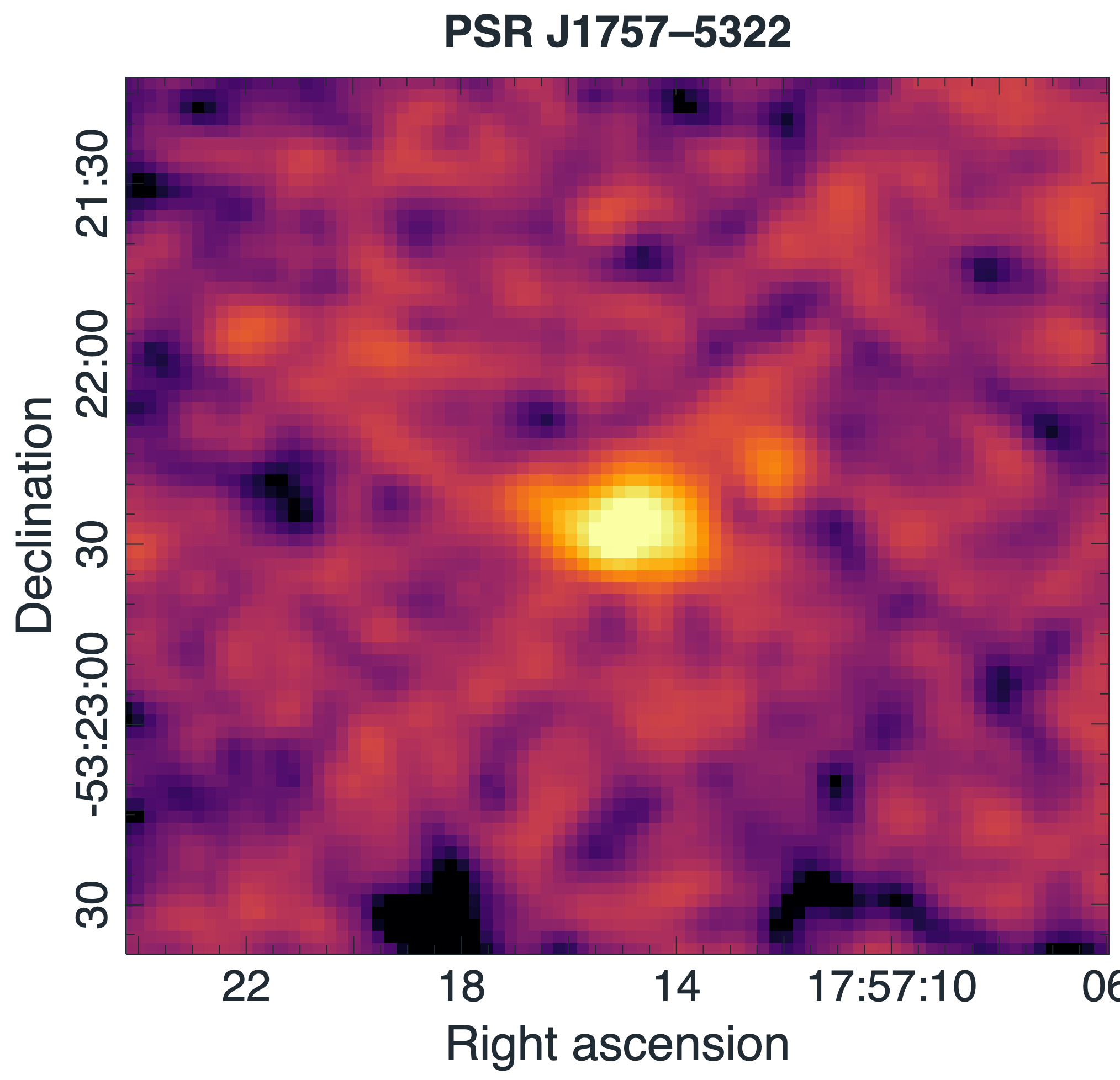}
\includegraphics[width=0.26\linewidth]{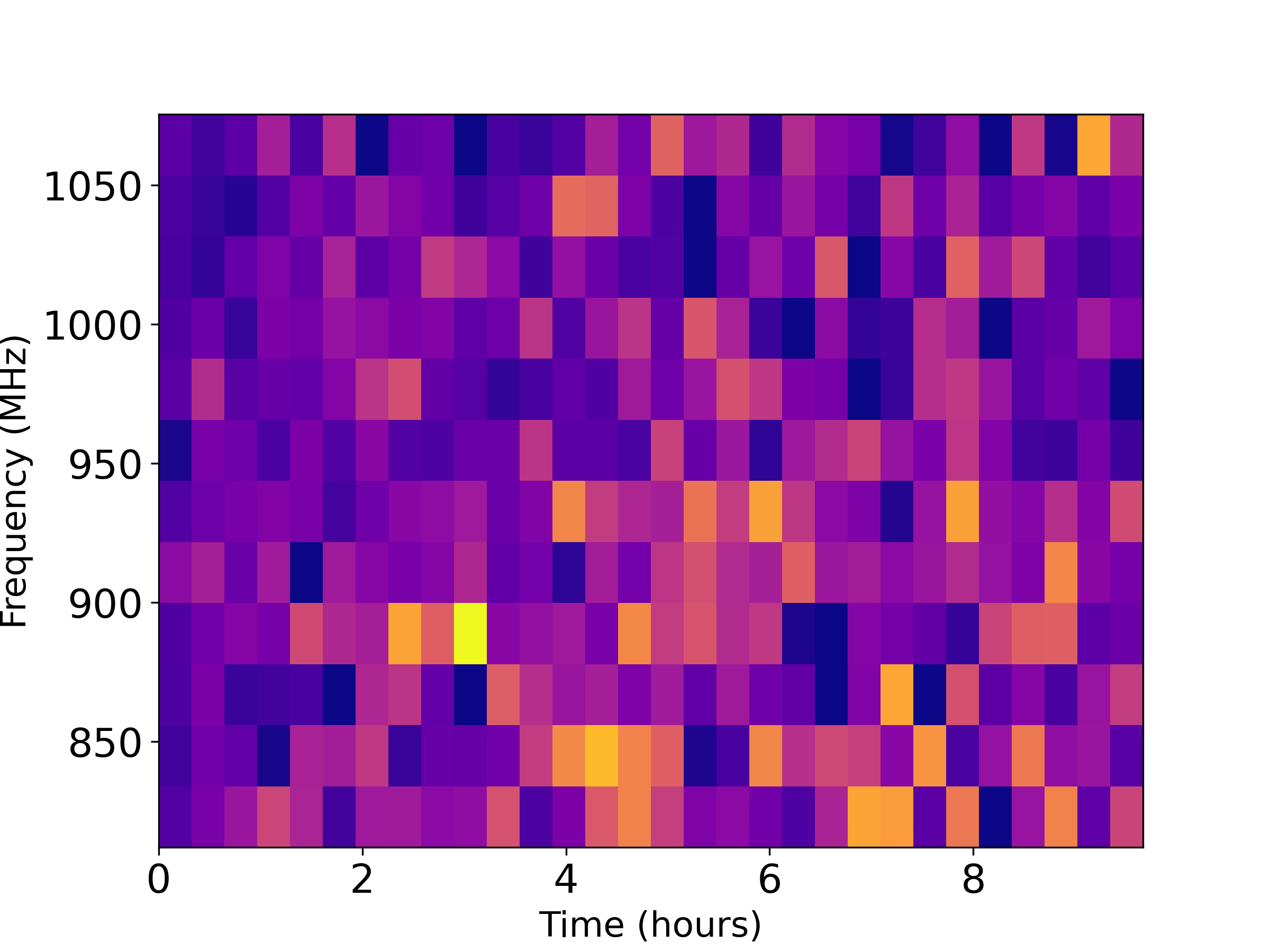}
\includegraphics[width=0.19\linewidth]{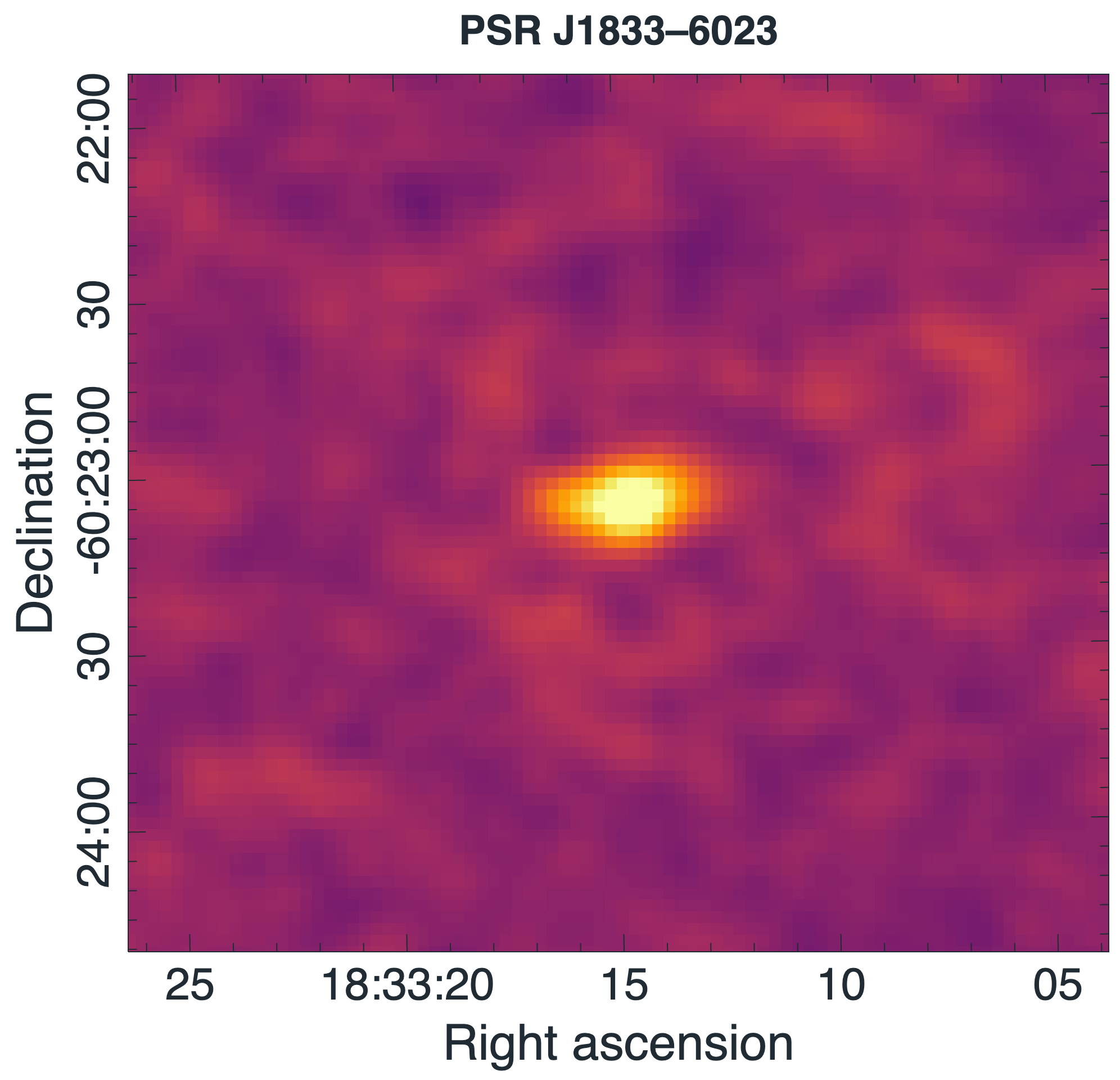}
\includegraphics[width=0.26\linewidth]{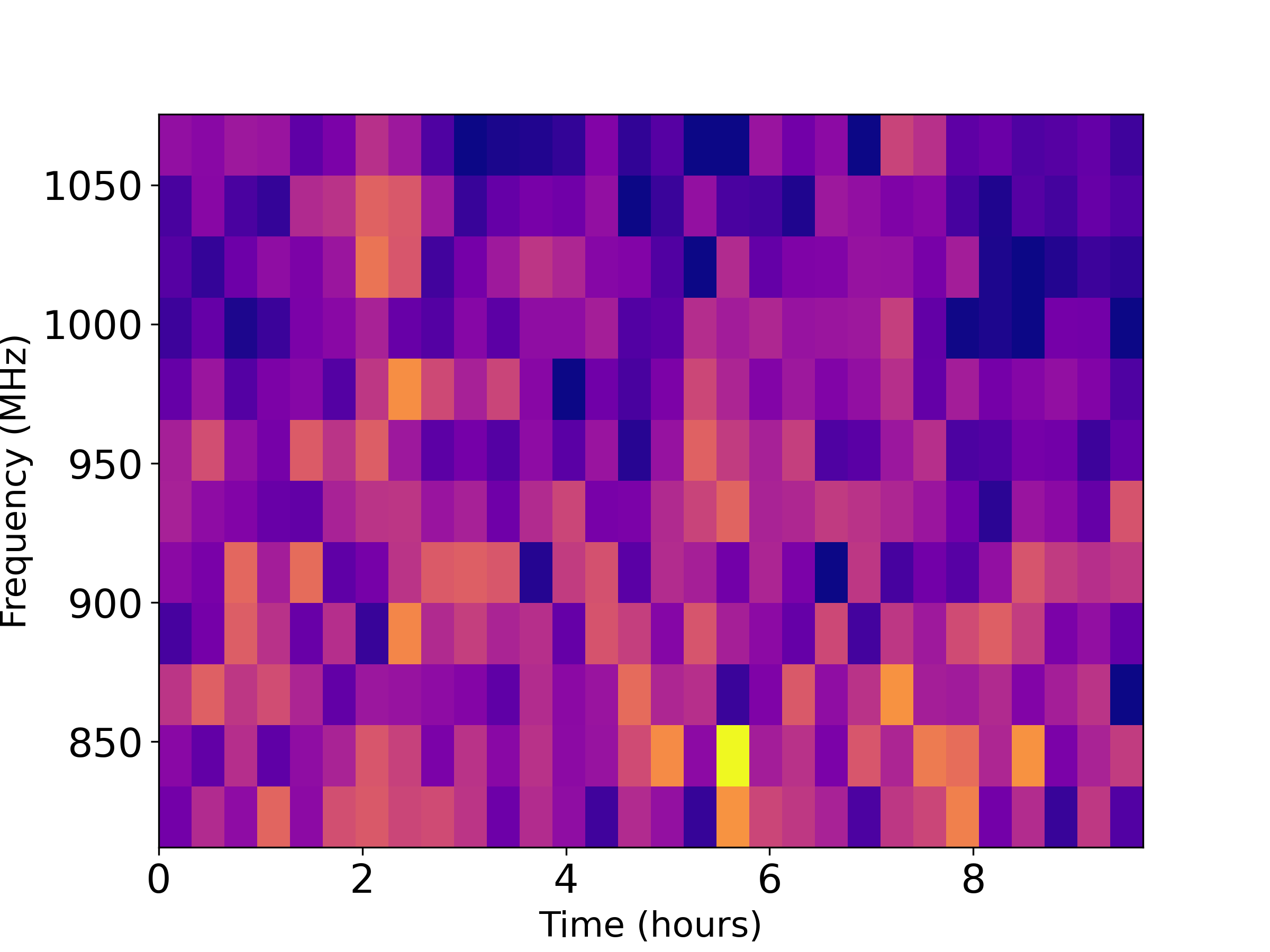}
\caption{Variance images (left) and dynamic spectra (right) of known pulsars, radio star (J1200--4929), and LPT source \citep[J1448--6856;][]{akr+25} detected in variance images. The continuum source J1704--6019 is potentially associated with PSR~J1704--6016 as reported in \citet{wng+23}.}
\label{fig_ds1}
\end{figure*}
\begin{figure*}
\centering
\includegraphics[width=0.19\linewidth]{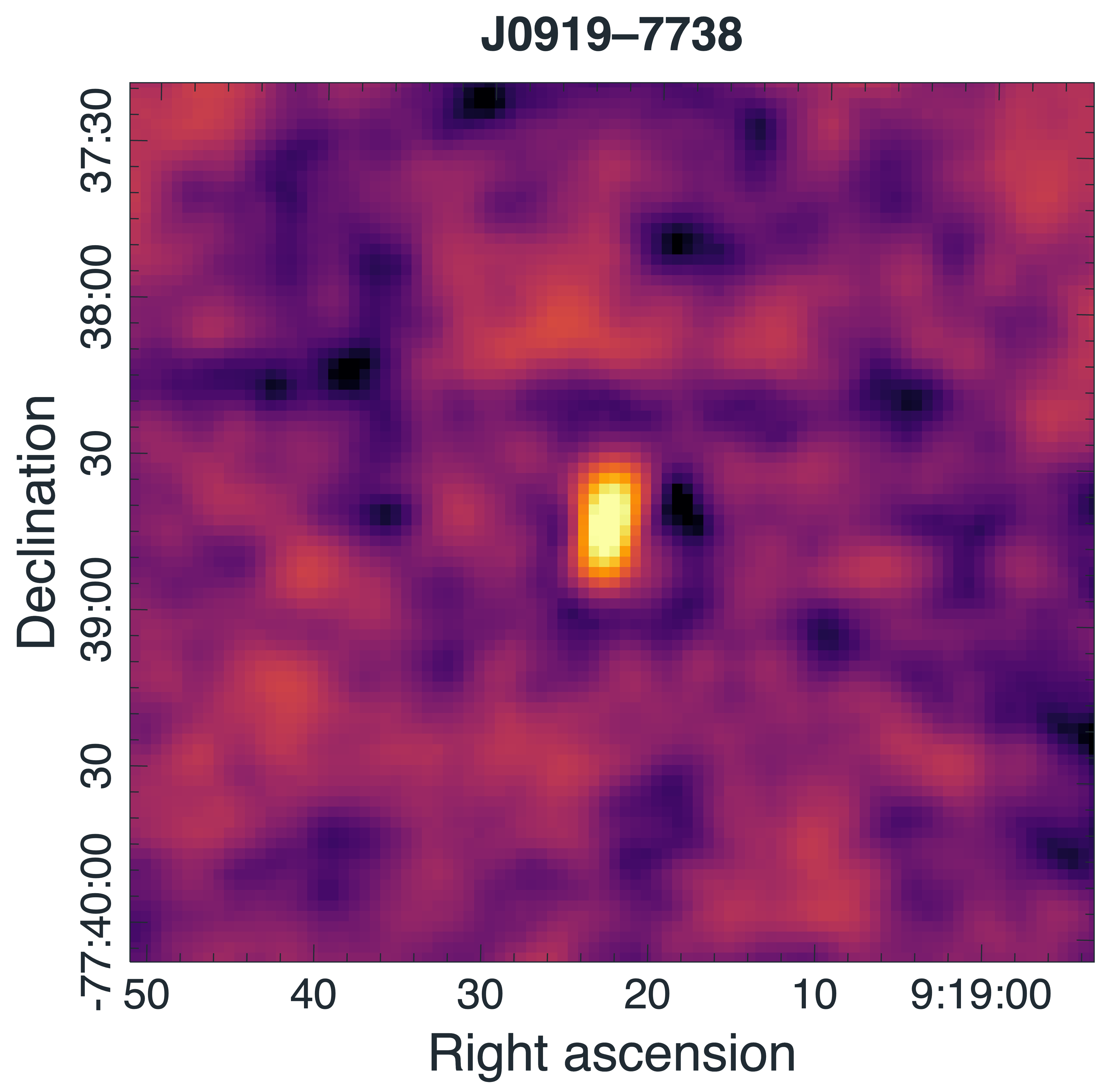}
\includegraphics[width=0.26\linewidth]{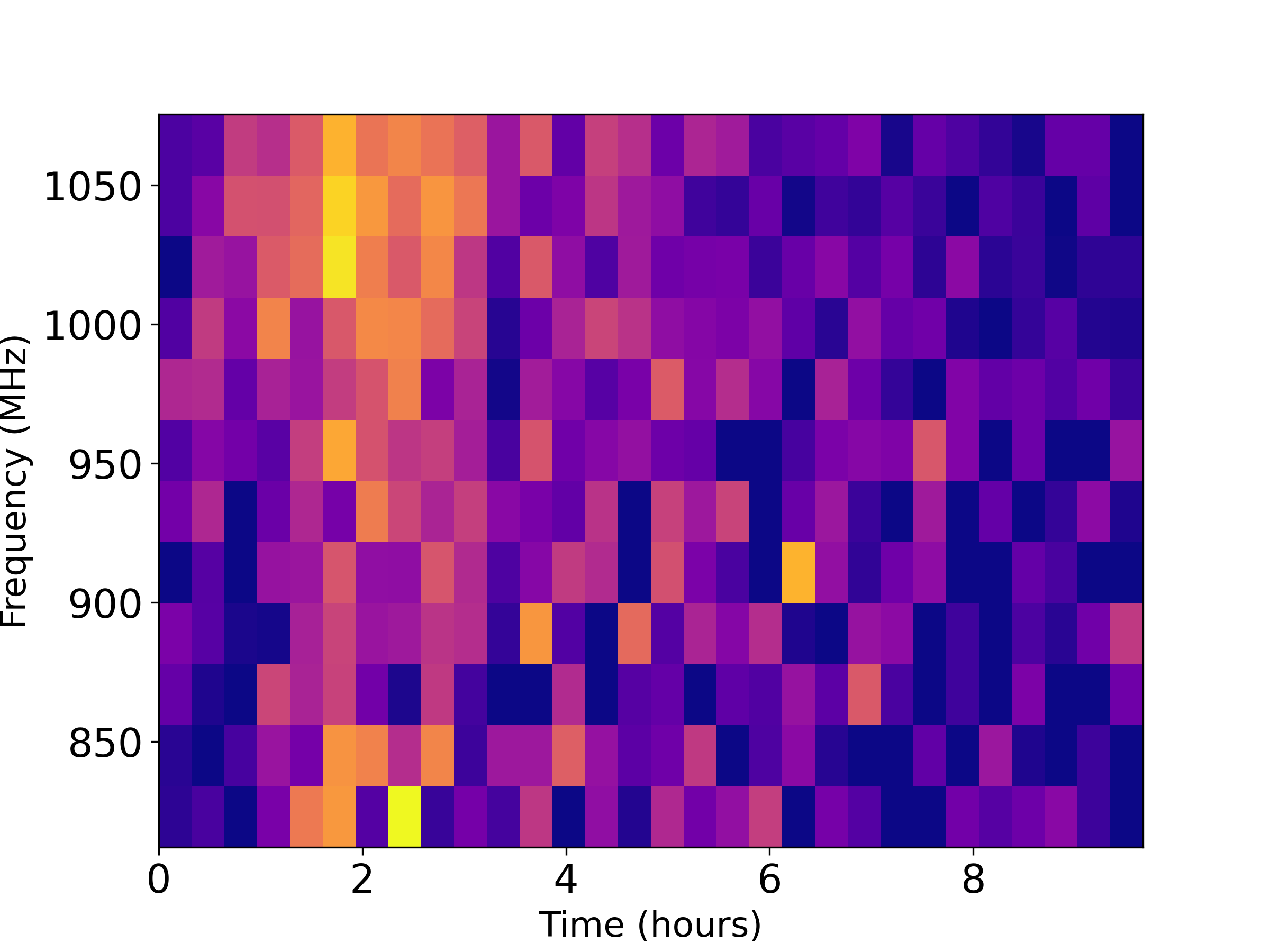}
\includegraphics[width=0.19\linewidth]{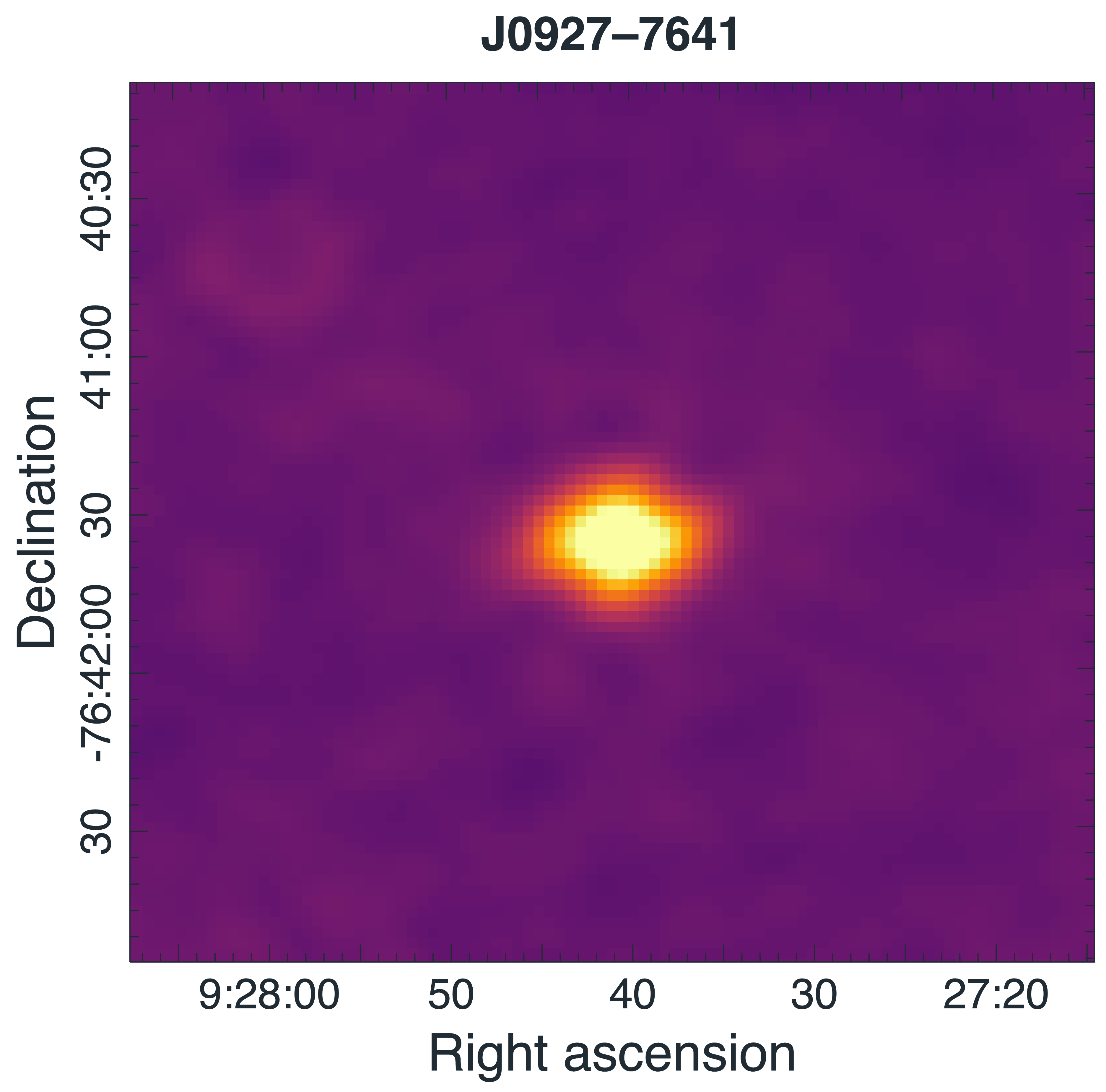}
\includegraphics[width=0.26\linewidth]{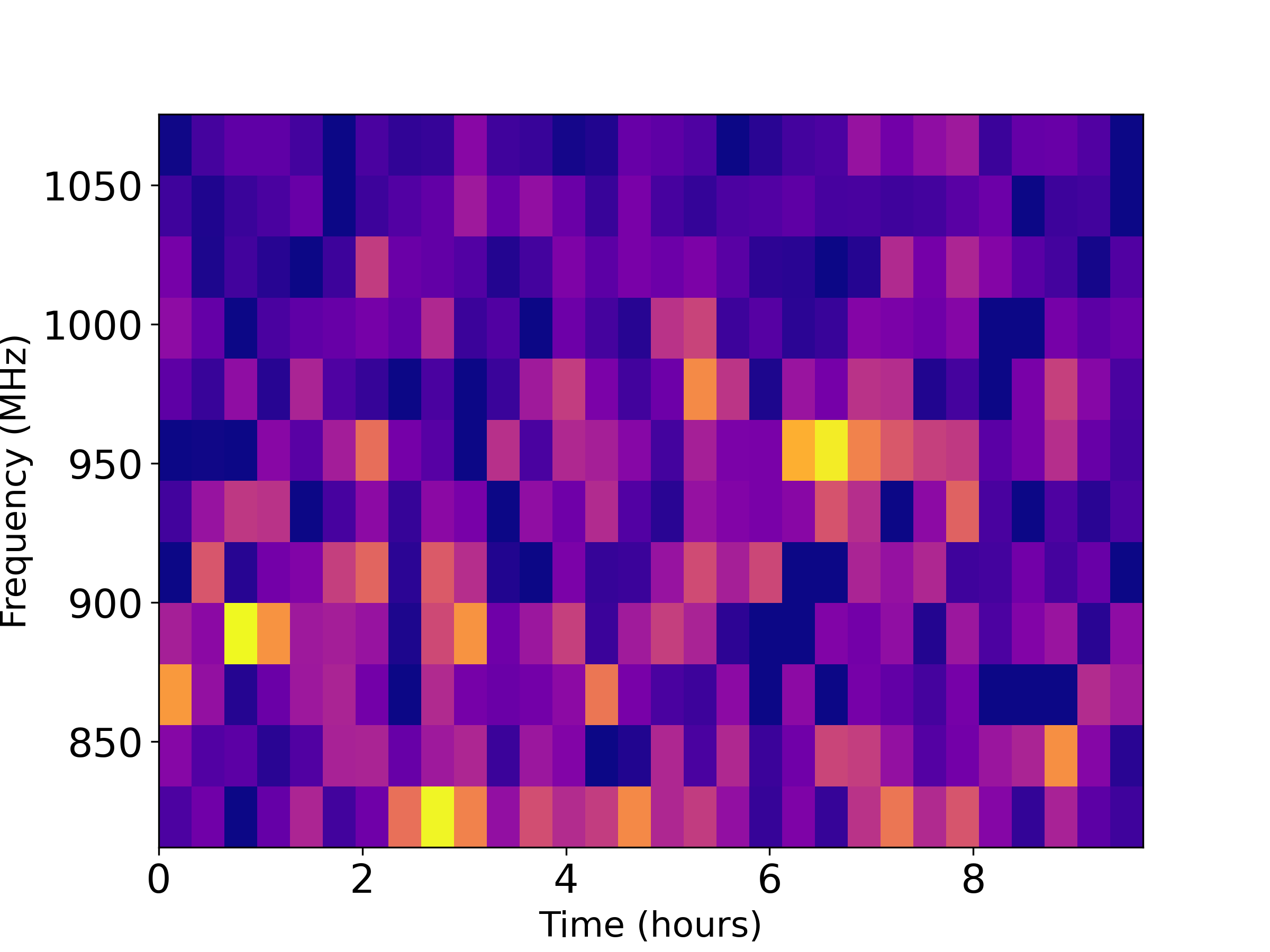}
\includegraphics[width=0.19\linewidth]{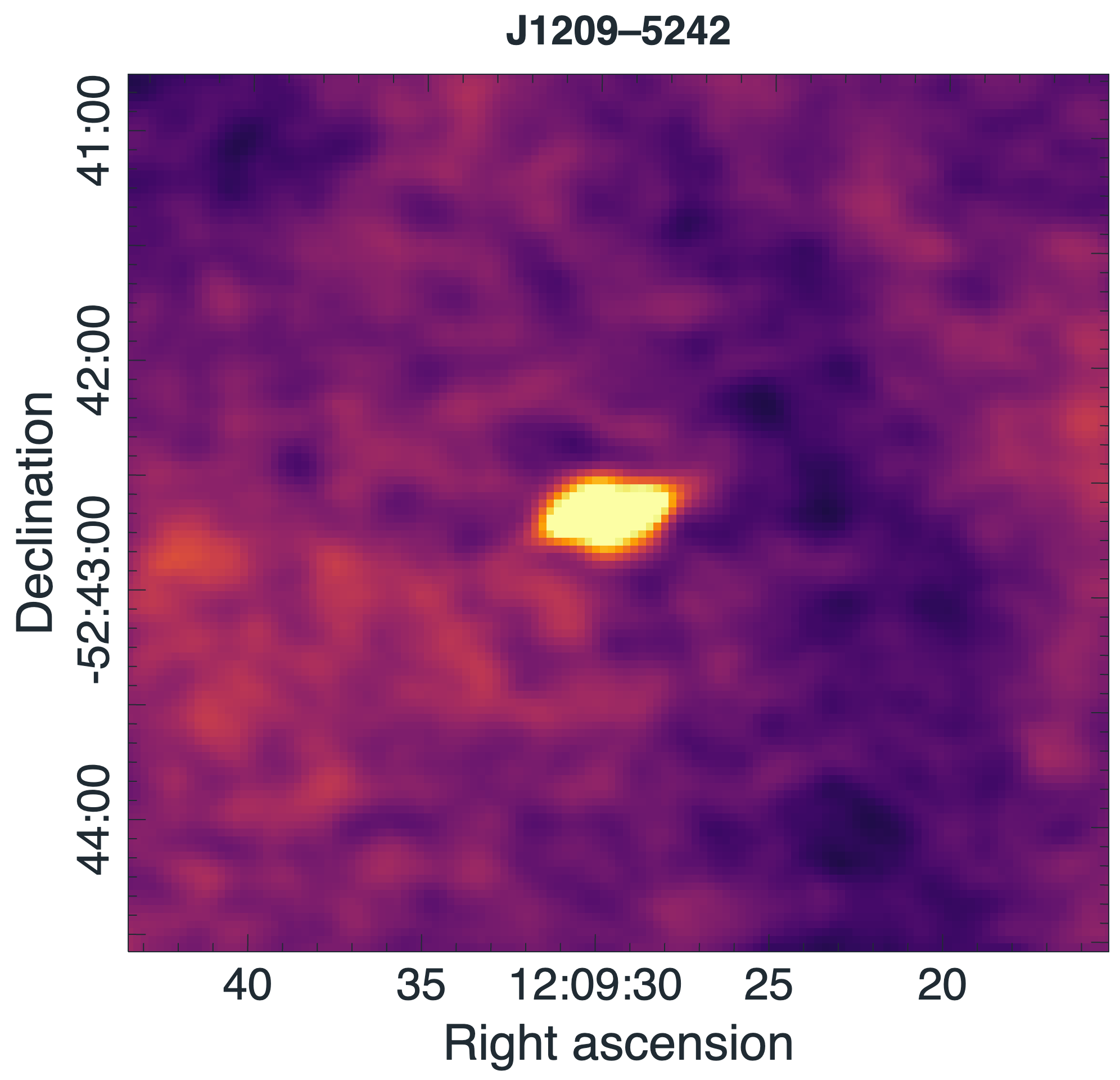}
\includegraphics[width=0.26\linewidth]{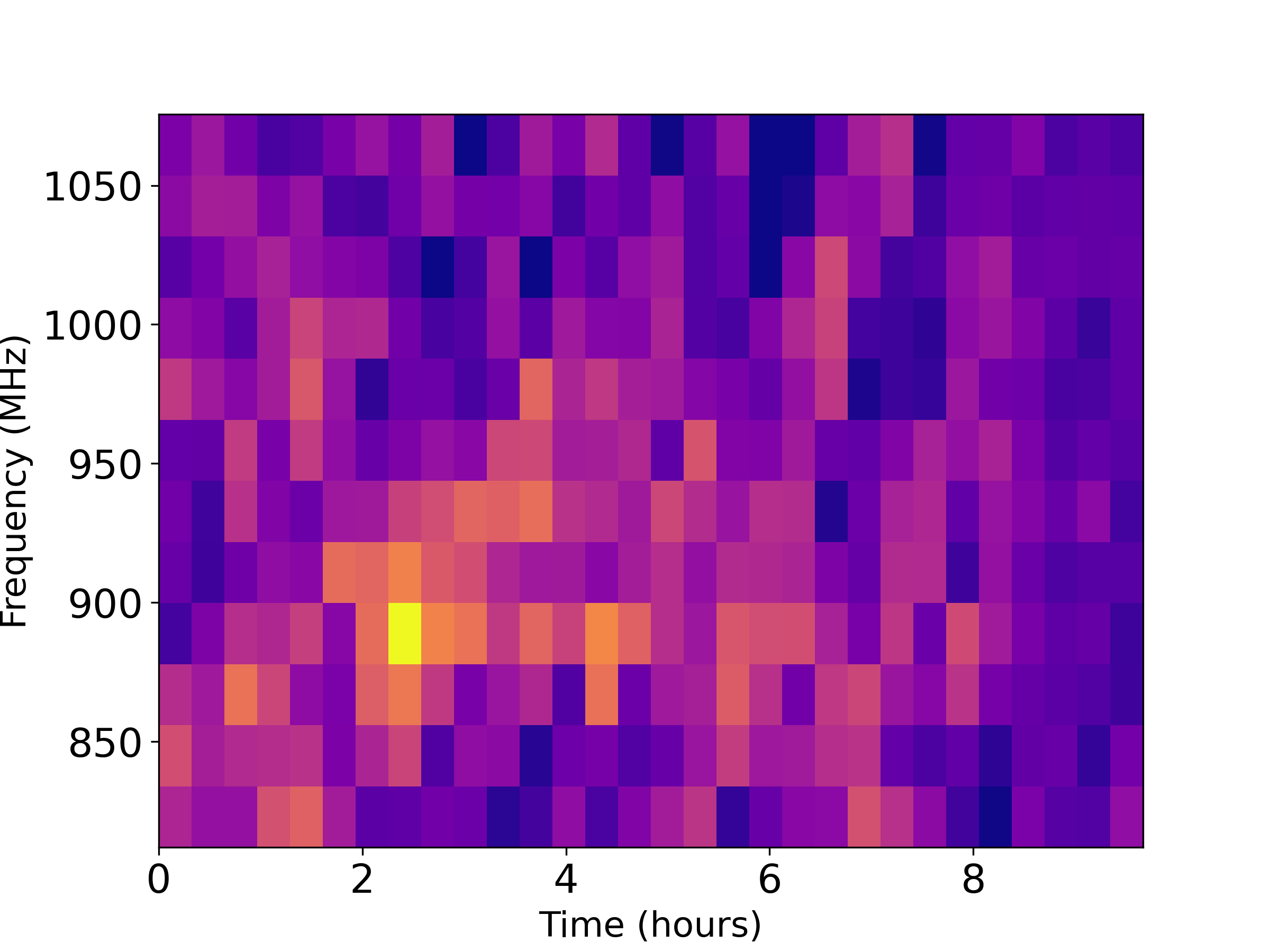}
\includegraphics[width=0.19\linewidth]{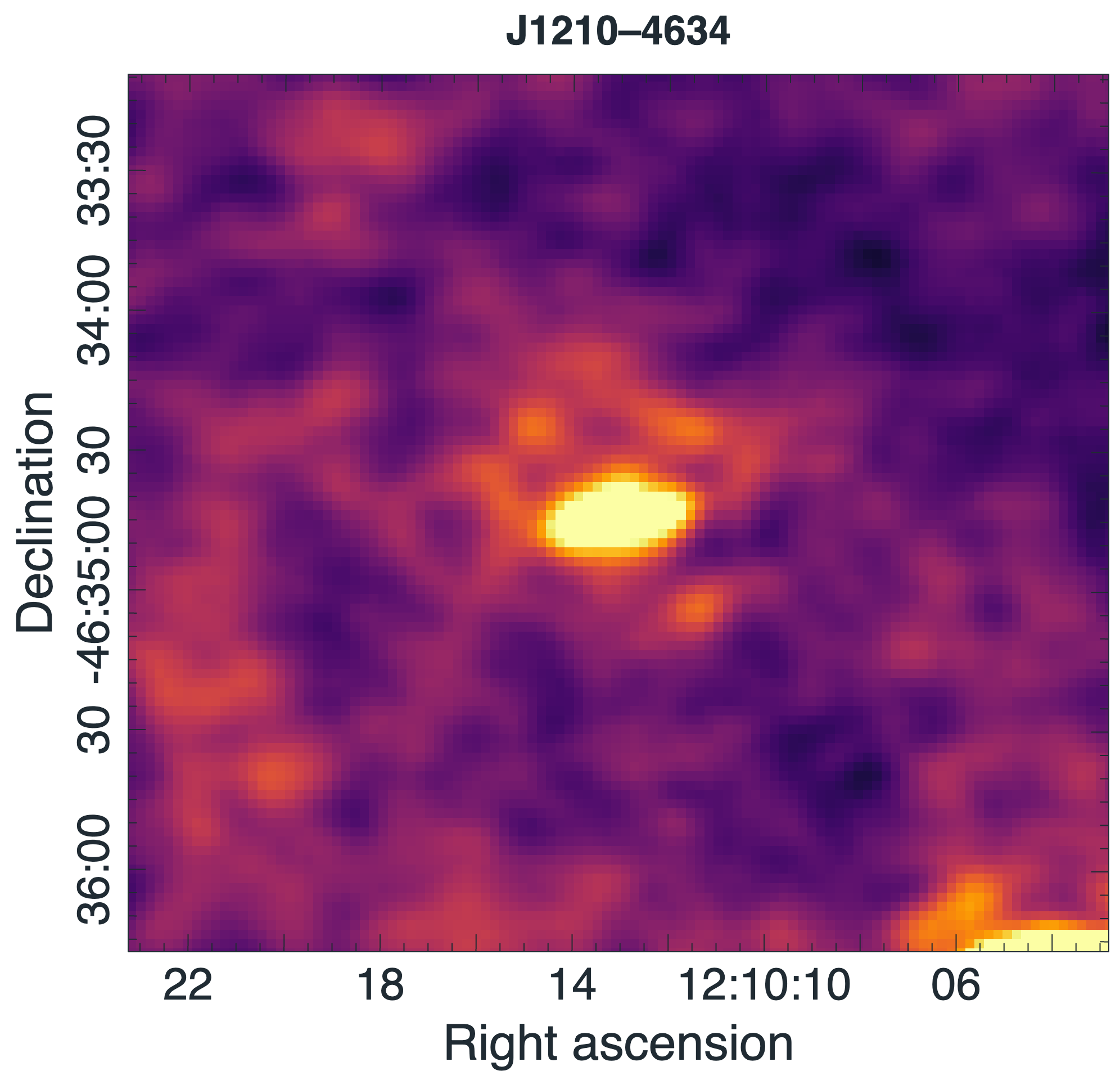}
\includegraphics[width=0.26\linewidth]{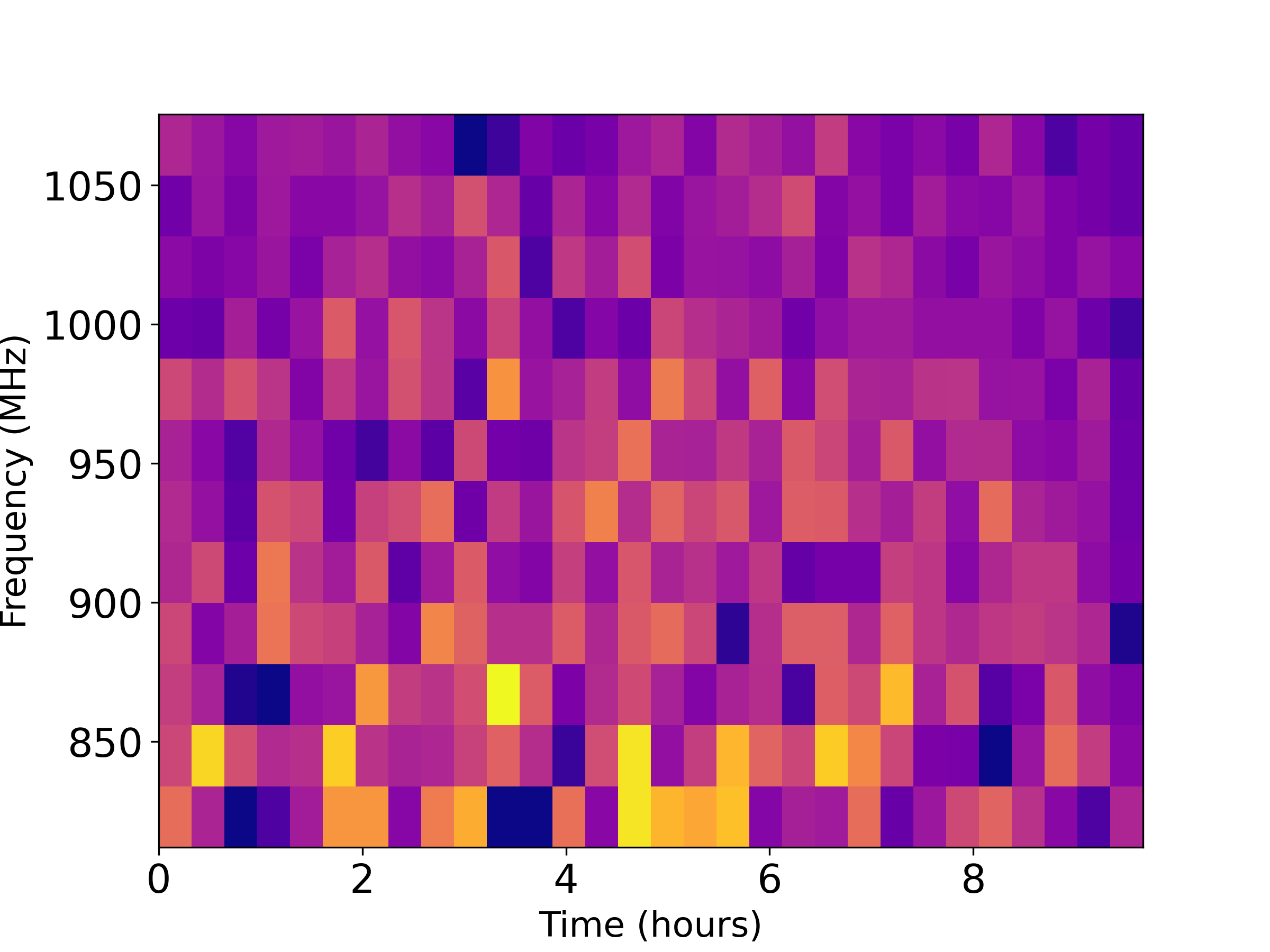}
\includegraphics[width=0.19\linewidth]{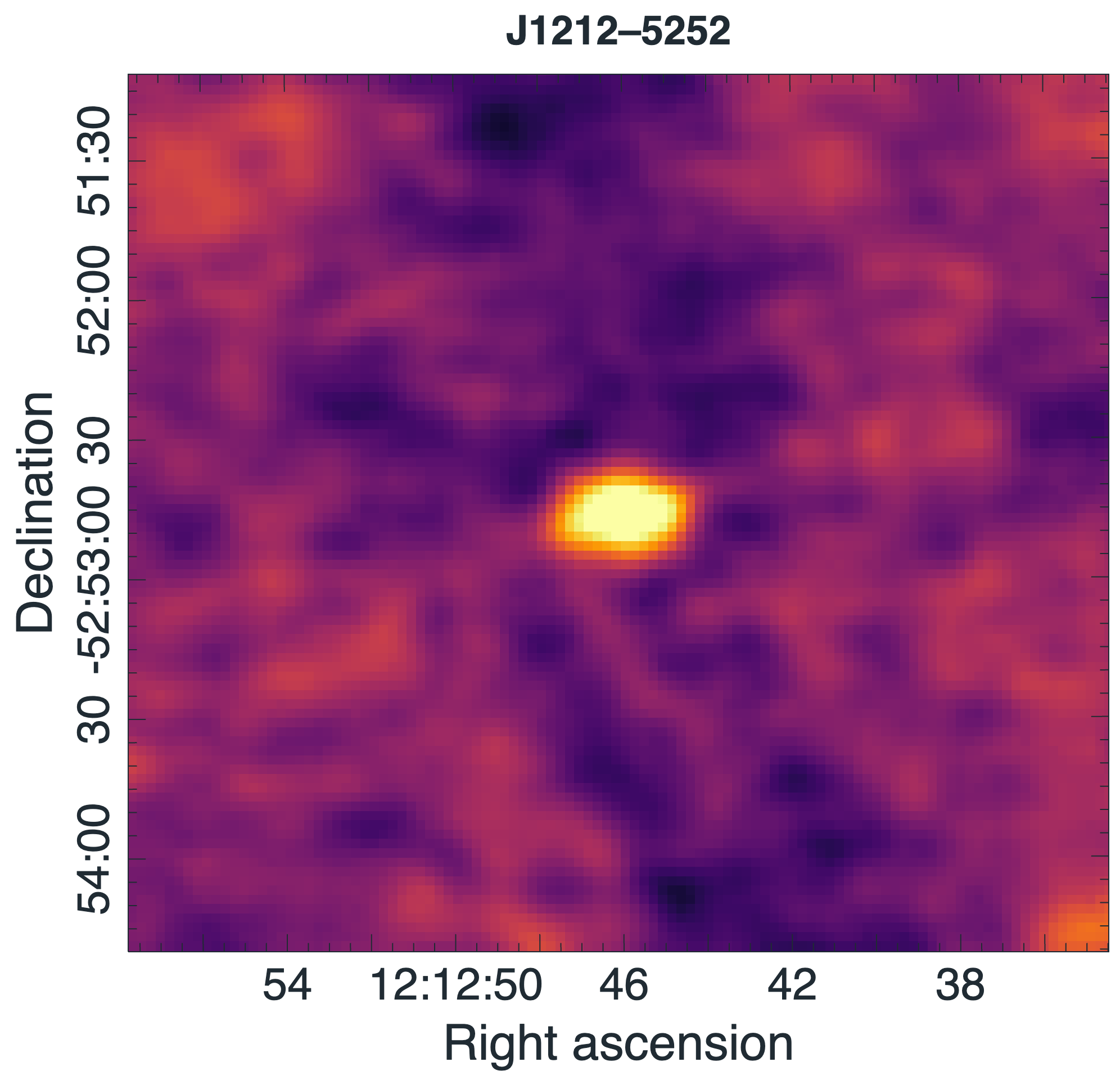}
\includegraphics[width=0.26\linewidth]{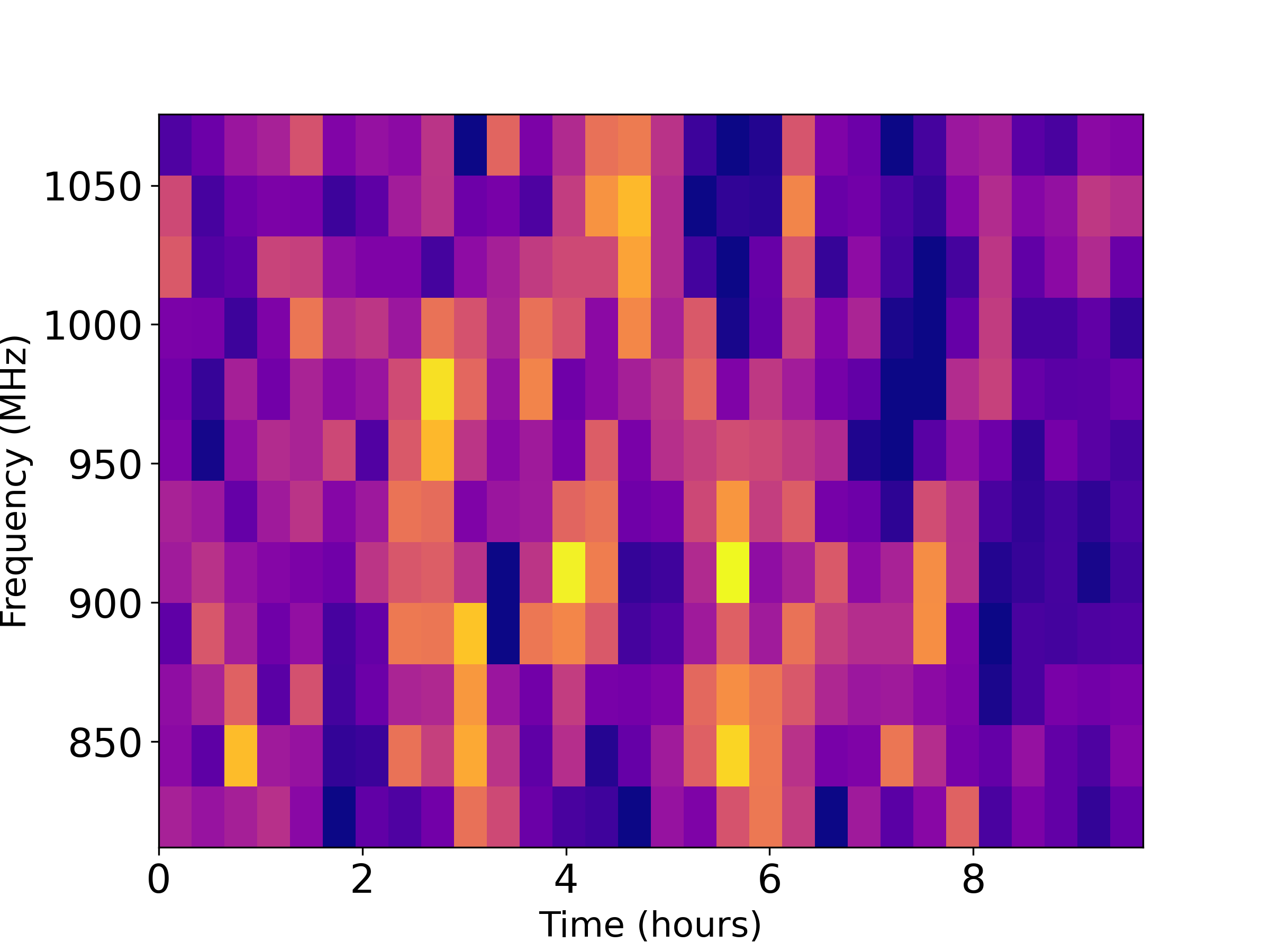}
\includegraphics[width=0.19\linewidth]{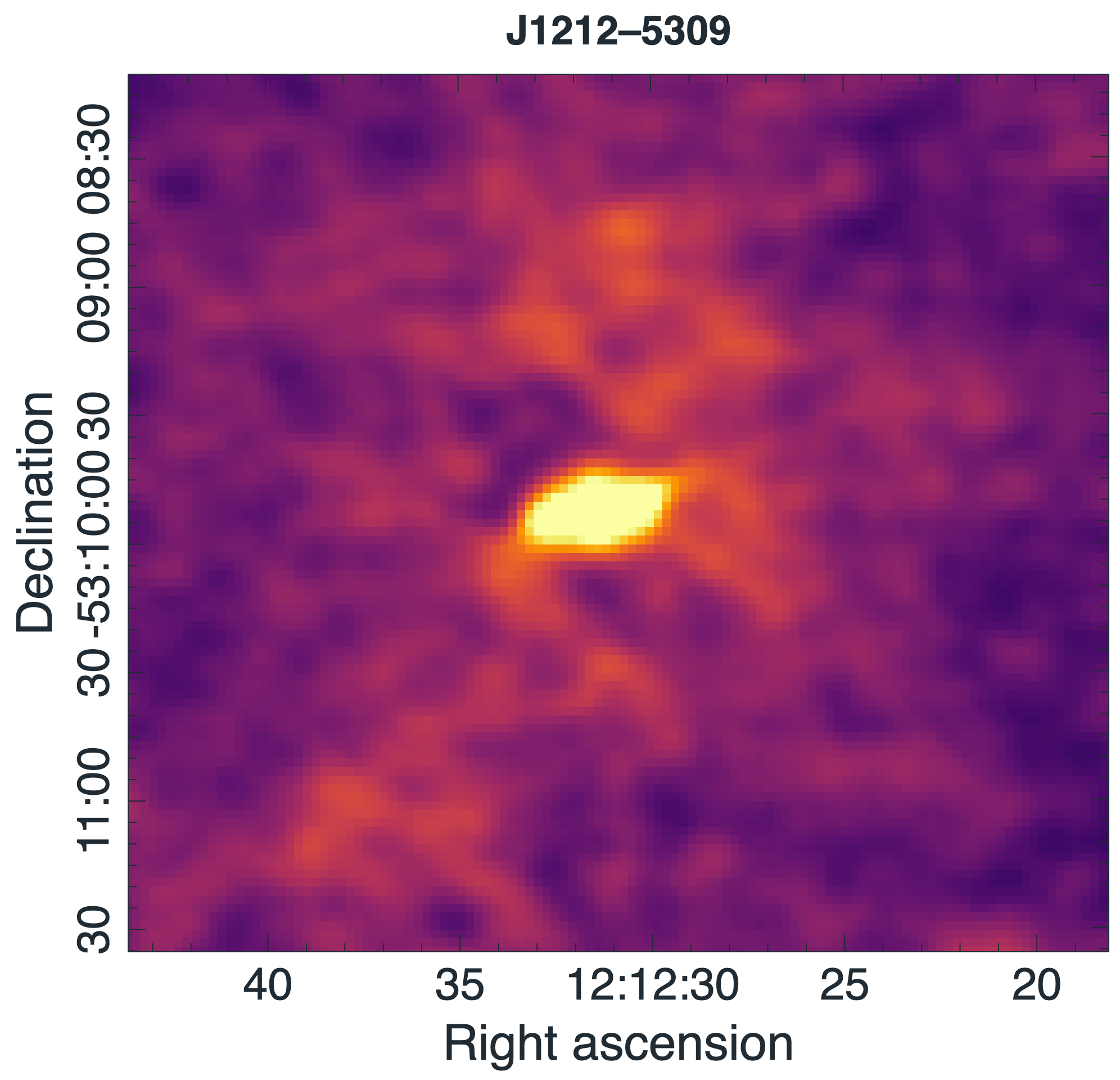}
\includegraphics[width=0.26\linewidth]{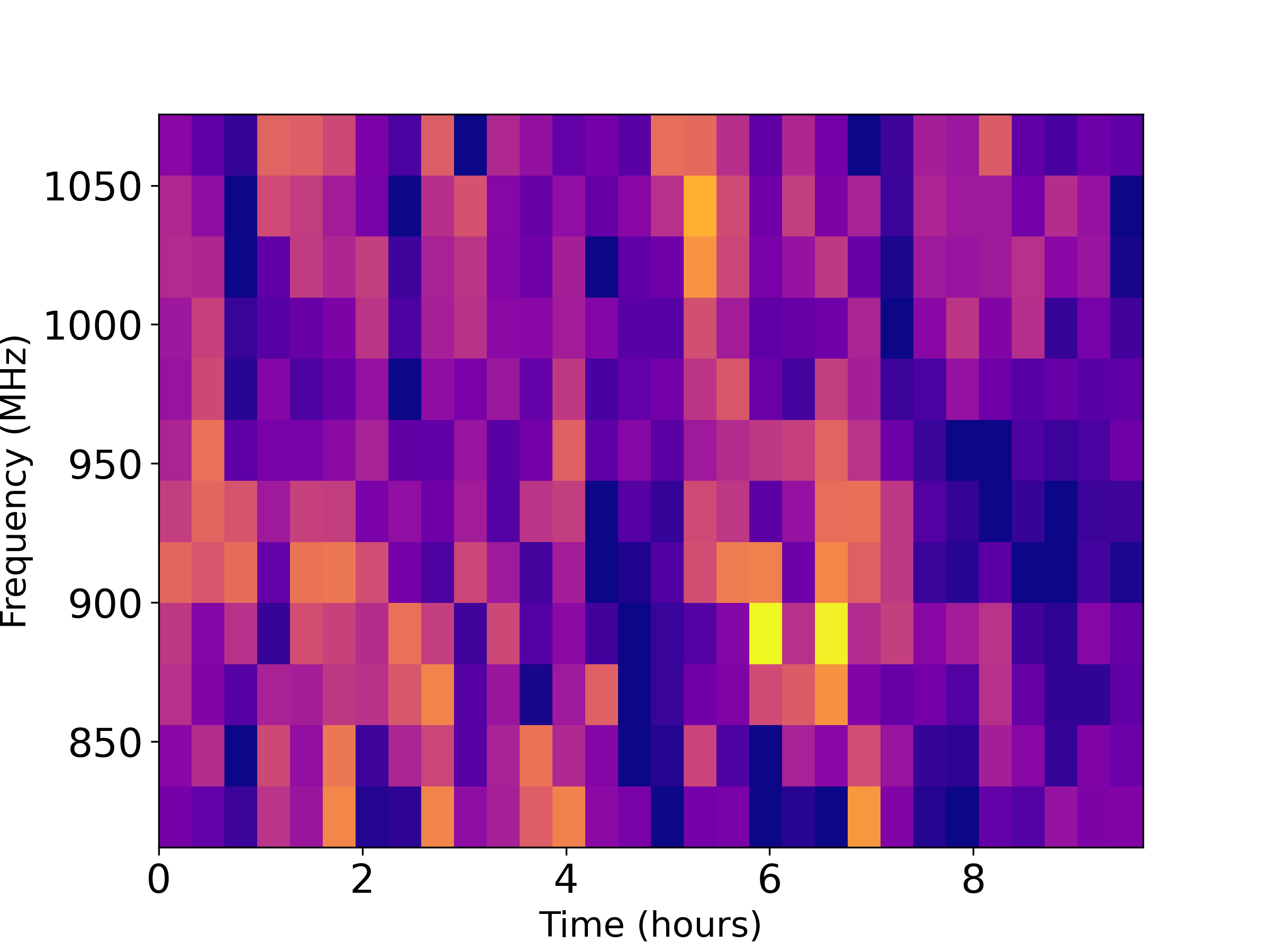}
\includegraphics[width=0.19\linewidth]{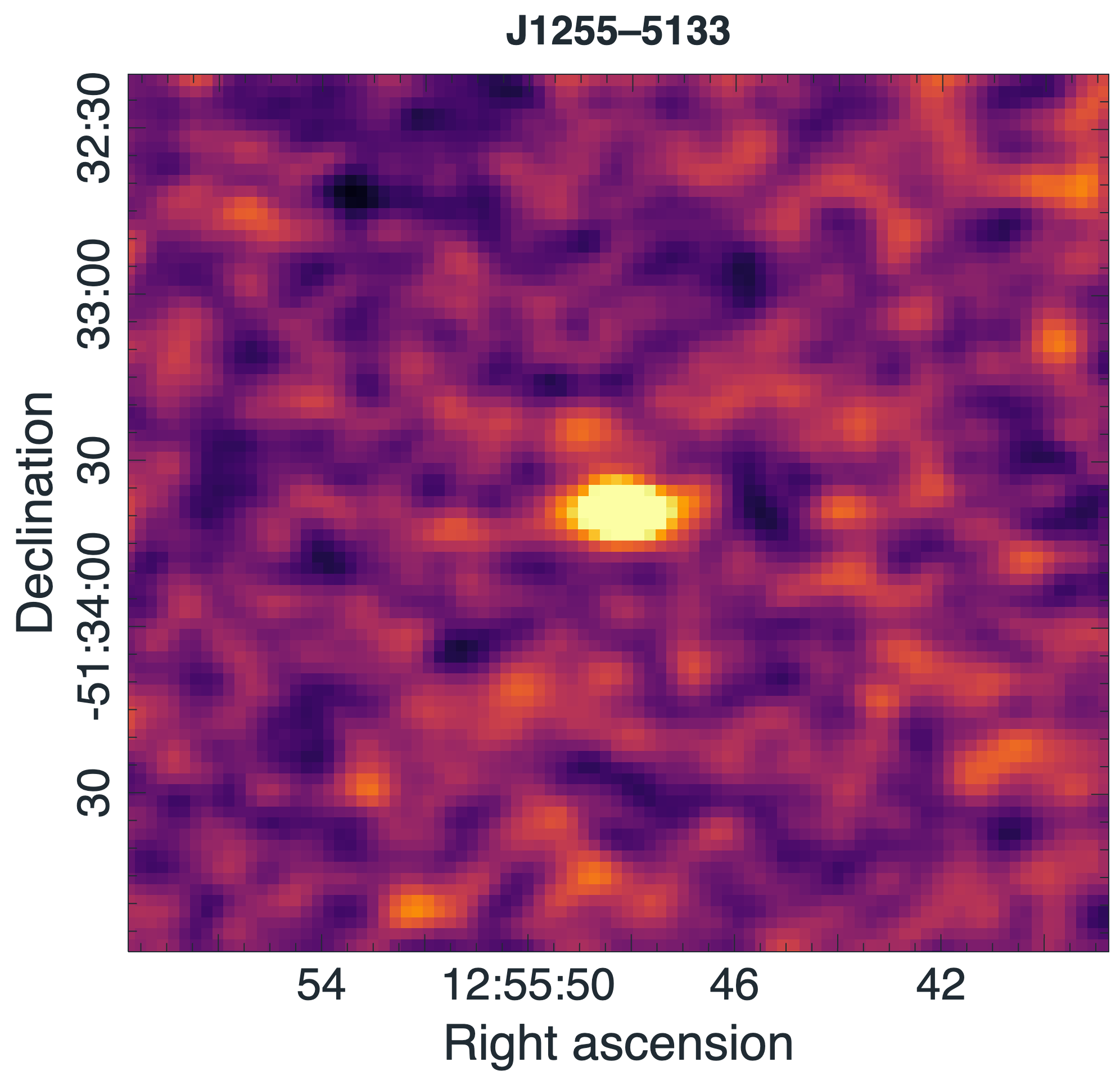}
\includegraphics[width=0.26\linewidth]{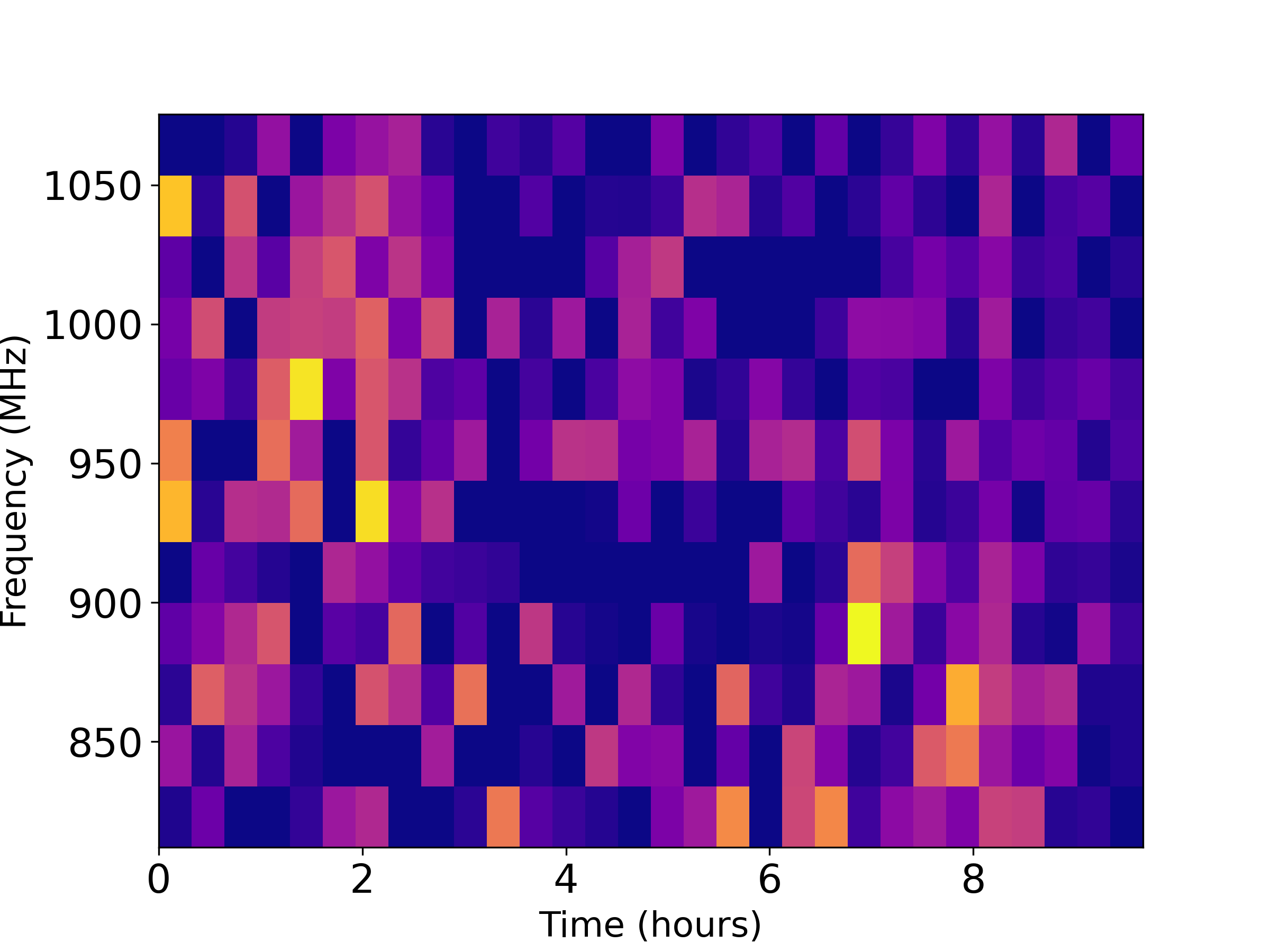}
\includegraphics[width=0.19\linewidth]{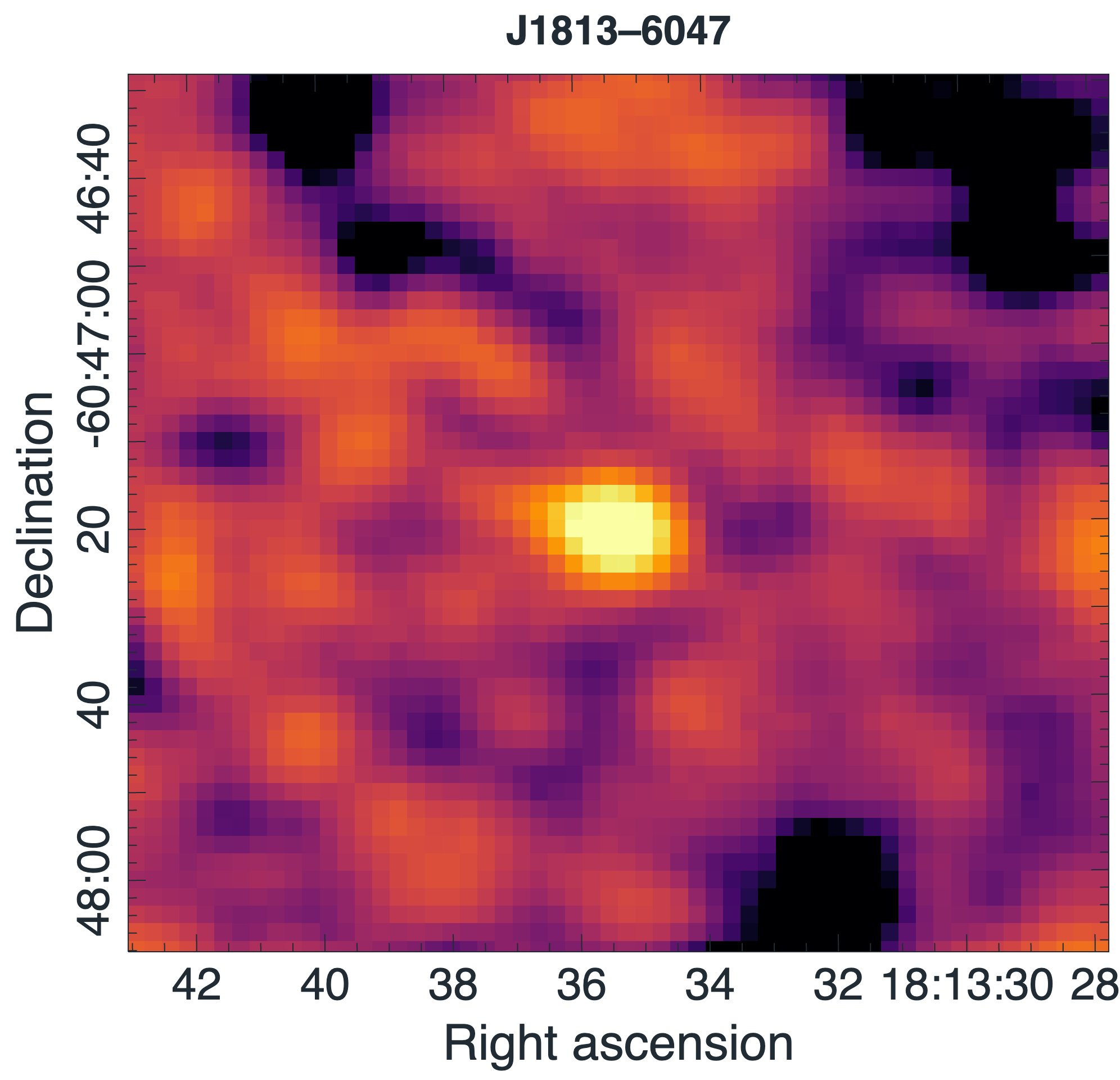}
\includegraphics[width=0.26\linewidth]{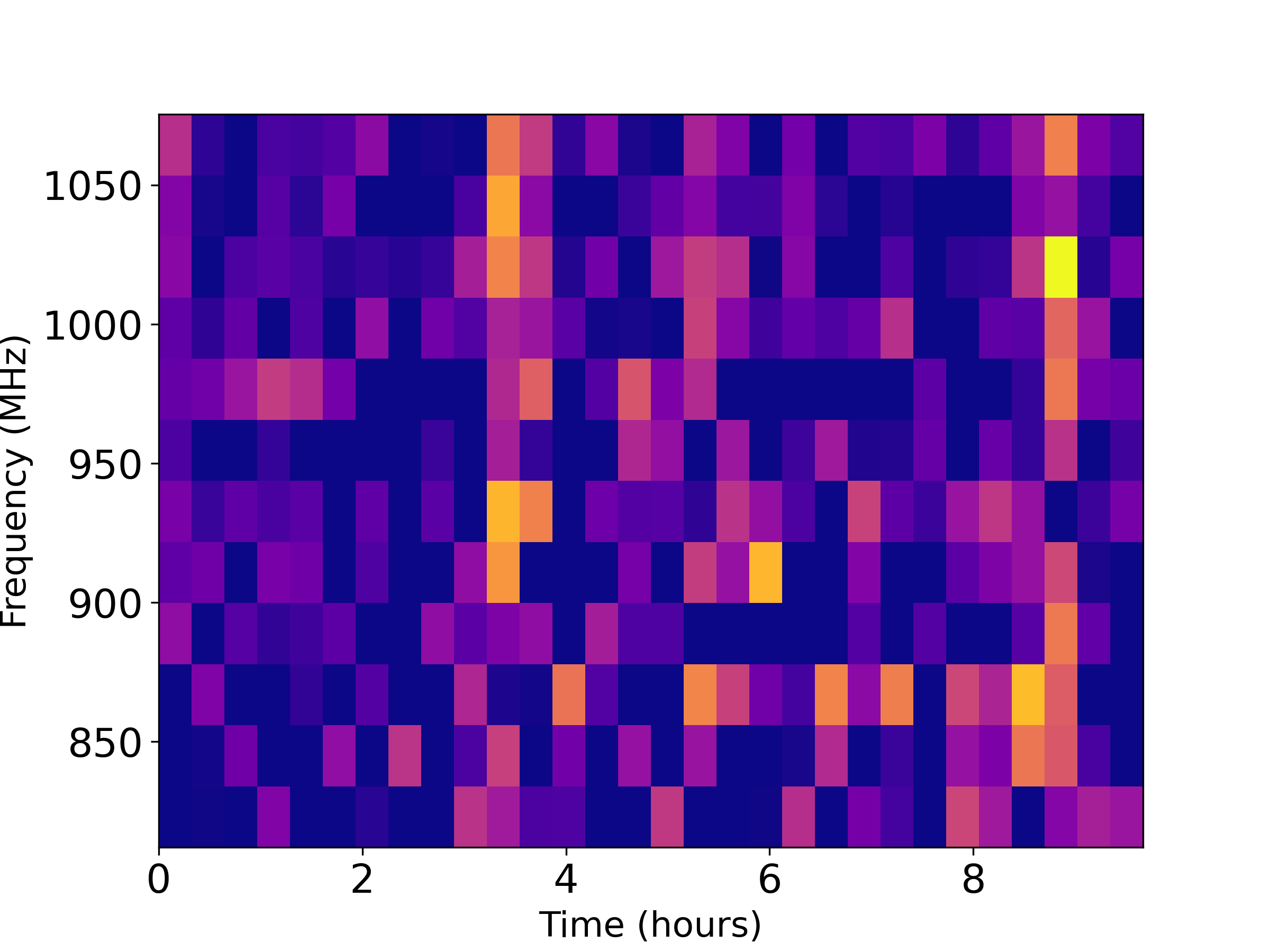}
\includegraphics[width=0.19\linewidth]{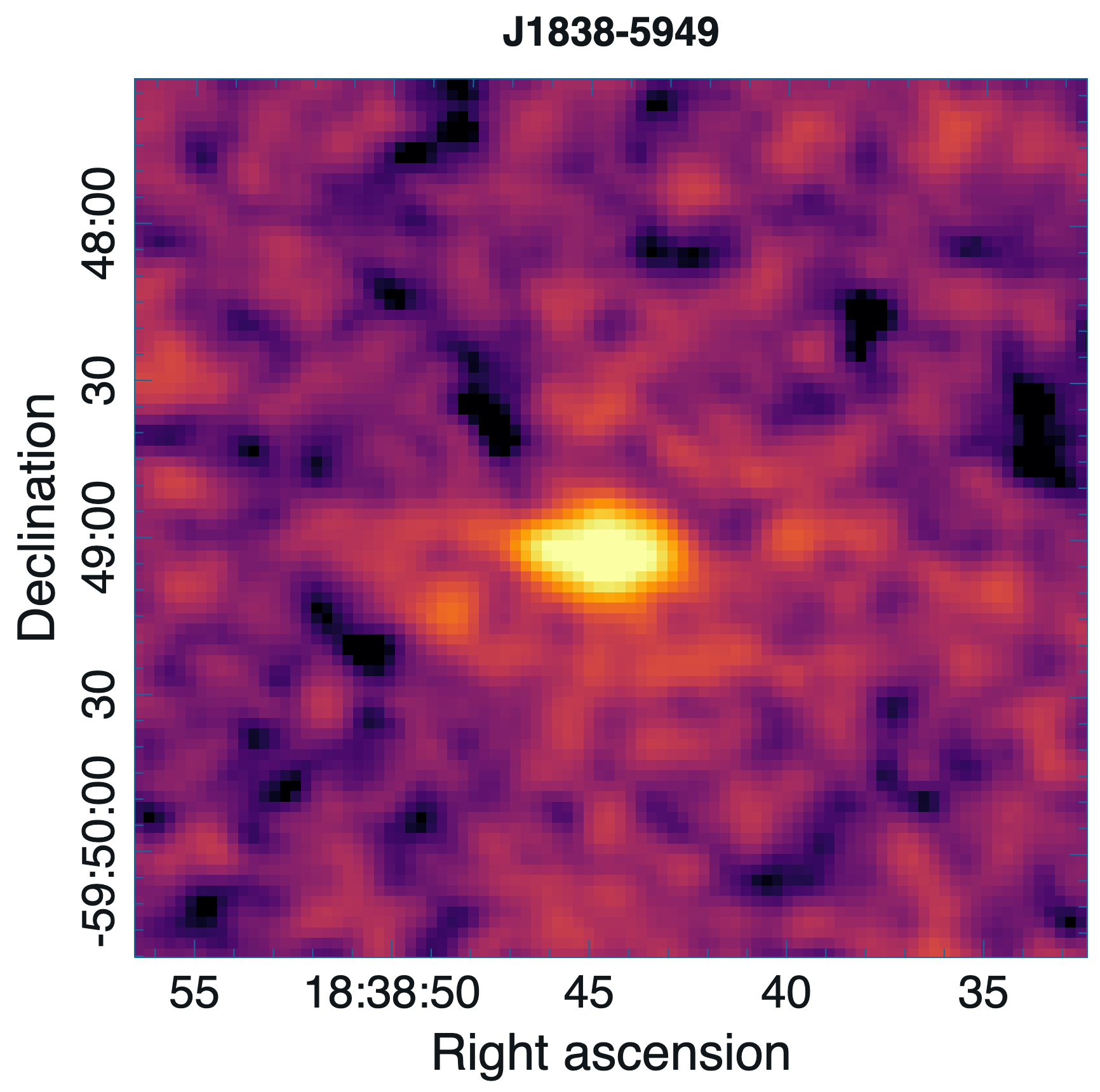}
\includegraphics[width=0.26\linewidth]{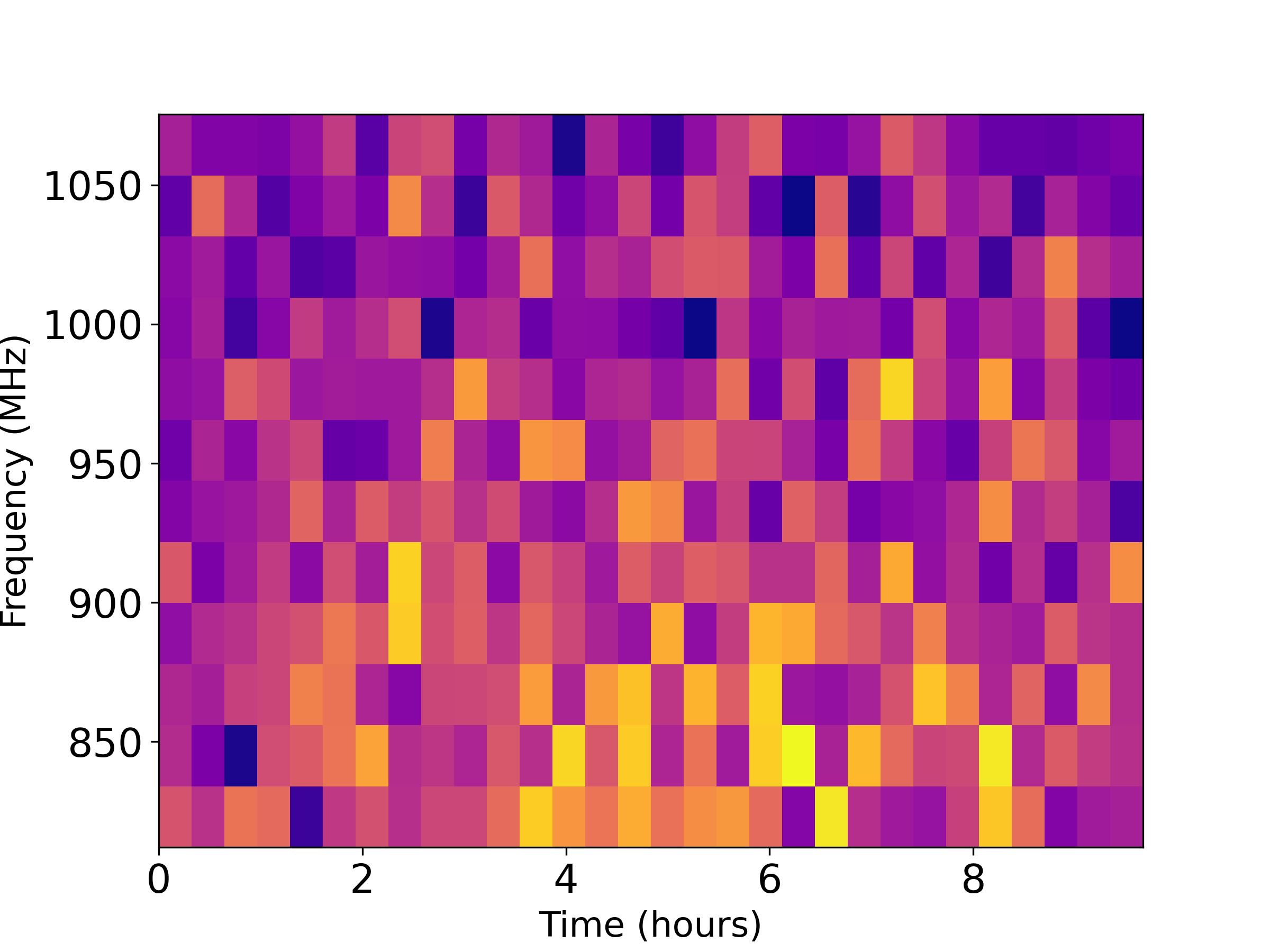}
\includegraphics[width=0.19\linewidth]{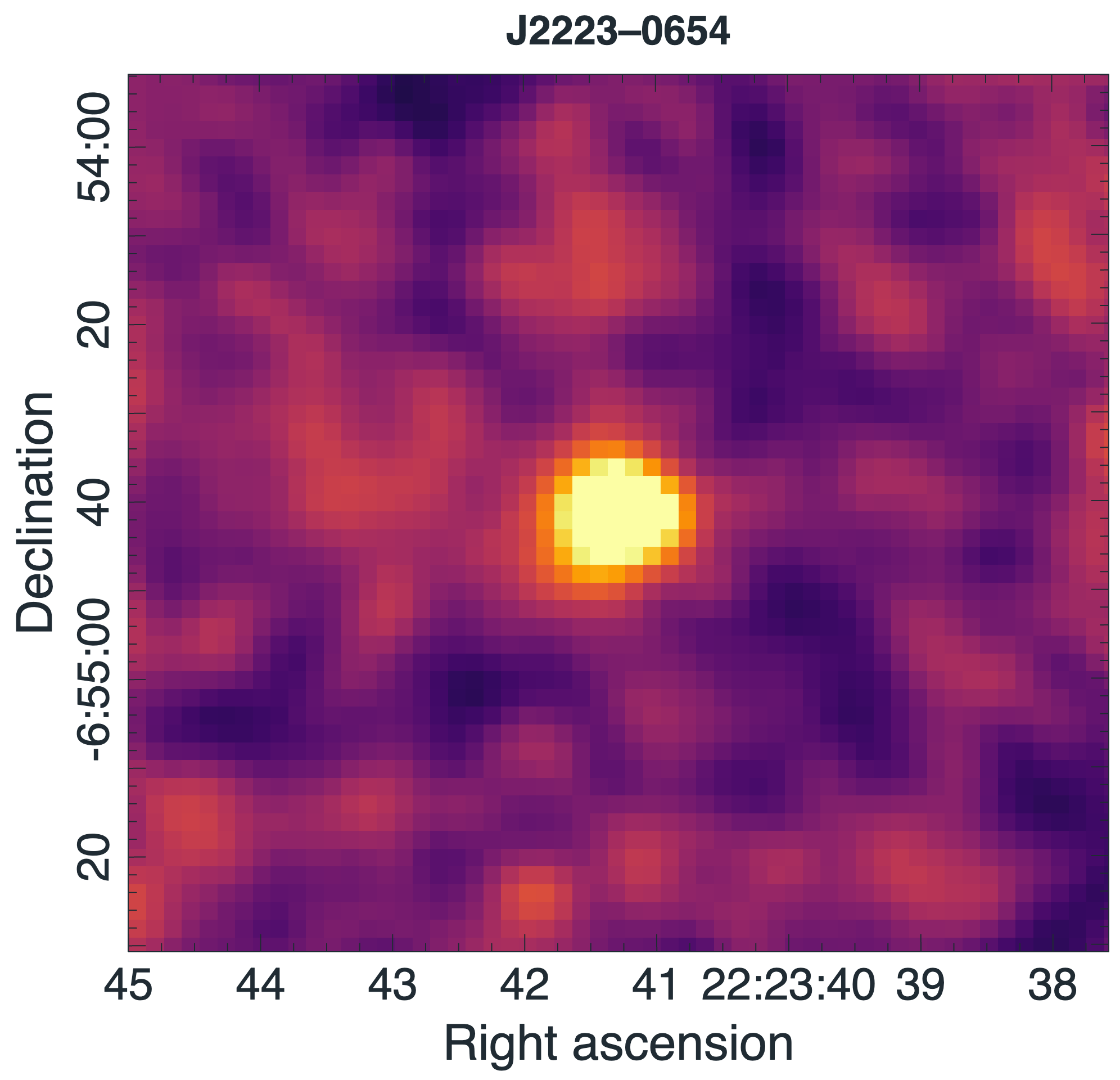}
\includegraphics[width=0.26\linewidth]{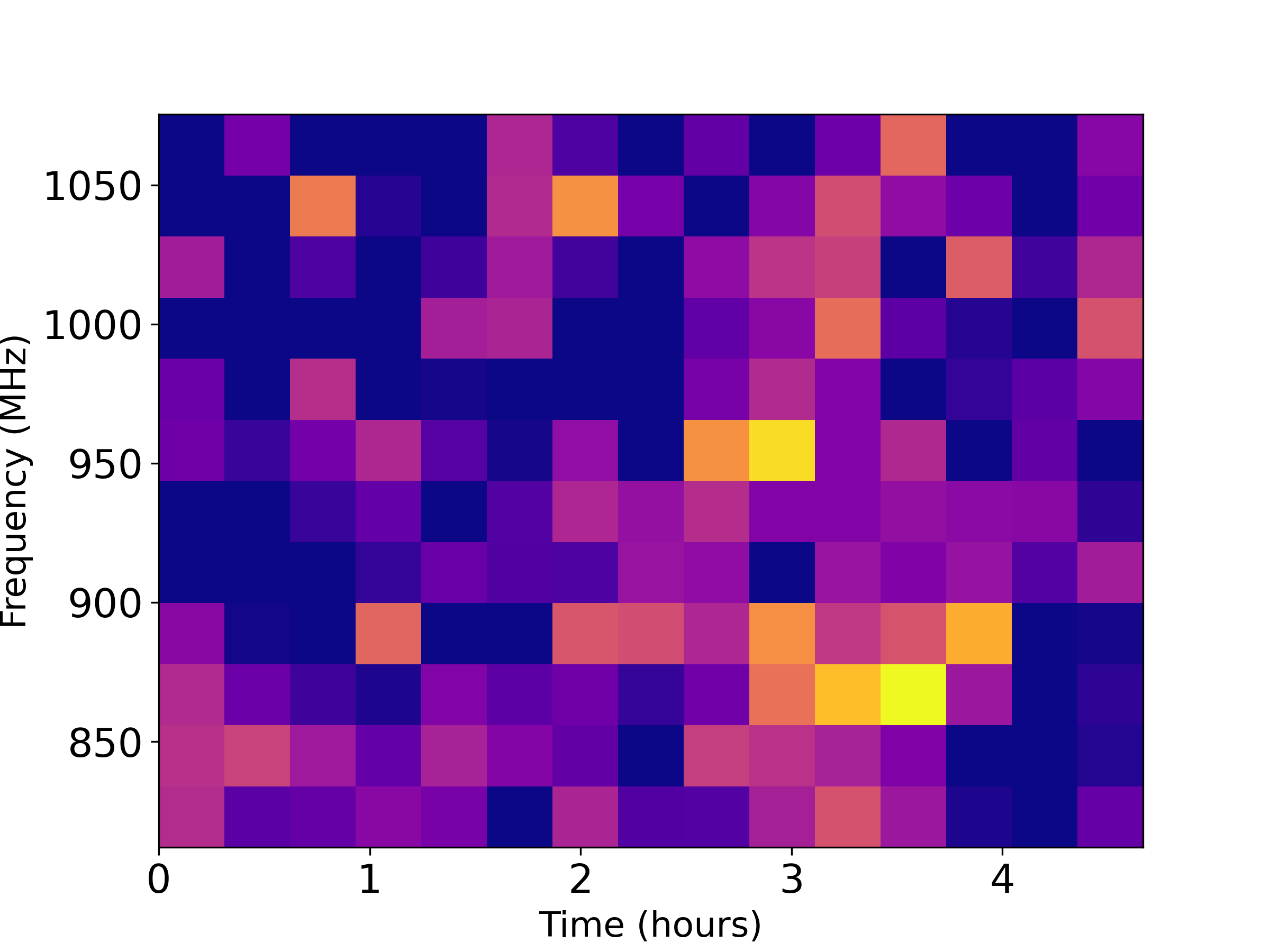}
\caption{Variance images (left) and dynamic spectra (right) of pulsar and radio star candidates (J0919--7738, J1255--5133, and J1813--6047) detected in variance images.}
\label{fig_DScand}
\end{figure*}

\begin{table*}
\caption{20 highly variable radio sources detected via variance imaging are listed in this table. Columns include source names, coordinates (RAJ $\&$ DECJ), schedule block ID (SBID), continuum flux densities at 943.5\,MHz from EMU (S$_{\rm 943.5}$), variance flux (S$_{\rm v}$), scintillation bandwidths ($\nu_{\rm diss}$) and timescales ($\tau_{\rm diss}$) of detected known and new pulsars at $\sim$1 GHz, variability (R) defined as the ratio of variance flux (S$_{\rm v}$) to continuum flux densities (S$_{\rm 943.5}$), and circular polarisation fractions (V/I) if detected.}     
\centering     
\setlength{\tabcolsep}{8pt}
\renewcommand{\footnoterule}{}  
\begin{tabular}{l c c c c c c c c c c}
\hline\hline                   
PSR & RAJ & DECJ & SBID & S$_{\rm 943.5}$ & S$_{\rm V}$ & $\nu_{\rm diss}$ & $\tau_{\rm diss}$ & R & |V|/I \\
 &  &  &  & (mJy) & (mJy) & (MHz) & (mins) & & (percent) \\
\hline

J0255--5304 & 02:55:56.28 & --53:04:20.46 & 49990 & 12.96 & 4.78 & 4.20 & 4.20 & 0.36 & \ldots \\

J0536--7543 & 05:36:30.70 & --75:43:54.00 & 46978 & 31.704 & 14.37 & 4.35 & 5.74 & 0.45 & 9 \\

J0904--7459$^{c}$ & 09:04:10.58 & --74:59:42.19 & 72176 & 8.60 & 1.46 & 0.15 & 1.22 & 0.17 & 2 \\

J1405--4656 & 14:05:51.40 & --46:56:01.31 & 62581 & 1.15 & 0.65 & 10.05 & 5.43 & 0.56 & \ldots \\

J1456--6843 & 14:55:59.80 & --68:43:40.00 & 50539 & 152.36 & 51.92 & 34.28 & 14.60 & 0.34 & 2 \\
J1704--6016 & 17:04:16.82 & --60:19:34.83 & 74230 & 14.84 & 3.02 & 0.11 & 1.31 & 0.20 & \ldots \\
J1732--5049$^{c}$ & 17:32:47.76 & --50:49:00.32 & 63789 & 3.06 & 0.71 & 0.09 & 1.46 & 0.23 & \ldots \\

J1757--5322 & 17:57:15.12 & --53:22:27.58 & 63789 & 1.40 & 0.62 & 0.84 & 6.58 & 0.44 & 18 \\

J1833--6023$^{c}$ & 18:33:14.84 & --60:23:04.43 & 74275 & 2.41 & 0.54 & 0.35 & 2.71 & 0.22 & 5 \\
\hline
ULP & & & & & & & & & \\
\hline
J1448--6856 & 14:48:34.00 & --68:56:44.00 & 50539 & 0.75 & 1.35 & \ldots & \ldots & 1.8 & 29 \\
\hline
Radio star & & & & & &  &  & & \\
\hline
J1200--4929 & 12:00:36.83 & --49:29:26.96 & 64419 & 1.20 & 0.61 & \ldots & \ldots & 0.48 & 25 \\
\hline
Radio star candidate & & & & & & &  & & \\
\hline
J0919--7738$^{a}$ & 09:19:22.90 & --77:38:42.00 & 72176 & 1.67 & 0.88 & \ldots & \ldots & 0.53 & 43 \\
J1255--5133$^{a}$ & 12:55:48.20 & --51:33:39.00 & 70283 & 0.70 & 0.62 & \ldots & \ldots & 0.89 & 74 \\
J1813--6047$^{a,b}$ & 18:13:35.54 & --60:47:20.12 & 74275 & 0.80 & 0.78 & \ldots & \ldots & 0.98 & 63 \\
\hline
Pulsar candidate & & & & & & & & & \\
\hline
J0927--7641$^{b}$ & 09:27:40.60 & --76:41:35.75 & 72176 & 4.63 & 1.43 & 0.98 & 1.50 & 0.31 & 7 \\
J1209--5242$^{a,b}$ & 12:09:29.83 & --52:42:38.84   & 64419 & 4.92 & 1.11 & \ldots & \ldots & 0.22 & \ldots\\
J1210--4634$^{b}$ & 12:10:13.00 & --46:34:44.26 & 72194 & 4.62 & 1.01 & \ldots & \ldots & 0.22 & \ldots \\
J1212--5252$^{b}$ & 12:12:46.10 & --52:52:46.00 & 64419 & 4.06 & 1.00 & \ldots & \ldots & 0.25 & \ldots \\
J1212--5309$^{a,b}$ & 12:12:31.53 & --53:09:52.49 & 64419 & 4.41 & 1.02 & \ldots & \ldots & 0.23 & \ldots \\
J1838--5949$^{b,c}$ & 18:38:44.74 & --59:49:02.12 & 74275 & 2.68 & 0.46 & 0.36 & 5.07 & 0.17 & 9 \\
J2223--0654$^{b}$ & 22:23:41.24 & --06:54:41.18 & 61946 & 1.97 & 0.67 & 3.75 & 8.87 & 0.34 & 57 \\
\hline
\end{tabular}
\label{tab:tab2}

{$a$: With potential optical counterpart. }
{$b$: Observed with Murriyang. }
{$c$: Not picked up by our selection criteria.}
\end{table*}

\subsection{Candidate verification}
The candidate selection criteria described in Section~\ref{sec:select} are shown to be highly efficient and effective. Several known pulsars stand out as outliers showing the highest level of variability. In each of our EMU tiles, we are able to reduce the number of candidates to less than five from a large number of compact sources (see Table~\ref{tab:tab1}). 

To verify pulsar candidates, we produced a dynamic spectrum for each of these candidates by measuring the flux density in each fine time and frequency resolution image. Examples of dynamic spectra for pulsars and pulsar candidates are shown in Fig.~\ref{fig_ds1} and Fig.~\ref{fig_DScand}, respectively. These dynamic spectra allow us to select candidates whose variability is most likely caused by scintillation rather than steep spectrum or imaging artifacts. 
%


{
Finally, we searched for multi-wavelength counterparts of each candidate. Sources that have been classified as active Galactic nuclei (AGN) were excluded as pulsar candidates. The presence of circular polarisation strongly suggests a pulsar or a radio star, and when combined with optical information, can help distinguish between the two. }

\begin{table}
\begin{center}
\caption{Measured and derived parameters of pulsar discoveries with current timing observations. }
\label{tab:psr}
\begin{tabular}{lccc}
\hline
\hline
\multicolumn{4}{c}{Measured parameters}     \\
\hline
RA (J2000)        & 09:27:40.7  & 18:38:44.7 & 22:23:41.1(1)          \\
DEC (J2000)       & $-$76:41:35.8 & $-$59:49:02.1 & $-$06:54:39(4)        \\
$\nu$ (Hz)         & 182.061310(2) & 67.4404130  & 11.6676520263 (2)        \\
EPOCH (MJD)       & 60909.983240 & 60946.501383 & 60525.7367852927    \\
Time span (MJD)    &   &  & 60525.74$-$60889.79   \\
DM (cm$^{-3}$\,pc) &  29.52(1) & 39.0(4) &  19.36(7)          \\
\hline
\multicolumn{4}{c}{Derived parameters}\\
\hline
$P$ (ms)                  &  5.49265521(6) & 14.82790443 & 85.707046949(2) \\
$\dot{P}$ (s\,s$^{-1}$)  &   &  & 1.17(3)$\times10^{-18}$\\
$\tau_{\rm c}$ (Gyr)     &   &  & 1.2 \\
$B_{\rm s}$ (G)        &   &  & 1.0$\times10^{10}$  \\
\hline
\end{tabular}
\end{center}
\end{table}

\begin{figure*}
\centering
\begin{subfigure}{0.45\textwidth}
\includegraphics[width=1\linewidth]{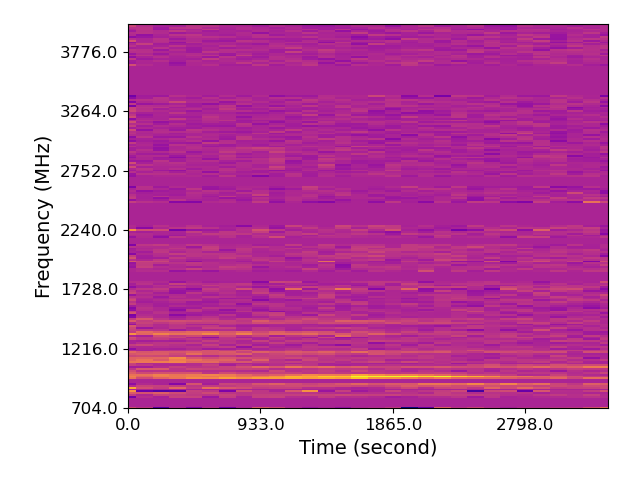}
\caption{J0927--7641, MJD=60909}
\end{subfigure}
\begin{subfigure}{0.45\textwidth}
\includegraphics[width=1\linewidth]{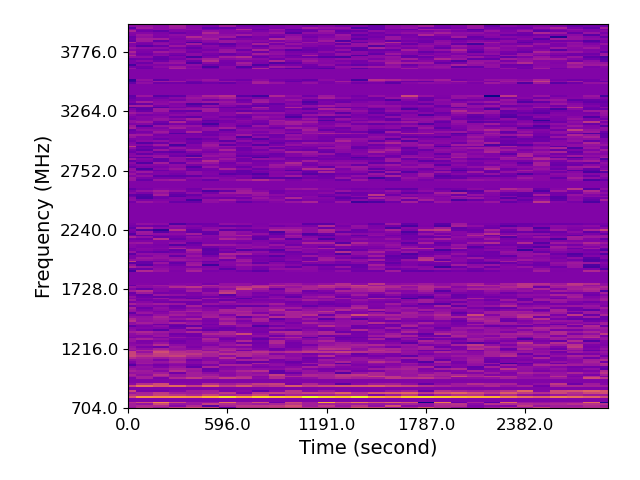}
\caption{J0927--7641, MJD=60925}
\end{subfigure}
\begin{subfigure}{0.45\textwidth}
\includegraphics[width=1\linewidth]{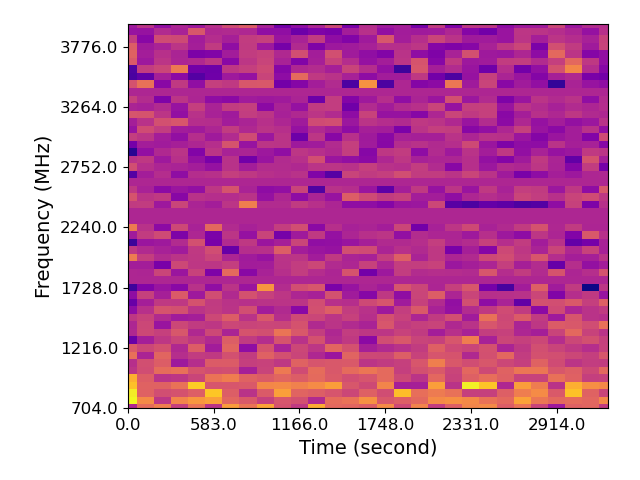}
\caption{J1838--5949, MJD=60936}
\end{subfigure}
\begin{subfigure}{0.45\textwidth}
\includegraphics[width=1\linewidth]{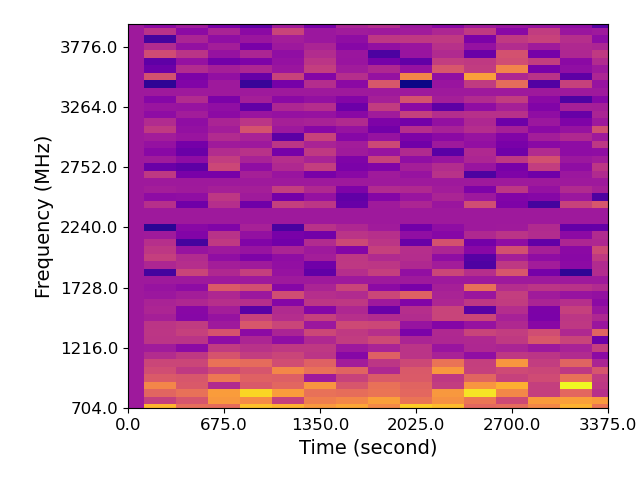}
\caption{J1838--5949, MJD=60946}
\end{subfigure}
\begin{subfigure}{0.45\textwidth}
\includegraphics[width=1\linewidth]{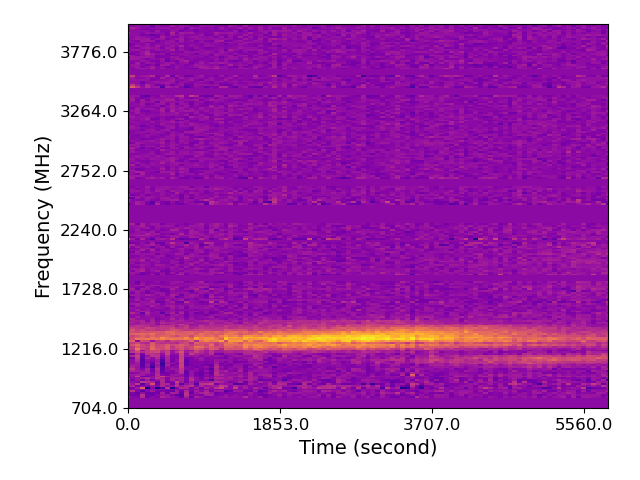}
\caption{J2223--0654, MJD=60573}
\end{subfigure}
\begin{subfigure}{0.45\textwidth}
\includegraphics[width=1\linewidth]{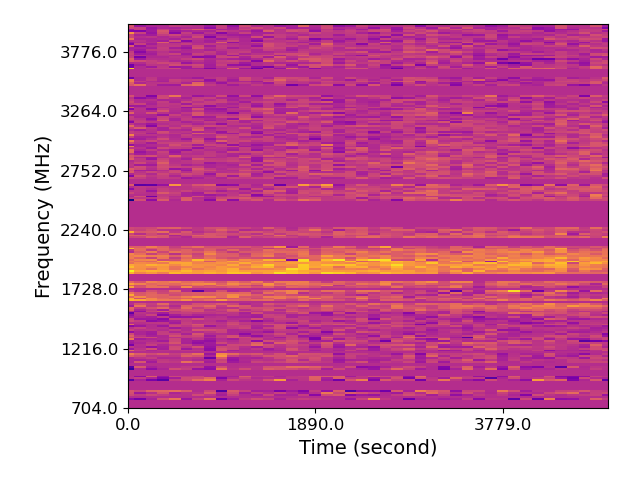}
\caption{J2223--0654, MJD=60643}
\end{subfigure}
\caption{Murriyang dynamic spectra of PSRs~J0927--7641, J1838--5949 and J2223--0654. 
}
\label{prof}
\end{figure*}

\begin{figure*}
\begin{center}
\includegraphics[width=0.33\linewidth]{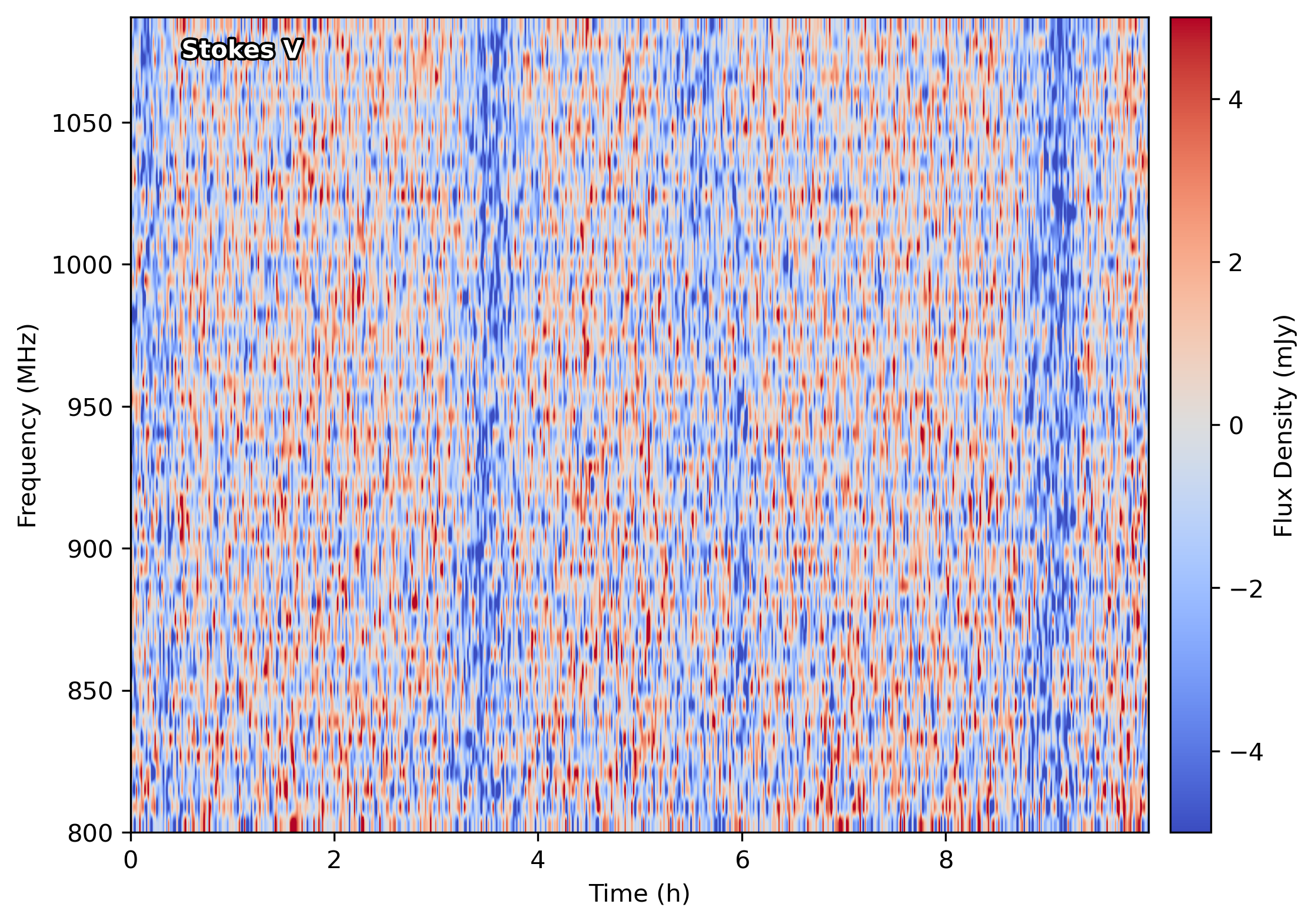}
\includegraphics[width=0.33\linewidth]{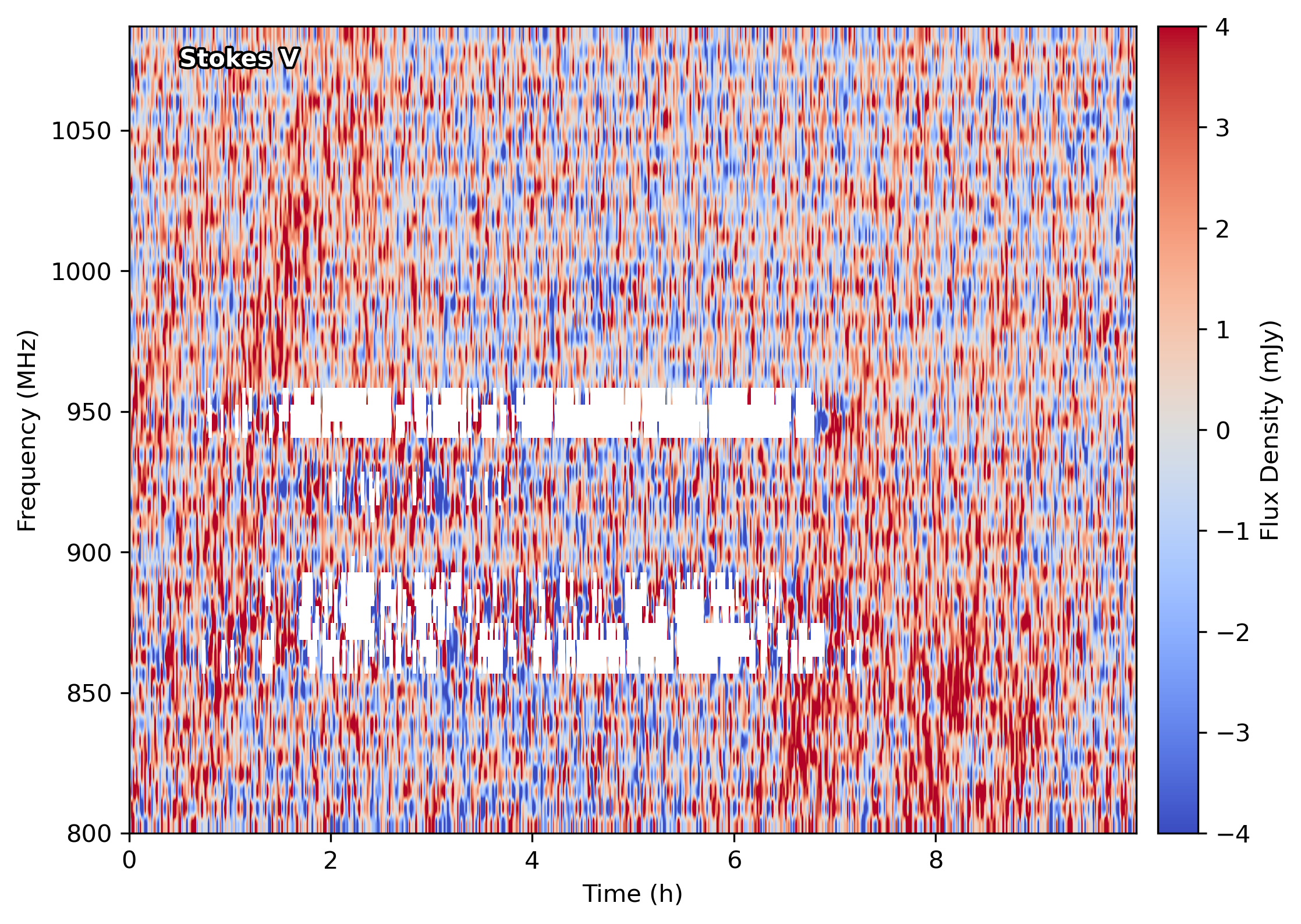}
\includegraphics[width=0.33\linewidth]{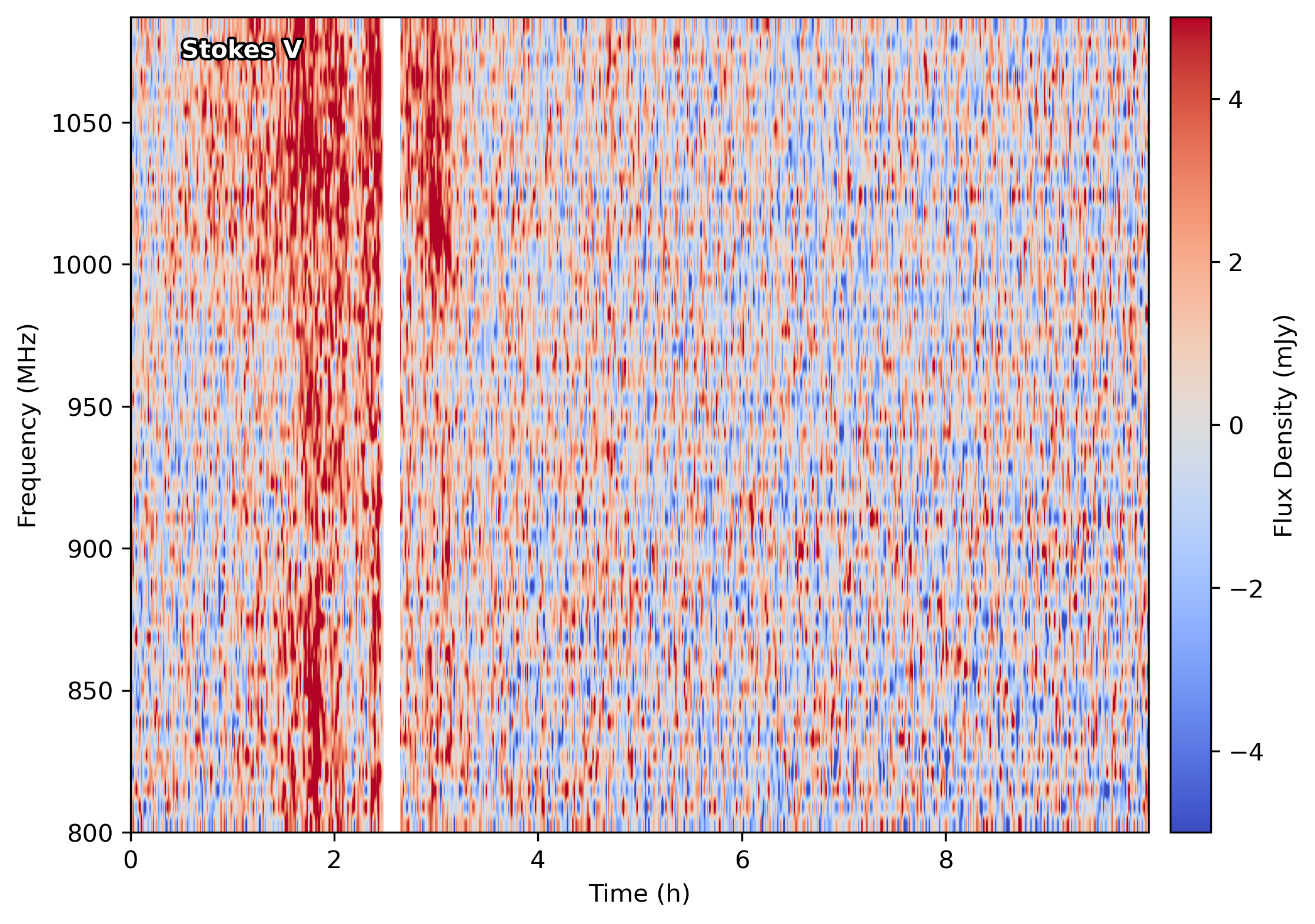}
\end{center}
\caption{Dynamic spectra of the circular polarisation of radio star candidates: J1813--6047 (left), J1255--5133 (middle), J0919--7738 (right).}
\label{fig:star}
\end{figure*}

\section{Results}\label{results}
\subsection{Detection of known pulsars and other radio sources}

Out of 27 known pulsars detected in all analysed EMU tiles, we successfully detected nine pulsars in their corresponding variance images. Their continuum flux density at 943.5\,MHz (S$_{\rm 943.5}$), variance flux (S$_{\rm V}$), and Variability (R) are given in the Table~\ref{tab:tab2}. In Fig.~\ref{fig_ds1}, we show their detection in variance images (left panel) and dynamic spectra (right panel). These dynamic spectra span a duration of 10\,hours, except the SB 61946, with a total observing duration of five hours (last panel in Fig.~\ref{fig_DScand}). The scintillation bandwidths $\nu_{\rm d}$ and timescales $\tau_{\rm d}$ of three known pulsars (J0255--5304, J0536--7543, and J1456--6843) have been previously measured \citep{johnston98}. {For PSRs~J0536--7543 and J1456--6843, the scintillation bandwidth and timescale observed in EMU (at $\sim$1 GHz) are given in Table~\ref{tab:tab2} and generally consistent with previous measurements. In contrast, PSR~J0255--5304 exhibits a significantly broader scintillation bandwidth and longer timescale.}
The continuum source J1704--6019 was detected at RAJ = 17:04:16.82, DecJ = --60:19:34.83. The positional offset between the continuum source J1704--6019 and the timing position of PSR~J1704$-$6016 has already been reported by \citet{wng+23}. The pronounced variability of the source is consistent with DISS at a DM of $\sim50$\,cm$^{-3}$\,pc, reinforcing the association of the continuum source with PSR~J1704$-$6016. 

In addition to known pulsars, we also detected the known long period transient (LPT) source J1448--6856~\citep{akr+25}, as well as one radio star. Along with pronounced temporal variability, J1448--6856 exhibits significant frequency-dependent variability, making it stand out clearly in the variance image.
{The detected radio star (J1200--4929) is listed in the Sydney Radio Star Catalogue\footnote{\url{https://radiostars.org/data/}} and was independently identified in previous studies through variability and circular polarisation searches \citep{lnd+24}. Their re-detection further supports the reliability of variance imaging in uncovering highly variable stellar radio sources.}
\subsection{Pulsar discoveries}
A strong pulsar candidate (J2223--0654) was identified in beam 05 of the tile SB 61946 (see Fig.~\ref{fig_var}). The dynamic spectrum of this candidate shows clear evidence of DISS. { The continuum source is $\sim60\%$ circularly polarised with no optical counterpart.} We observed this candidate using the UWL receiver on Murriyang, the Parkes radio telescope (see Section~\ref{sec:pks} for details) and detected a periodic signal with a period of $\sim$85.7\,ms and a DM of 19.4\,pc\,cm$^{-3}$. Follow-up observations with Murriyang confirmed the discovery of an isolated pulsar (hereafter designated as PSR\,J2223--0654), and an initial timing solution has been obtained (see Table~\ref{tab:psr}).

The wideband dynamic spectra of PSR~J2223--0654 at two different Murriyang observing epochs are shown in Fig.~\ref{prof}. These observations at different epochs show that the flux density of the pulsar varies significantly across a wide band (704 to 4032\,MHz) and exhibits clear frequency-dependent intensity variations, which is a characteristic of strong DISS. 
For example, in one observation (Fig.~\ref{prof}e), the pulsar can only be detected within a narrow spectral window ($\sim$1000--1472\,MHz), while in another (Fig.~\ref{prof}f), it appears to be brighter at higher frequencies and within a wider bandwidth ($\sim$1500--2250\,MHz). 

Another pulsar candidate (J0927--7641) was identified in beam 32 of the tile SB 72176. It exhibits strong variability ($>30\%$) and is clearly distinct from the other sources in the beam (Fig.~\ref{fig_var}). The continuum source is also circularly polarised ($|V|/I\approx7$\%) and has no optical counterpart. We observed this candidate with Murriyang using the same observational setup as described above and discovered a pulsar with a spin period of 5.492\,ms and a DM of 29.5\,cm$^{-3}$\,pc. The wideband dynamic spectrum of PSR~J0927--7641 is shown in Fig.~\ref{prof}a and Fig.~\ref{prof}b, which shows strong scintillation at $\sim1$\,GHz. Whether this is an isolated MSP or an MSP in a binary system is to be determined. So far, we have conducted four observations (at MJD 60910, 60913, 60920, and 60925), during which we observed an almost linear decrease in the pulsar's measured spin period. This indicates that PSR~J0927--7641 is likely in a binary system with an orbital period on the order of several tens of days. Follow-up observations of this pulsar are underway to measure its orbital parameters and obtain a coherent timing solution. 
%

A third pulsar candidate (J1838--5949) was identified in beam 14 of tile SB74275. With targeted observations using Murriyang, we discovered a pulsar with a spin period of 14.828\,ms and a DM of 39.0\,cm$^{-3}$\,pc. Unlike PSRs~J0927--7641 and J2223--0654, J1838--5949 did not meet the $5\sigma$ variability threshold. Instead, it was identified as a candidate because it was detected in both the variance and circular polarisation images (Section~\ref{sec:select}). The associated continuum source is approximately 9\% circularly polarised. Follow-up observations with Murriyang (Fig.~\ref{prof}c and d) indicate that J1838--5949 is not strongly scintillating. Rather, it is likely a pulsar with an extremely steep radio spectrum, consistent with its dynamic spectrum observed in the EMU data. Current Murriyang observations also suggest that J1838--5949 is likely in a binary system, and follow-up observations are ongoing.

To estimate the scintillation bandwidth ($\nu_{\rm diss}$) and the scintillation time-scale ($\tau_{\rm diss}$), we calculated the expected scattering time ($\tau_{\rm s}$) using the relation obtained by \cite{krishna15} and scaled to 1\,GHz assuming $\tau_{\rm s}$ $\approx$ $\nu^{\rm -4}$ dependence:
\begin{equation}
\label{eqn:eqn5}
    \tau_{\rm s,1\,GHz} (\mu\,\rm s) = 4.1 \times 10^{-6} DM^{2.2} (1 + 0.00194 DM^{2}),
\end{equation}
Assuming a homogeneous medium with a Kolmogorov spectrum \citep{cordes98}, $\nu_{\rm diss}$ can be obtained by,
\begin{center}
\begin{equation}
\label{eqn:eqn6}
  2\pi\nu_{\rm diss}\tau_{\rm s}=C_{1},
\end{equation}
where $C_{1}=1.16$ for a uniform medium with a Kolmogorov wavenumber spectrum.
\end{center}
The scintillation time-scale, $\tau_{\rm diss}$, can be estimated as~\citep{johnston98},
\begin{equation}
\label{eqn:eqn7}
  \tau_{\rm diss} = \frac{3.85 \times 10^{4} \sqrt{D \nu_{\rm diss}}}{\nu V} ,
\end{equation}
where $\nu$ is the observing frequency in GHz, and $V$ is the speed of the interstellar diffraction pattern relative to the Earth in units of km\,s$^{-1}$, dominated by the pulsar transverse velocity. Estimated scintillation bandwidth and timescale for all known and new pulsars are presented in Table~\ref{tab:tab2}. For pulsars whose proper motions are not known, we have used a nominal transverse velocity of 100\,km\,s$^{-1}$. Scintillation parameters of PSRs~J0255--5304, J0536--7543, and J1456--6843 measured by \citet{johnston98} were quoted in Table~\ref{tab:tab2}.
The scintillation bandwidths of PSRs~J0927--7641 and J2223--0654 are broadly consistent with the scintillation structure observed in EMU at 943.5\,MHz.

\subsection{Radio star candidates}

Three of our candidates, J0919--7738, J1255--5133 and J1813--6047, exhibit strong circular polarisation and are associated with optical counterparts. For J1813--6047, a possible periodic double-peaked structure with a period of $\sim5.8$\,hr is evident in the dynamic spectrum of the circular polarisation (Fig.~\ref{fig:star}), suggesting it may be a flaring star. J0919--7738 and J1255--5133 also show clear variability in their dynamic spectra, although no periodicity can be established with the current data. 
A detailed investigation of the optical companions to these candidate radio stars is beyond the scope of this work. Follow-up radio observations will be necessary to better characterise the nature of their radio emission.

{Although unlikely to be a pulsar, we conducted follow-up observations of candidate J1813--6047 using Murriyang. A single observation of J1813--6047 was carried out with an integration time of 2204\,s.
No periodic signal was detected from the source across a DM range of 0 to 500\,cm$^{-3}$\,pc. }

\subsection{Other pulsar candidates}

In addition to PSRs~J0927--7641 and J2223--0654, we conducted Murriyang follow-up observations of pulsar candidates: J1209--5242, J1210--4634, J1212--5252 and J1212--5309. The observational setup and pulsar searching strategy were the same as described above, with an integration time of approximately 1\,hr. No periodic signals were detected from these sources over a DM range of 0--500\,cm$^{-3}$\,pc. Compared with PSRs~J0927--7641 and J2223--0654, these pulsar candidates show lower variabilities, although DISS-like structures can be observed in their dynamic spectra (Fig.~\ref{fig_DScand}). Repeated observations might be necessary to reveal their periodic signals.

\section{Discussion and conclusion}\label{discuss}

In this pilot study, we applied the variance imaging technique to EMU data observed at high Galactic latitude regions, where pulsars are expected to have lower DM and broader scintillation bandwidth and timescales, enabling the creation of variance images using a relatively small number of frequency channels and time integrations. 
{Within the total FoV, nine of 27 known pulsars detected in continuum images were detected in variance images.}
The targeted pulsation search of six compact, highly scintillating pulsar candidates identified through variance imaging led to the discovery of two new pulsars: {one with a period of 85.707\,ms and a DM of 19.4\,cm$^{-3}$\,pc, and another with a period of 5.492\,ms and a DM of 29.5\,cm$^{-3}$\,pc.}. Additionally, a third pulsar was discovered in the variance images, with a period of 14.828\,ms and a DM of 39.0\,cm$^{-3}$\,pc. This source shows a steep radio spectrum and a high degree of circular polarisation. 

Our discoveries demonstrated the effectiveness of variance imaging in distinguishing highly scintillating pulsars from other compact radio sources. This method eliminates the need for pixel-by-pixel blind pulsar surveys with either single dishes or tied-array beams, as well as the computationally intensive processing of large volumes of high-time-resolution data. Instead, it enables efficient, targeted deep and wideband searches that maximize the likelihood of detecting elusive pulsars such as those with large duty cycles, strong scintillation, or in compact binary systems. Beyond pulsars, variance imaging also holds promise for identifying ULP candidates and highly variable stellar radio sources (e.g., flare stars).

The next step is to apply the variance imaging technique across the full EMU sky at intermediate and high Galactic latitudes. We also plan to generate variance images at both higher and lower time and frequency resolutions to explore a broader range of DM.
In addition to lower frequency and time resolution required to create variance images at high Galactic latitudes, we expect lower sky temperature and reduced source confusion due to the lack of extended radio emission, which further simplifies the process and enhances the sensitivity. 
%
%
This potentially enables the detection of MSPs and binaries, which were likely missed by previous shallow and single-epoch pulsar surveys at high Galactic latitudes \citep[e.g.,][]{keith10,kbj+18}. 

In the Galactic plane, the dense ISM leads to high DMs, which reduce scintillation bandwidths and timescales, often below the spectral and temporal resolution of a given continuum survey. This suppresses detectable variability due to DISS at around 1\,GHz, limiting the effectiveness of variance imaging for typical high-DM pulsars. Additionally, the crowded sky and diffuse radio emission (for e.g., SNRs, PNe, HII regions) increase the likelihood of false positives due to image artifacts and source confusion.
%
%
High-frequency continuum surveys such as the Very Large Array Sky Survey \citep[VLASS, 2--4\,GHz;][]{lacy+20}, Deep Synoptic Array 2000-antenna \citep[DSA-2000, 0.7--2\,GHz;][]{dsa+19}, and future SKA-mid surveys \citep{ska+15} offer better avenues for pulsar discoveries with variance imaging in the Galactic plane.

While our pilot survey demonstrated that variance imaging and our source selection strategy are effective and efficient for identifying pulsars, it also revealed several challenges.
A key challenge in variance imaging is that snapshot images are constructed across multiple frequency channels and time integrations, each associated with a different point spread function (PSF) due to the frequency- and time-dependent nature of the uv-coverage. 
%
As a result, compact sources may appear elliptical and rotate across successive snapshots. When pixel-wise variance is computed from such images, these PSF variations can lead the sources to appear ring-like or winged-structured in the variance image, often characterised by suppressed variance at around the source central pixels and elevated variance in the outer wings. 
%
%
Such morphological artifacts can result in false positives and reduce our sensitivity to faint sources. {Correcting for the time- and frequency-dependent PSF is challenging, as it varies in complex ways due to flagging and baseline projection. In future pipelines, we plan to test convolving snapshot images to a common beam and assess its effectiveness for identifying scintillating sources compared with our current approach.}

Another critical factor affecting the fidelity of variance images is the presence of bright sources, either within or outside the primary beam. Even when a bright source lies outside the imaged tile, its response can leak into the snapshot images through the sidelobes of the primary beam. {Although these sidelobes are not expected to scintillate over time, they can exhibit strong frequency-dependent variations}, introducing spurious compact and extended structures in the variance image that could be interpreted as artificial scintillation/variable sources (see Fig.~\ref{fig_psr}). 
Additionally, bright sources within the primary beam can also produce residual sidelobes if not accurately modelled and subtracted from the visibilities. These residual sidelobe patterns introduce artefacts in the variance image around bright sources and can act as false positives. 
%
{To address this, we are developing additional steps in our pipeline. For bright out-of-beam sources, we will apply the source peeling technique \citep{wiliam+19} to subtract their contributions from the visibilities, with source locations identified from shallow surveys such as the Rapid ASKAP Continuum Survey \citep[RACS;][]{racs20, racs+25}. We also plan to apply phase and amplitude self-calibration to mitigate sidelobes from bright in-beam sources.}

%
%
%

Finally, we plan to transition from CASA's tclean to the WSClean imager \citep{ofringa+14}, which offers significant computational advantages. By replacing the w-projection algorithm (tclean) with the w-stacking CLEAN algorithm (WSClean), we anticipate three major benefits: (1) improved scalability and multinode parallelism for large datasets, (2) execution efficiency and reduced processing times, particularly in the deep sky model and snapshot imaging stages, and (3) high fidelity images with reduced noise \citep[e.g.,][]{tao+23}.

\begin{acknowledgement}
We thank Joshua Pritchard, Andrew Zic, and Tim Galvin for the useful discussions. Murriyang, CSIRO's Parkes radio telescope, is part of the Australia Telescope National Facility\footnote{\url{https://www.atnf.csiro.au}} (ATNF) which is funded by the Australian Government for operation as a National Facility managed by the Commonwealth Scientific and Industrial Research Organisation (CSIRO). 
The ATNF Pulsar Catalogue (\url{https://www.atnf.csiro.au/research/pulsar/psrcat/}) was used for this work.
%
This scientific work uses data obtained from Inyarrimanha Ilgari Bundara, the CSIRO Murchison Radio-astronomy Observatory. We acknowledge the Wajarri Yamaji People as the Traditional Owners and native title holders of the Observatory site. CSIRO’s ASKAP radio telescope is part of the Australia Telescope National Facility (\url{https://ror.org/05qajvd42}). Operation of ASKAP is funded by the Australian Government with support from the National Collaborative Research Infrastructure Strategy. ASKAP uses the resources of the Pawsey Supercomputing Research Centre. Establishment of ASKAP, Inyarrimanha Ilgari Bundara, the CSIRO Murchison Radio-astronomy Observatory and the Pawsey Supercomputing Research Centre are initiatives of the Australian Government, with support from the Government of Western Australia and the Science and Industry Endowment Fund.
The data processing was performed on the OzSTAR\footnote{\url{https://supercomputing.swin.edu.au/ozstar}} national facility at Swinburne University of Technology. The OzSTAR program receives funding in part from the Astronomy National Collaborative Research Infrastructure Strategy (NCRIS) allocation provided by the Australian Government, and from the Victorian Higher Education State Investment Fund (VHESIF) provided by the Victorian Government.
\end{acknowledgement}

\paragraph{Data Availability Statement}
The observations from Murriyang are publicly available from \url{https://data.csiro.au/domain/atnf} after an 18 month embargo period. This study made use of archival ASKAP data obtained from the CASDA \url{https://data.csiro.au/domain/casdaObservation}.  
\printendnotes
\printbibliography
\end{document}